\documentclass{aa}  

\usepackage{graphicx}
\usepackage{txfonts}
\usepackage{caption}
\usepackage{color}
\usepackage{natbib}
\bibpunct{(}{)}{;}{a}{}{,} 

\begin{document} 

\newcommand{\MM}[1]{\textsf{M{#1}}}
\newcommand{\CC}[1]{\textsf{C{#1}}}
\newcommand{\FF}[1]{\textsf{F{#1}}}
\newcommand{\DD}[1]{\textsf{D{#1}}}

\title{Spectral shifting strongly constrains molecular cloud disruption by radiation pressure on dust}

\author{Stefan Reissl\inst{\ref{inst1}} \and Ralf S. Klessen \inst{\ref{inst1},\ref{inst2}}  \and Mordecai-Mark Mac 
Low\inst{\ref{inst1},\ref{inst3}} \and Eric W. Pellegrini \inst{\ref{inst1}} 
}

\institute{
Universit\"{a}t Heidelberg, Zentrum f\"{u}r Astronomie, Institut f\"{u}r 
Theoretische Astrophysik, Albert-Ueberle-Str. 2, 69120 Heidelberg, Germany 
\label{inst1}
\and
Universit\"{a}t Heidelberg, Interdisziplin\"{a}res Zentrum f\"{u}r 
Wissenschaftliches Rechnen, Im Neuenheimer Feld 205, 69120 Heidelberg, Germany 
\label{inst2}
\and
Department of Astrophysics, American Museum of Natural History, Central Park 
West at 79th Street, New York, NY 10024-5192, USA \label{inst3}
}
                                                

\abstract{
{\bf Aim:} 
We aim to test the hypothesis that radiation pressure from young star clusters
acting on dust is the dominant feedback agent disrupting the largest
star-forming molecular clouds and thus regulating the star-formation
process.
{\bf \\ Methods:} We performed multi-frequency, 
3D, radiative transfer calculations including both scattering and
absorption and re-emission to longer wavelengths for model clouds with masses of 
$10^4$--$10^7\,$M$_{\odot}$, containing embedded clusters with 
star formation efficiencies of 0.009\%--91\%, and varying 
 maximum
grain sizes up to $200\,\mu$m. We calculated the ratio between 
radiative and gravitational forces to determine whether 
radiation pressure can disrupt clouds. 
{\bf \\ Results:} We find that radiation pressure acting on dust almost never 
disrupts star-forming clouds. Ultraviolet and optical photons from young stars 
to which the cloud is optically thick do not scatter much. Instead, they 
quickly get absorbed and re-emitted by the dust at thermal wavelengths. 
As the cloud is typically optically thin to far-infrared 
radiation, it promptly escapes, depositing little momentum in the cloud.
The resulting 
spectrum is more narrowly peaked than the corresponding Planck function, and 
exhibits an extended tail at longer wavelengths. As the 
opacity drops significantly across the sub-mm and mm wavelength regime, the 
resulting radiative force is 
   even 
smaller than for the corresponding single-temperature 
blackbody. 
We find that the force from radiation pressure falls below the
strength of gravitational attraction by
   an order of magnitude or more
   for either Milky Way or moderate starbust conditions.
Only for unrealistically large
   maximum
grain sizes, and star 
formation efficiencies far exceeding 50\% do we find that the strength of 
radiation pressure can exceed gravity. 
        {\bf \\ Conclusions:} We conclude that radiation pressure acting on dust 
    does not
disrupt star-forming molecular clouds
in 
   any Local Group galaxies.
Radiation pressure 
    thus appears unlikely to
regulate the star-formation 
process on either local or global scales. 
}
  \maketitle
%

\section{Introduction}
\label{sec:Intro}
Identifying and characterizing the physical processes that control the 
formation 
of stars on local scales within molecular clouds as well as on global scales of 
galaxies as a whole is one of the key open problems in astronomy and 
astrophysics \cite[see, e.g., the reviews by][]{maclow2004, mckee2007, 
klessen2016}. A convincing and comprehensive answer is yet to be found.
We know that stars form only within the densest and coldest regions of the 
complex, multi-phase, interstellar medium (ISM). Observations indicate that 
only 
a small fraction of the available gas mass actually gets converted into stars: 
the star formation efficiency
   per free-fall time 
in molecular clouds is very low, typically of 
order of a few percent. A similar conclusion holds for galactic scales, where 
the overall gas depletion timescale is between one and two orders of magnitude longer 
than the dynamical time scale of star-forming clouds, again indicating the 
processes that convert ISM gas into stars are relatively inefficient. We do not 
fully understand why this is the case.  
At first sight, this low efficiency is surprising, because molecular
clouds are massive and cold, with masses of
$10^3$--$10^6\,$M$_{\odot}$ and temperatures in the range
10--30~K \cite[e.g.,][]{blitz2007}. Therefore, when considering only
 the competition between self-gravity and gas pressure
\citep{jeans1902}, we find that the dense parts of the ISM should be
highly unstable against gravitational contraction and should form
stars efficiently and on short timescales. These are comparable to the sound
crossing or free-fall times of the cloud. This is not
observed. Consequently, additional physical agents have been suggested
to strongly delay star formation. One of these is the momentum input from stellar radiation \citep[e.g.,][]{tenorio1988}. \\
The same problem occurs on galactic scales, where observations indicate gas 
depletion times, or equivalently the inverse of star formation rate, is rather 
large ranging, from several hundred million years to several tens of 
billion of years  (see e.g., \citealt{bigiel2008}; \citealt{leroy2008}; 
\citealt{bigiel2011}, for a critical discussion, see \citealt{shetty2013, 
shetty2014}). This is significantly longer than the typical 
dynamical timescale in a galaxy. Again, we arrive at the conclusion that star 
formation on galactic scales is inefficient, converting only a few percent of 
the available gas reservoir into stars in each dynamical timescale, so there must 
be some physical agents that are causing this behavior, possibly the same ones 
responsible for inefficient star formation on local cloud scales.\\
The question remains whether stellar feedback mechanisms can do the job or not.
The proposal is that the energy and momentum input from stars into the ISM
can regulate stellar birth within an individual star-forming
region as well as within a galaxy as a whole, and that this can explain
the observed low efficiency. Stars can affect their environment in
several ways including supernovae, winds, and radiative pressure 
(\citealt{maclow2004,mckee2007,bp2007,klessen2016}; also 
\citealt{hill2012,  hennebelle2014}, \citealt{walch2015}, 
\citealt{girichidis2016} for high-resolution simulations).\\
 In this paper we focus on radiation pressure on dust grains. The energy in
the radiation field produced by a typical stellar population 
exceeds the mechanical energy input from stellar winds and supernovae
by an order of magnitude \citep{abbott1982}.  
The key question, however, is whether that radiation can effectively be converted into kinetic energy and counteract gravity to regulate the star formation process.\\
 Ultraviolet and optical photons carry momentum that can effectively be 
absorbed  and scattered by dust grains in the ISM 
\citep{tielens2010,draine2011}. In the 
case of young star-forming regions, the deposition of momentum as radiation is 
absorbed can be shown to produce a response in the ISM pressure consistent with 
rough hydrostatic equilibrium across expanding structures 
\cite[e.g.,][]{pellegrini2007,pellegrini2009}. One can draw a direct connection 
between gas densities of ${\rm H^+/H^0/H_2}$ interface regions measured with 
density sensitive nebular line ratios and stellar feedback. Because dust is 
dynamically coupled to the gas, the momentum transferred from photons to dust 
grains can drive gas
flows, and so can 
   cut off further accretion, as well as 
contribute to driving ISM turbulence. This momentum transfer can thus help to
stabilize molecular clouds on large scales against further collapse,
or even drive outflows, while at the  
same time it might trigger the build-up of new stars in compressed shells of 
swept up ISM material. 
It has been proposed that the momentum absorbed by 
dust can disrupt star-forming molecular clouds in a wide range of 
different environments, and that this sets the overall star formation 
efficiency. It has even been suggested that radiation pressure from massive 
stars and star clusters is the dominant feedback mechanism regulating the 
global star formation process in galaxies ranging from normal disk
galaxies like the Milky Way to starburst systems 
\citep{thompson2008,krumholz2009a,murray2010,andrews2011,krumholz2012}. 
Radiation pressure has also been invoked to explain the galactic fountains and outflows 
observed in many galaxies  \citep[see, e.g.,][]{thompson2005, murray2011,  
krumholz2012, zhang2012,coker2013, rosdahl2015, 
ishibashi2015,thompson2015,seo2016}.  
We note that the impact of radiation pressure has also been studied
for a wide variety of other systems and scales, such as accretion disks
around black holes \citep{begelman1978, gu2012}  high-mass stars
\citep{yorke2002, krumholz2009, krumholz2010, kuiper2010, tanaka2010,
  kuiper2013}, low-mass cores, filaments, or shell structures in the
ISM \citep{verdolini2013, ochsendorf2014, seo2016}, and 
galaxy formation in a cosmological context \citep{wise2012, moody2014,
  sales2014, hopkins2014}.  
Here, we critically assess the hypothesis that radiation pressure is
the main regulatory agent for star formation. 
Its effectiveness has already been called into question by
   \citet{2012ApJ...760..155K}, who showed that acceleration of dense
   gas by radiation pressure produces Rayleigh-Taylor instabilities
   that allow photon escape. \citet{davis2014}, however, argued that a
   more accurate treatment of the radiation transfer (RT) showed that
   outflows could still be driven.
We examine the detailed spectral evolution of escaping radiation
    rather than the dynamics, and 
conclude that the impact of radiation on molecular cloud evolution and global ISM
dynamics may still be
dramatically overestimated. We note that
similar conclusions have been reached, for example,  by
\citet{silich2013} and \citet{martinez2014}. 
In our study we focus on three key points: the efficiency of the coupling between radiation and matter; the fraction of the available radiative energy that can readily be converted into kinetic energy; and taken together, the environmental conditions under which radiation pressure can disrupt star-forming molecular clouds and thus contribute to the self-regulation of stellar birth.
When photons scatter, they just change 
direction, but keep their original energy. When a photon gets absorbed 
by a dust grain, on the other hand, reemission occurs at the 
temperature of the dust grain, in the thermal infrared. Stellar photons in the 
ultraviolet and optical bands are thus quickly shifted to much longer 
wavelength.  The dust absorption and 
scattering cross sections for ultraviolet or optical photons are large, 
so molecular clouds are generally optically thick to them. 
However, at lower energies and longer wavelength both cross 
sections drop dramatically, so eventually the cloud 
becomes transparent to the radiation. 
   The importance of this effect to correctly calculating the effect
   of radiation pressure on outflows was first appreciated by
   \citet{Habing1994} in the context of winds from cool giants.

The exact frequency at which this 
transition occurs depends on the actual absorption and scattering cross 
sections, which in turn are subject to the chemical composition and size 
distribution of dust grains, and on the density, velocity and temperature 
structure of the cloud out of which the star or star cluster has formed. 
The shift of photons to longer wavelengths as they interact with dust 
inevitably 
causes them to reach a wavelength at which the remaining column
density out of the cloud is transparent, 
at which point they escape.  The question of how quickly this occurs ultimately 
determines the answers to the above questions of the impact of radiation 
pressure on the dynamics of the ISM and the self-regulation of star formation. 
We have approached this problem by means of three-dimensional (3D), RT 
simulations. We scan a wide range of cloud and cluster masses, star 
formation efficiencies, and dust size distributions,
with specific focus on the detailed 
microphysics of the interplay between radiation and matter. Our models
are applicable to a variety of star-forming environments, ranging
from small groups of stars forming in low-mass clouds similar to those observed 
in the solar neighborhood up to very massive clusters embedded in
high-mass clouds representative of the most extreme star
formation conditions seen in the Local Group.        
In Section~\ref{sect:Method} we introduce our one-dimensional cloud and cluster 
radial profiles. Then, we describe how we calculated the forces acting on different 
radial shells in the cloud. As we focus on the detailed interplay between radiation 
and dust, we provide the details of our dust model and describe our method to 
self-consistently model absorption and scattering processes using our 
Monte-Carlo radiative transfer code {\sc Polaris}. We consistently account for 
the shift of the spectral energy distribution toward longer wavelengths as the 
radiation propagates outwards through the cloud.
Our results are presented in Section~\ref{sect:Results}, where we compute  
the ratio of radiative to gravitational forces at 
any given cloud radius, for a wide range of model parameters, varying the mass 
and size of the molecular cloud as well as of the embedded cluster. We 
considered 
different  grain size distributions and dust temperature calculations. Our 
set of models 
covers conditions typical to the solar neighborhood and extends to more extreme 
starburst systems. In Section~\ref{sect:Discussion} we discuss observational 
implications, compare our models with the existing literature, and comment on 
the limitations of our approach. Finally, we summarize and conclude in 
Section~\ref{sec:summary}.

\section{Method}

\label{sect:Method}
\subsection{Initial conditions}
\subsubsection{Cloud model}

\begin{table*}[t]
  \centering
    \begin{tabular}{lcccccc}

      \multicolumn{6}{c}{\bf Molecular cloud models (mass)} \\
      \hline 
      Cloud mass [$\rm{M_{\odot}}$] & $10^{4}$ & $10^{5}$ & $10^{6}$ & $10^{7}$ 
&\\
      \hline
      Notation & {\MM4} & {\MM5} & {\MM6} & {\MM7} &\\
      \hline\\
      
      \multicolumn{6}{c}{\bf Cluster models (mass)} \\
      \hline
      Cluster mass [$\rm{M_{\odot}}$] & $10^{3}$ & $10^{4}$ & $10^{5}$ & 
$10^{6}$ & $10^{7}$ \\
      Cluster luminosity [$\rm{L_{\odot}}$]  & $1.02 \times 10^{6}$ & $1.02 
\times 10^{7}$ & $1.02 \times 10^{8}$ & $1.02 \times 10^{9}$ & $1.02 \times 
10^{10}$ \\
      \hline
      Notation & {\CC3} & {\CC4} & {\CC5} & {\CC6} & {\CC7} \\
      \hline\\
      
      \multicolumn{6}{c}{\bf Cluster models (concentration)} \\
      \hline
      FWHM [$\rm{pc}$] & 0 & 0.1 & 1 & 2 & 4 &  \\
      \hline
      Notation & {\FF0} & {\FF0.1} & {\FF1} & {\FF2} & {\FF4} &  \\
      \hline\\
      
      \multicolumn{6}{c}{\bf Dust models (grain size)} \\
      \hline
      $a_{\rm{max}}$ [$\rm{\mu m}$] & 0.02& 0.2 & 2.0 & 20.0 & 200.0 \\
      \hline
      Notation & {\DD1} & {\DD2} & {\DD3} & {\DD4} & {\DD5} \\
      \hline
      
    \end{tabular}
    
    \caption{ Physical quantities and notation of the applied models. 
Our clouds are spherically symmetric and follow a Plummer-like density profile 
(Eq. \ref{eq:CloudDensity}) with a core radius of 1\% of the outer radius, 
$R_{\rm c} = 0.01 R_{\rm{out}}$. 
    The standard cloud radius is
$R_{\rm{out}} = 5\,$pc, but we also consider values of 20$\,$pc and 150$\,$pc for 
selected cases.}
    \label{tab:parameter}
\end{table*}
We studied spherically symmetric clouds (cloud quantities are 
subscript MC)
with an outer 
radius of $R_{\rm{out}} \in [ 5, 20, 150]\,$pc. We adopted 
a \cite{Plummer1911} profile for the density
\begin{equation}
  \rho_{\rm{MC}}(r)  = \rho_{\rm{MC,0}} \left(
\frac{R_{\rm{c}}^2}{r^2+R_{\rm{c}}^2} \right)^{\eta/2}
\label{eq:CloudDensity}
,\end{equation}
where $r$ is the distance from the cloud's center, $R_{\rm{c}}$ fixes the size 
of the flat inner region, and the exponent $\eta$ controls the steepness of the 
power-law fall-off outside. This is a good description for many star-forming 
regions, which are frequently observed to be centrally condensed
\citep[e.g.,][]{lin2016}. We used values of $R_{\rm{c}} = 0.01
R_{\rm{out}}$ and  $\eta = 2$ typical for star-forming molecular clouds 
\citep[][]{Motte1998,Mueller2002,Pirogov2009,Motte2015,lin2016} for all models.
Since our models have a sharp outer radius $R_{\rm{out}}$, they contain a well 
defined cloud mass. Integrating 
Eq.~(\ref{eq:CloudDensity}), 
the mass enclosed within a certain radius $r$ is 
\begin{equation}
  M_{\rm{MC}}(r) =4\pi \rho_{\rm{MC}}R_{\rm{c}}^2 \left[ r -R_{\rm{c}} \arctan\left( \frac{r}{R_{\rm{c}}}\right)\right]\;.
\label{eq:CloudMass}
\end{equation}
We considered four different values for the central density $\rho_{\rm{MC,0}}$ 
such that the resulting cloud mass  $M_{\rm{MC}} \in [10^4, 10^5, 10^6, 
10^7]\ M_{\odot}$ (see Table~\ref{tab:parameter}). For the dust mass we apply 
the canonical dust to gas ratio ($DGR$) of $\xi = m_{\rm{dust}}/m_{\rm{gas}}  = 
 0.01$ \citep[e.g.,][]{Boulanger2000} in all molecular cloud models.

\subsubsection{Cluster model}
\label{sect:ClusterModel}
 In order to model radiation scattering and absorption, we used a realistic 
incident stellar energy distribution (SED) scaled from a zero age 
main-sequence, $10^{5}\,$M$_{\odot}$, solar metallicity cluster (cluster 
quantities are subscript CL) using {\sc 
Starburst99} \citep{leitherer1999}. The shape of the spectrum therefore 
remained 
identical for all cluster  models, while the luminosity scales with total 
cluster mass (see Table~\ref{tab:parameter}). We chose a cluster of 
$10^{5}\,$M$_{\odot}$ as reference in order to ensure a full sampling of the 
initial mass function (IMF). This is relatively safe to scale down to the 
lowest 
cluster mass of $10^3 M_{\odot}$, where stochasticity becomes important. In 
lieu of fully exploring stochastic clusters, we note that our approach of scaling from a massive 
cluster produces a $10^3 M_{\odot}$ luminosity consistent with  approximately five O 
stars, which is five times that of the Orion Trapezium.
In a statistical sense, we considered this as an upper limit 
of the expected feedback from our lowest mass clusters.  
We considered clusters that are either point-like or have finite radial extent. In the 
first case, all photons are emitted from a single position, while in the second 
case, mass and emissivity are spatially distributed following a 3D Gaussian 
distribution
\begin{equation}
  M_{\rm{CL}}(r) = M_{\rm{CL,0}} - M_{\rm{CL,0}}\exp \left( -8 \ln 2  
\frac{ r^2}{\beta^2_{\rm{FWHM}} }  \right) \;,
  \label{eq:ClusterMass}
\end{equation}
where $M_{\rm{CL,0}}$ is the the total mass of the 
cluster and $\beta_{\rm{FWHM}}$ is the full width at half maximum (FWHM) of the 
Gaussian mass distribution.

\subsection{Forces}
\label{sect:Force}

\subsubsection{Gravity}
\label{subsec:gravity}
\begin{figure}[ht]
      \includegraphics[width=0.49\textwidth]{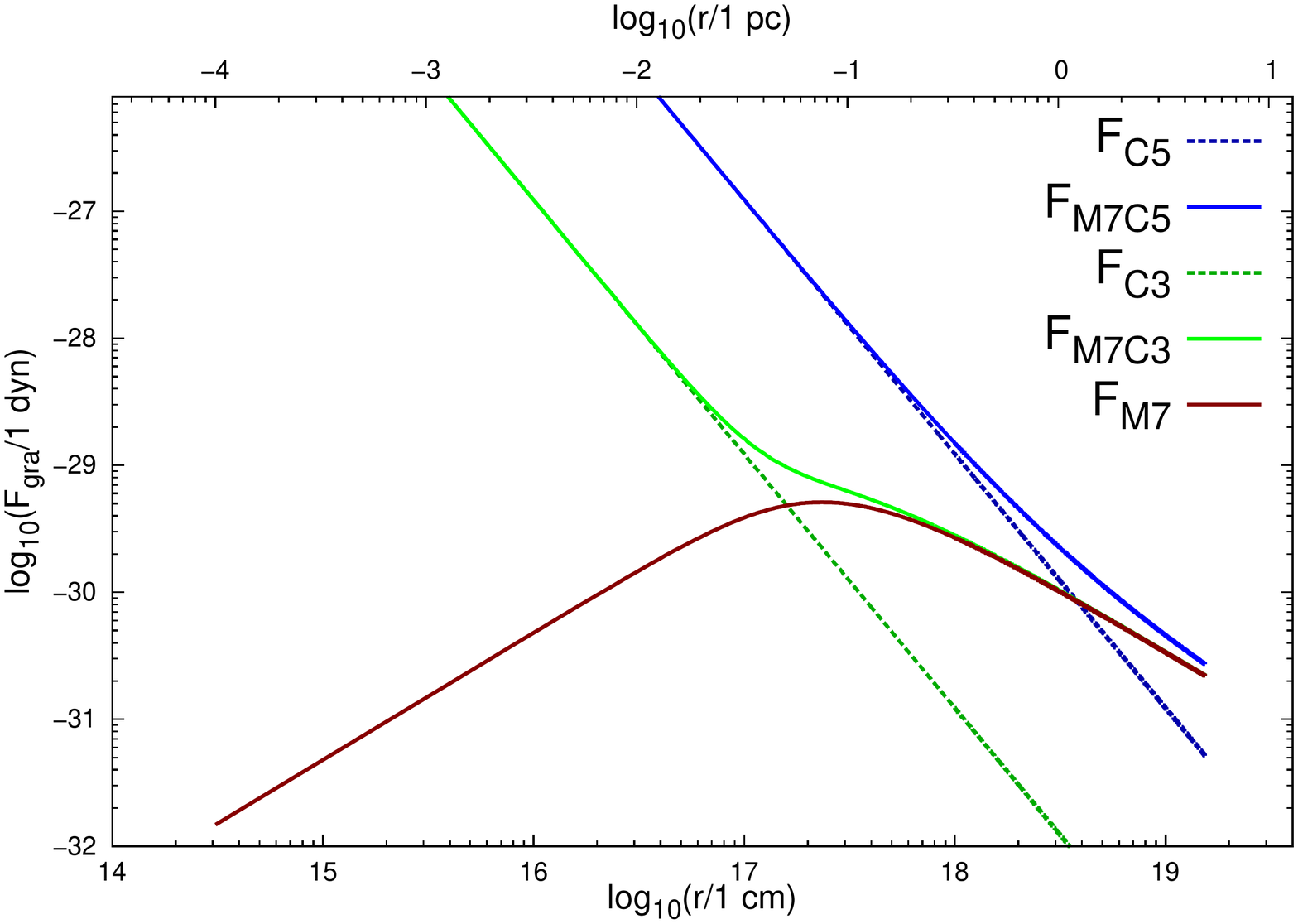}
      \includegraphics[width=0.49\textwidth]{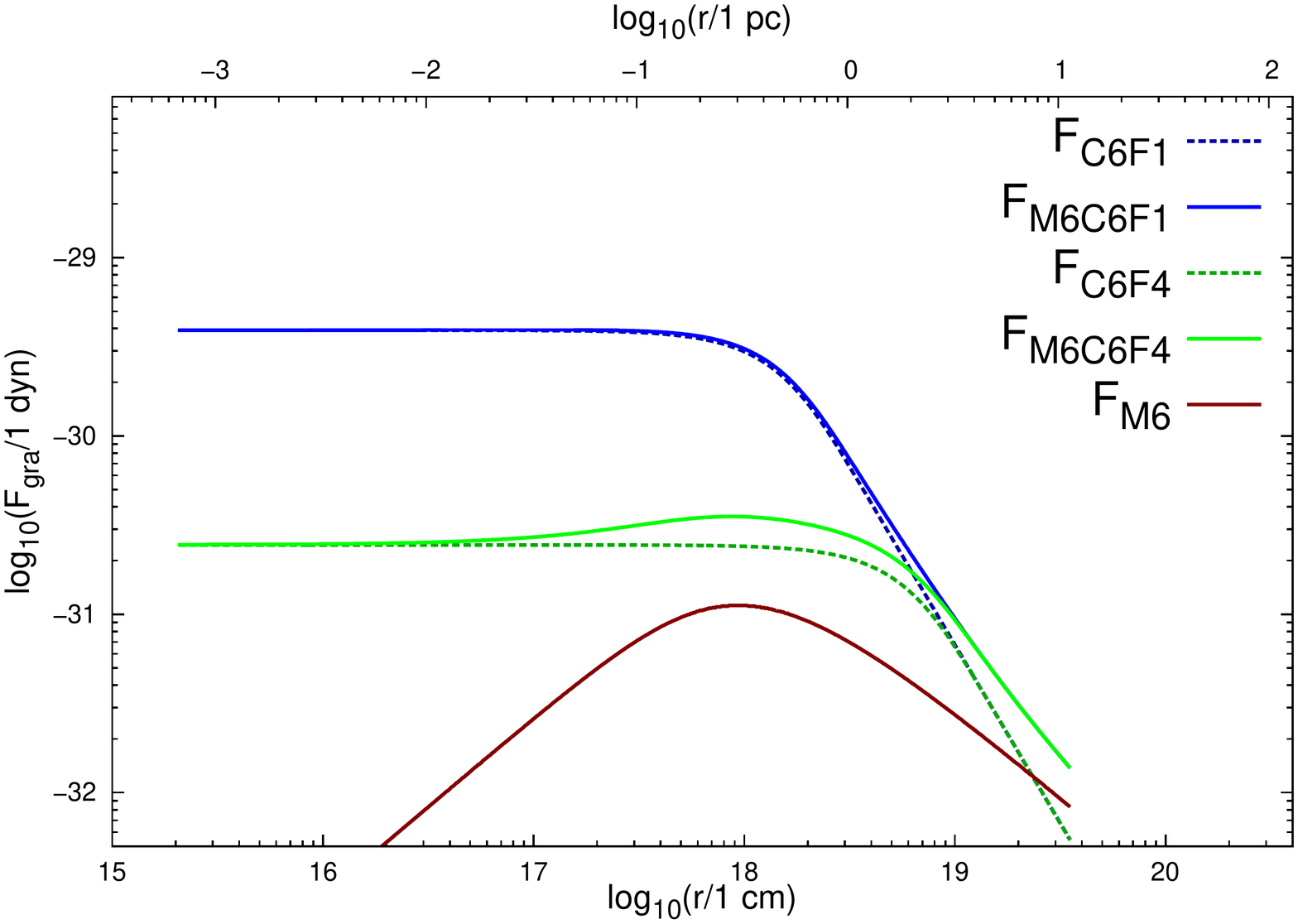}
\caption{Top panel: Gravitational force, 
$F_{\rm{gra}}$, for an ${\MM7}$ molecular cloud model with an outer radius of 
$R_{\rm{out}} = 5\ \rm{pc}$, the two point-like cluster models ${\CC3}$ and 
${\CC7}$, and their combinations. Bottom  panel: The same combinations as the 
top  
panel for an ${\MM6}$ molecular cloud with an outer radius of $R_{\rm{out}} = 
20\  
\rm{pc}$ and a ${\CC6}$ cluster model with a 3D, Gaussian, 
spatial extension having FWHM of $1\ \rm{pc}$ or $4\ \rm{pc}$.}
\label{fig:Gravity}
\end{figure}

At each position along the radial distance $r$ from the center of the 
molecular 
cloud, the forces of gravity $F_{\rm{gra}}$ and radiation $F_{\rm{rad}}$ act in 
opposition. The gravitational force experienced by the cloud at every radius 
includes the contribution from the assumed cluster and that of the gas and dust 
that compose the cloud. Assuming a perfect 
dynamical 
coupling between the two components, we can write  
\begin{equation}
  \vec{F}_{\rm{gra}}(\vec{r}) = -G 
\frac{m_{\rm{d}}}{\xi}\frac{\vec{r}}{|\vec{r}|^3}
\left[M_{\rm{CL}}(|\vec{r}|)+M_{\rm{MC}}(|\vec{r}|)\right]\;,
  \label{eq:Fgrav}
\end{equation}
where we have scaled the gravitational force per dust grain by $\xi^{-1}$ to 
account for the dust to gas  ratio. Only the gravitational force from a 
point-like cluster decreases with distance $r$ with a simple $r^{-2}$ law. 
Gaussian clusters have additional correction terms as defined in 
Eq.~(\ref{eq:ClusterMass}), and the same holds for the molecular cloud 
material where the mass depends on the radius as defined in 
Eq.~(\ref{eq:CloudMass}).  For small clusters and massive clouds, this 
term 
dominates the gravitational force at large radii. 
Figure~\ref{fig:Gravity} shows in the top panel the contributions of cloud mass 
and cluster mass to the net gravitational force experienced by the dust grain, 
assuming a point-like cluster. Here, the central cluster strictly follows a 
$r^{-2}$ power law while the chosen mass distribution of the molecular cloud 
gas 
leads to a clearly defined maximum in gravitational force from the gas near the 
edge of the cloud. Consequently, the net gravitational force declines steeply 
near the cloud's center until the cloud mass becomes dominant toward the edge 
of the molecular 
cloud.  The bottom panel of Figure~\ref{fig:Gravity} shows the net gravity 
assuming a cluster with a spatially extended cluster mass distribution in a 
smaller cloud. The mass distribution of the cluster is Gaussian  
(Eq.~\ref{eq:ClusterMass}) resulting in an almost constant contribution to net 
gravity close to the center and a decline steeper than $r^{-1}$ in the outer 
parts. The cloud mass produces a small increase in the net 
gravity in the outer regions.

\subsubsection{Radiation}
\label{subsec:radiation}
The two ISM components that absorb radiation are gas and dust, and it is 
important to determine which dominates.  For non-ionizing 
radiation the dominant gas opacity is $H_2$ Lyman-Werner 
absorption, but these bands cover only a small fraction of the total momentum 
output of the massive clusters considered here. Thus,
for the gas densities considered, dust dominates the absorption of non-ionizing 
radiation. For ionizing radiation, the dominant gas opacity is
photoionization. We  were able to estimate the impact of neglecting gas absorption of ionizing radiation 
(which carries the majority of the momentum) by comparing the relative fraction 
of radiation absorbed by gas or dust. The change of a photon flux $\phi$ 
through a finite path length $dr$ due to absorption is 
\begin{equation}
\frac{d\phi}{dr} = -n \sigma \phi - \frac{\alpha 4\pi r^2 n^2}{Q_0(H)}\;,
\end{equation}
where $n$ is the gas density, $\sigma$ is the dust absorption cross section, 
$\alpha$ the H recombination rate, $r$ the radius form the source, and $Q_0(H)$ 
is the rate ionizing photons are produced by the cluster. We see that the 
importance of dust increases as the photoionization rate of the cluster 
increases, and decreases with increasing gas density because of the quadratic nature of 
recombination. Taking characteristic values for our clusters,  we find that 
absorption by dust is almost three orders of magnitude greater than by gas at a 
radius corresponding to the Str\"{o}mgren radius of our low-mass star clusters. 
For the most massive clusters in the most dense clouds in our suite of models, 
the ratio of gas to dust
absorption increases by two orders of magnitude but is 
still smaller than the contribution from dust. Thus, gas absorption is at most a 
10\% correction to the models considered here. This is additional 
justification for our investigation of  the impact of radiative pressure  acting 
on dust. For the calculation of the outward radiative force, we recall that at a 
distance 
$r$ from a source of radiation with a bolometric luminosity $L_{*}$, a dust 
grain with a radius $a$ and an average cross section of $\langle C_{\rm{pr}} 
\rangle$ will be accelerated by a force \citep{Hulst1981}
\begin{equation}
\vec{F}_{\rm{rad}}(\vec{r}) =  \frac{\langle C_{\rm{pr}} \rangle L_{*}}{4 
\pi c} \frac{\vec{r}}{|\vec{r}|^3} \;.
\label{eq:Frad}
\end{equation}
The averaged radiative pressure cross section  $\langle C_{\rm{pr}} \rangle$ is 
defined as a flux mean with
\begin{equation}
  \langle C_{\rm{pr}} \rangle =  \frac{1}{F}\int   C_{\rm{pr,\lambda}} 
F_{\rm{\lambda}} d\lambda \;,
  \label{eq:FluxMeanCrossSection}
\end{equation}
where $F=\int F_{\rm{\lambda}} d\lambda$ is the overall 
radiative flux. Applying flux conservation,
\begin{equation}
F_{\rm{\lambda}} = \frac{L_{\rm{*,\lambda}}}{4 \pi r^2}\;,
  \label{eq:FluxConservation}
\end{equation}
 yields
\begin{equation}
  \vec{F}_{\rm{rad}}(\vec{r}) = \frac{\vec{r}}{4 \pi c |\vec{r}|^3} \int  
C_{\rm{pr,\lambda}} 
L_{\rm{\lambda}}  d\lambda \;,
  \label{eq:Frad1}
\end{equation}
where the simplest assumption is that $L_{\rm{*}} = \int L_{*,\rm{\lambda}} 
d\lambda$ 
represents the integrated spectral luminosity of the cluster. For point-like 
clusters, the radiative force decreases as $r^{-2}$, while for Gaussian 
clusters 
the radial dependence is more complicated, as discussed above.
Additional complications come from the heating of the dust by cluster 
radiation. 
This changes the cross section, as well as the frequency of the thermal 
radiation re-emitted by the dust grains. Simply assuming a fixed value for the 
dust temperature independent of radius and the properties of the cluster and 
cloud introduces errors and biases. Assessing the impact of radiative pressure 
therefore requires a fully self-consistent model of the local radiation field, 
where the quantity  $\vec{L}_{\rm{\lambda}}(\vec{r})$ represents the 
directed local spectrum of radiation after reprocessing by the dust. We combined 
a physically motivated dust grain model with a state of the art radiative 
transfer code in order to calculate this reprocessed spectrum, and thus obtain 
the correct radiative force acting on the dust grain at each radius.  By 
including this process, we go beyond any previous model in this
field. 
Star-forming molecular clouds 
  have
widely varying grain size distributions
     depending on accreted grain properties and local grain growth.
This 
variation strongly influences the coupling of the dust with the radiation 
field, and thus needs to be modeled consistently to assess the competition 
between gravitational and radiative forces. In order to account for the 
wavelength-dependent optical properties of different 
grain size distributions, we calculated the total radiative force 
\begin{equation}
  \vec{F}_{\rm{rad}}(\vec{r}) =\frac{1}{4 \pi c |\vec{r}|^2} \sum_i \chi_i 
\int_{a_{\rm{min}}}^{a_{\rm{max}}} \int_{0}^{\infty} n(a)  
C_{\rm{pr,\lambda,i}}(a) \vec{L}_{\rm{\lambda}}(\vec{r})  d\lambda da.
\label{eq:Frad2}
\end{equation}
Here, $\chi_i$ is the fractional abundance of distinct dust grain materials 
$i$, 
such that $\sum \chi_i = 1$, $n(a)$ is the grain size distribution, and the 
parameters $a_{\rm{min}}$ and $a_{\rm{max}}$, are the smallest and largest 
grain 
sizes. The various dust models we consider in this paper are
introduced in the next section.

\begin{figure*}[th]
  \begin{minipage}[c]{0.48\linewidth}
    \begin{center}
      \includegraphics[width=1.0\textwidth]{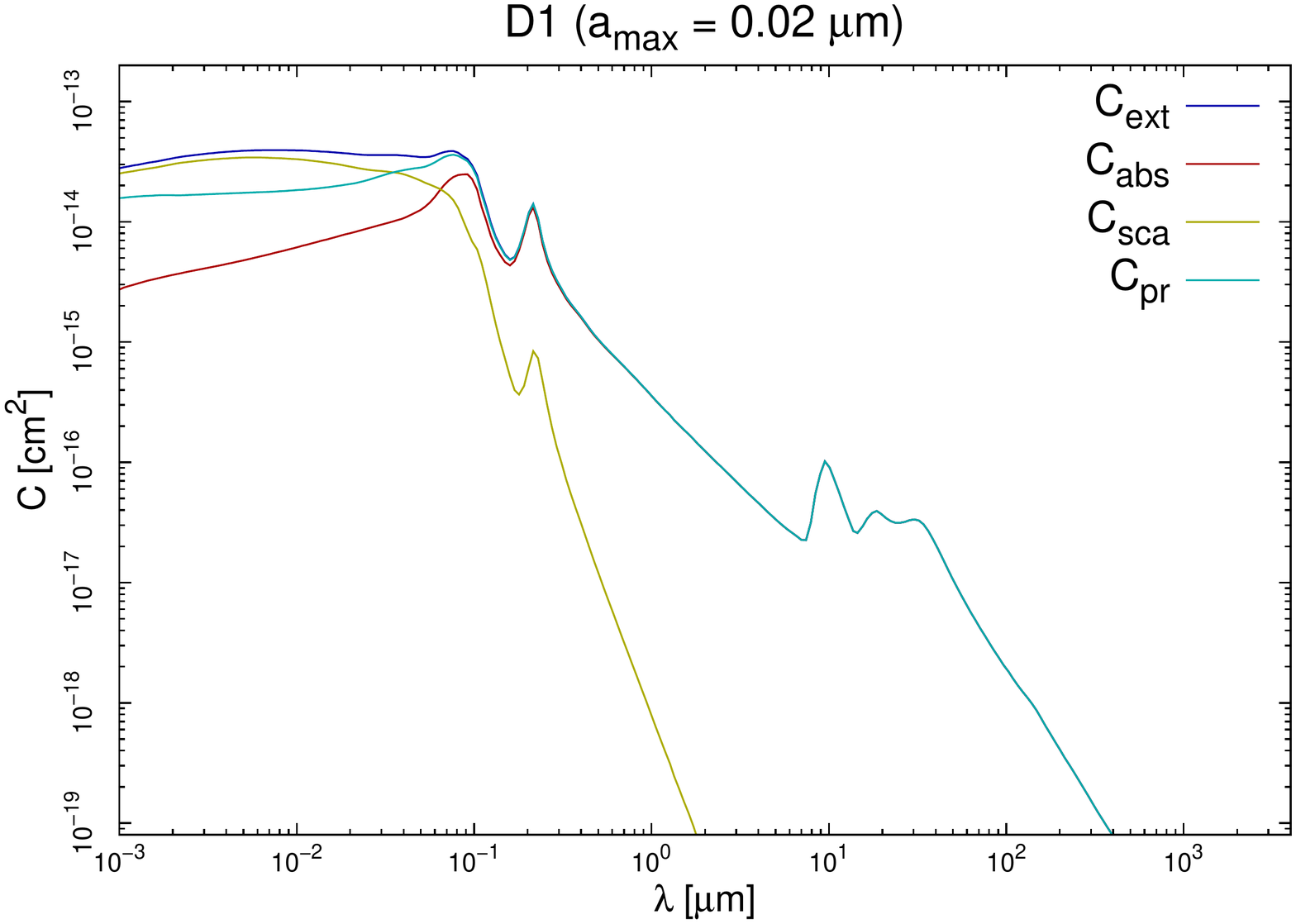}\\
      \includegraphics[width=1.0\textwidth]{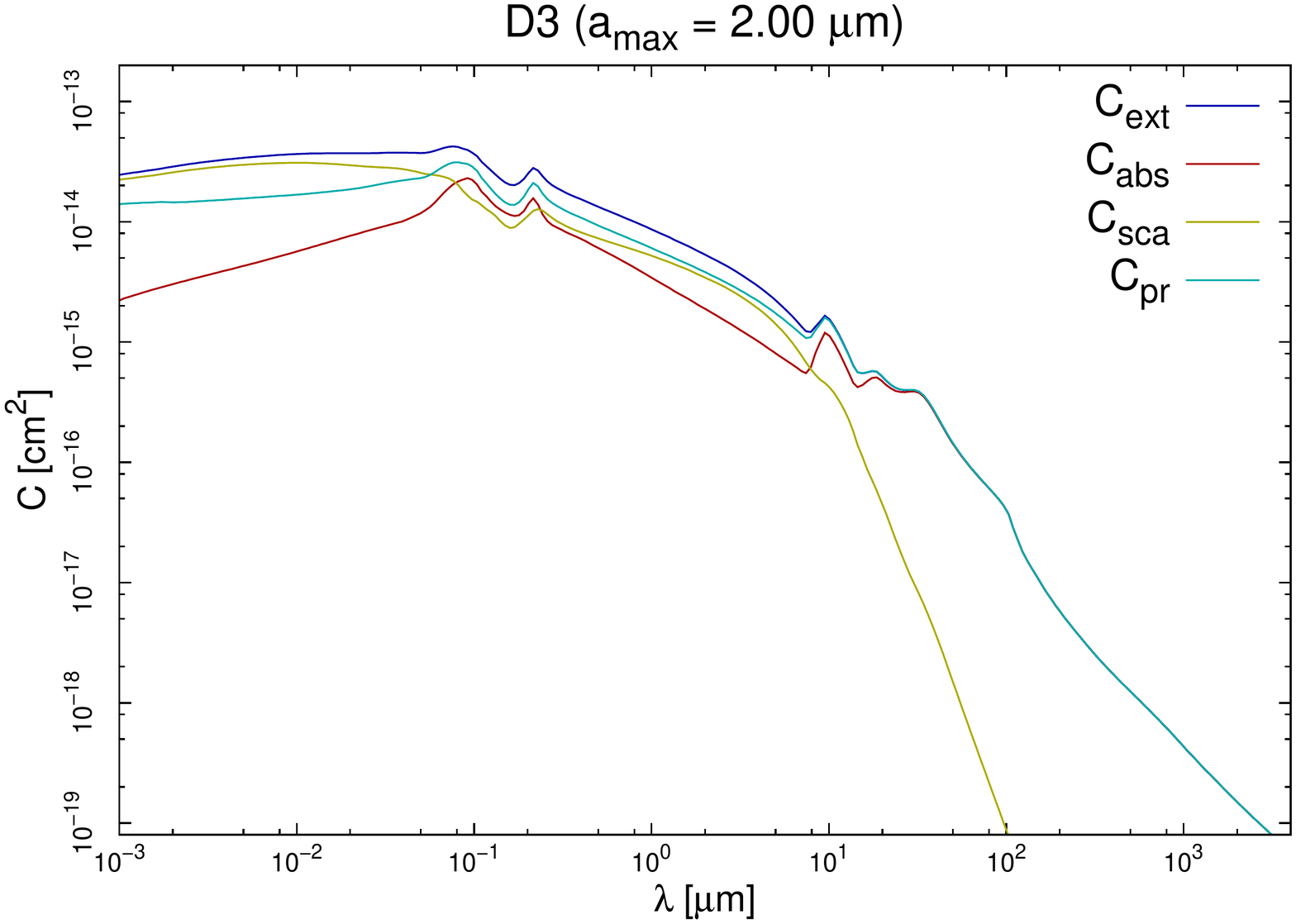}\\
      \includegraphics[width=1.0\textwidth]{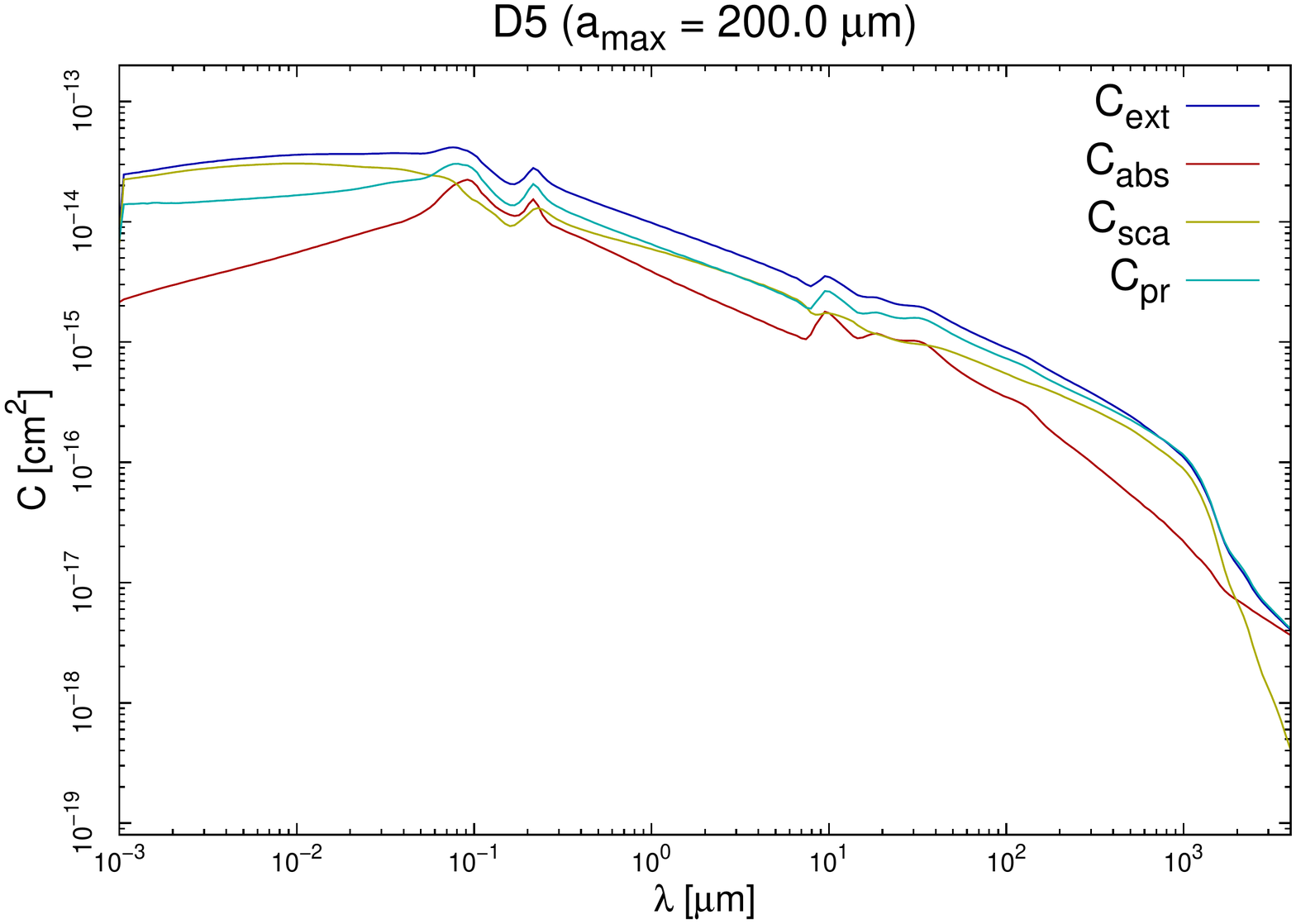}
    \end{center}
  \end{minipage}
  \begin{minipage}[c]{0.48\linewidth}
    \begin{center}
      \includegraphics[width=1.0\textwidth]{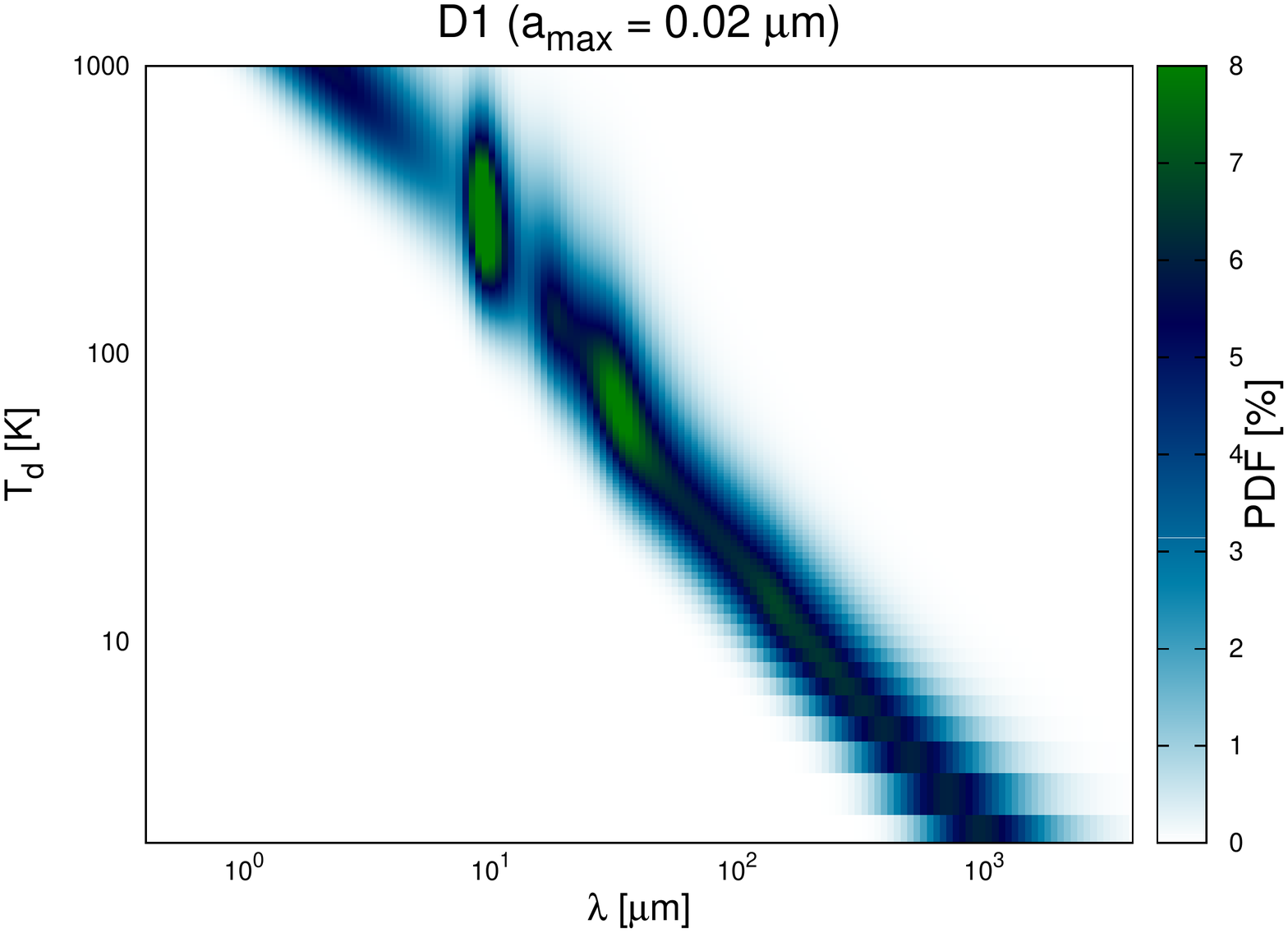}\\
      \includegraphics[width=1.0\textwidth]{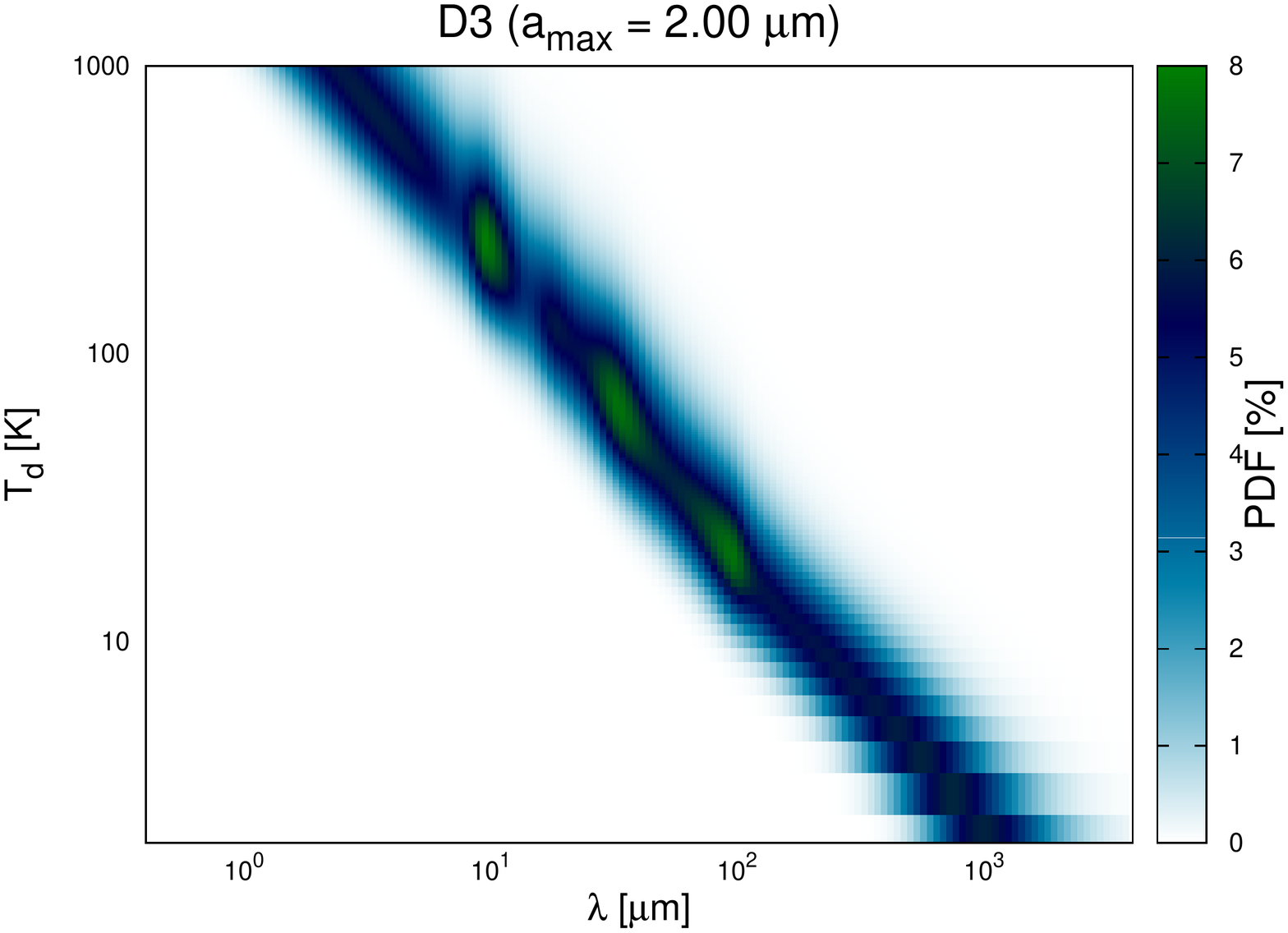}\\
      \includegraphics[width=1.0\textwidth]{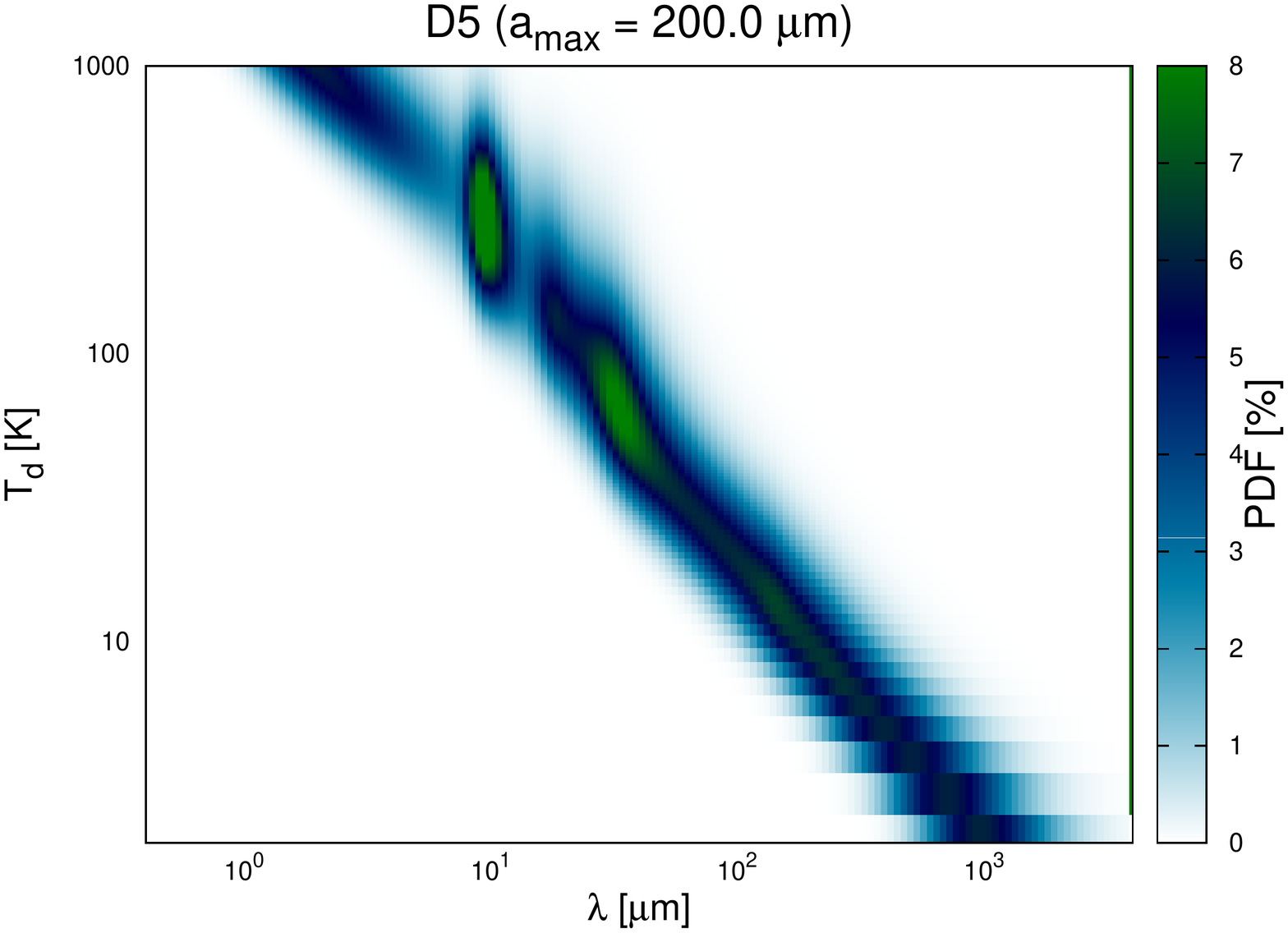}
    \end{center}
  \end{minipage}
  \caption{Left panels: Size averaged cross section of extinction 
$C_{\rm{ext}}$, scattering $C_{\rm{sca}}$, absorption $C_{\rm{abs}}$, and 
pressure $C_{\rm{pr}}$ over wavelength. Right panels: Probability distribution 
of re-emitted wavelength for different dust temperatures. Different rows are 
for 
distinct maximum grain radii corresponding to the dust grain models ${\DD1}$ 
(top row), ${\DD3}$ (middle row), and ${\DD5}$ (bottom row).}
\label{fig:DustProperties}
\end{figure*}

\subsection{Dust modeling}
\label{sect:DustModel}
The interstellar extinction curve varies only gradually 
in the ultraviolet  and optical wavelength regime, but
decreases dramatically from near-infrared  to 
sub-millimeter (sub-mm) and millimeter (mm) wavelengths, with well-defined 
silicate features occurring at $\lambda = 9.7\ \rm{\mu m}$ and $\lambda = 18\ 
\rm{\mu m}$. The standard \citet[][hereafter MRN]{Mathis1977} model captures 
these characteristics. It has a power-law grain size distribution $n(a) \propto 
a^{-q}$, with a lower and upper cut-off radius $a_{\rm{min}}$ and 
$a_{\rm{max}}$, respectively. The distribution of dust grain sizes arises from 
the competing processes of dust grain growth by accretion and nucleation and 
reprocessing toward smaller dust 
grains by sputtering, and the collision of dust grains with the
surrounding gas and each other. 
The dust is taken to be a mixture with $62.5\%$ astronomical silicate (taken to 
be olivine) and $37.5\%$ graphite \citep{Draine2001}. The size limits 
for silicate and graphite in the ISM are $[a_{\rm{min}}, a_{\rm{max}}] = 
[10\;$nm$, 100\ $nm$]$ with a size exponent of $q = 3.5$ 
\citep[][]{Weingartner2000}. For the silicate and graphite  components we use 
sublimation temperatures of $1200\ \rm{K}$ and $3915\ \rm{K}$, respectively. We 
defined the sublimation radius in each molecular cloud model as the distance at 
which the smallest silicate grains of 
our dust model start to evaporate. Consequently, the photons of the cluster 
have no interaction with the dust up to this radius. 
Later, in Section~\ref{sect:ResultsDustModel} we successively increase the 
maximum grain radius up to $a_{\rm{max}} = 200\ \mu$m in our radiative force 
calculations to determine the influence of dust growth in dense environments 
such as molecular clouds. For simplicity we assumed the dust grains to be 
compact 
spheres. We calculated the required optical dust properties for $250$ distinct 
wavelength bins over an ensemble of $200$ dust grain sizes with the {\sc miex} 
code \citep{Wolf2004}. This code calculates the cross sections of extinction 
$C_{\rm{ext,\lambda}}(a)$, scattering $C_{\rm{sca,\lambda}}(a)$, absorption 
$C_{\rm{abs,\lambda}}(a)$, and the average scattering angle $\left\langle 
\cos(\psi) \right\rangle$ on the basis of Mie scattering theory 
\citep[][]{Mie1908} with material specific refractive indices as input. We note 
that the average scattering angle is also a function of wavelength and dust 
grain size. For the astronomical silicate and graphite components we used the 
refractive indices provided by \citet{Draine1994} and \citet{Weingartner2000} 
and extrapolated them logarithmically to cover a broader spectral range of 
$\lambda \in [1\ \rm{nm}, 4\ \rm{mm}]$.
A beam of photons contributes to the radiative force on dust grains by 
transferring momentum.
At each scattering event a part of the
incident beam is redirected by an angle $\psi$ and the remaining
momentum of the incident beam drops on average by a
factor of $\left\langle \cos(\psi) \right\rangle$. Therefore, the cross
section of radiative pressure required to evaluate Eq.~(\ref{eq:Frad}) is
\begin{equation}
        C_{\rm{pr,\lambda}}(a) =  C_{\rm{ext,\lambda}}(a) - \left\langle 
\cos(\psi) \right\rangle C_{\rm{sca,\lambda}}(a).
\label{eq:Cpr}
\end{equation}
In the left column of Figure~\ref{fig:DustProperties} we show the resulting 
size averaged cross sections of 
extinction, absorption, scattering, and radiative pressure versus wavelength 
for 
three of our chosen maximal dust grain radii $a_{\rm{max}}$ (see also Table 
\ref{tab:parameter}). 
At the smallest considered 
$a_{\rm{max}} = 0.02\ \mu\textrm{m}$ (model ${\DD1}$, 
Figure~\ref{fig:DustProperties} top left 
panel), scattering decreases quickly for $\lambda > 0.1\
\mu\textrm{m}$ and can be neglected at $\lambda > 1\ \mu\textrm{m}$, so 
extinction as well as 
the radiative pressure cross section
are completely dominated by the absorption behavior of the dust. Since a 
wavelength of $1\ \mu\textrm{m}$ corresponds to a temperature of 
$T_{\rm{d}} \approx 2900\;$K and most of the dust embedded in a molecular cloud 
can be expected to be much cooler, the radiative force quickly becomes 
irrelevant
in an environment with such a narrow dust grain size distribution. Keeping the 
total dust mass constant while redistributing the grain sizes toward larger 
dust grains results in an  increase of the scattering cross section. For a 
maximum of $a_{\rm{max}} = 2.0\ \mu\textrm{m}$ 
(${\DD3}$, Figure~\ref{fig:DustProperties} middle left panel) scattering 
remains 
relevant up to a wavelength of $\approx 10\ \mu\textrm{m}$. The relevance of 
scattering extends even to the sub-mm and mm 
wavelength regime for dust grains with sizes up to $a_{\rm{max}} = 200.0\ 
\mu\textrm{m}$ (${\DD5}$, Figure~\ref{fig:DustProperties} bottom left panel). 
In 
molecular 
clouds enriched with large dust grains 
the radiative force can be enhanced by several orders of magnitude compared to 
clouds with smaller dust grains. This holds even if the large dust grains are 
so cold that 
the spectrum of the cluster is shifted toward mm wavelengths.
\begin{figure*}[ht]
\begin{center}
  \begin{minipage}[c]{0.49\linewidth}
    \begin{center}
      \includegraphics[width=1.0\textwidth]{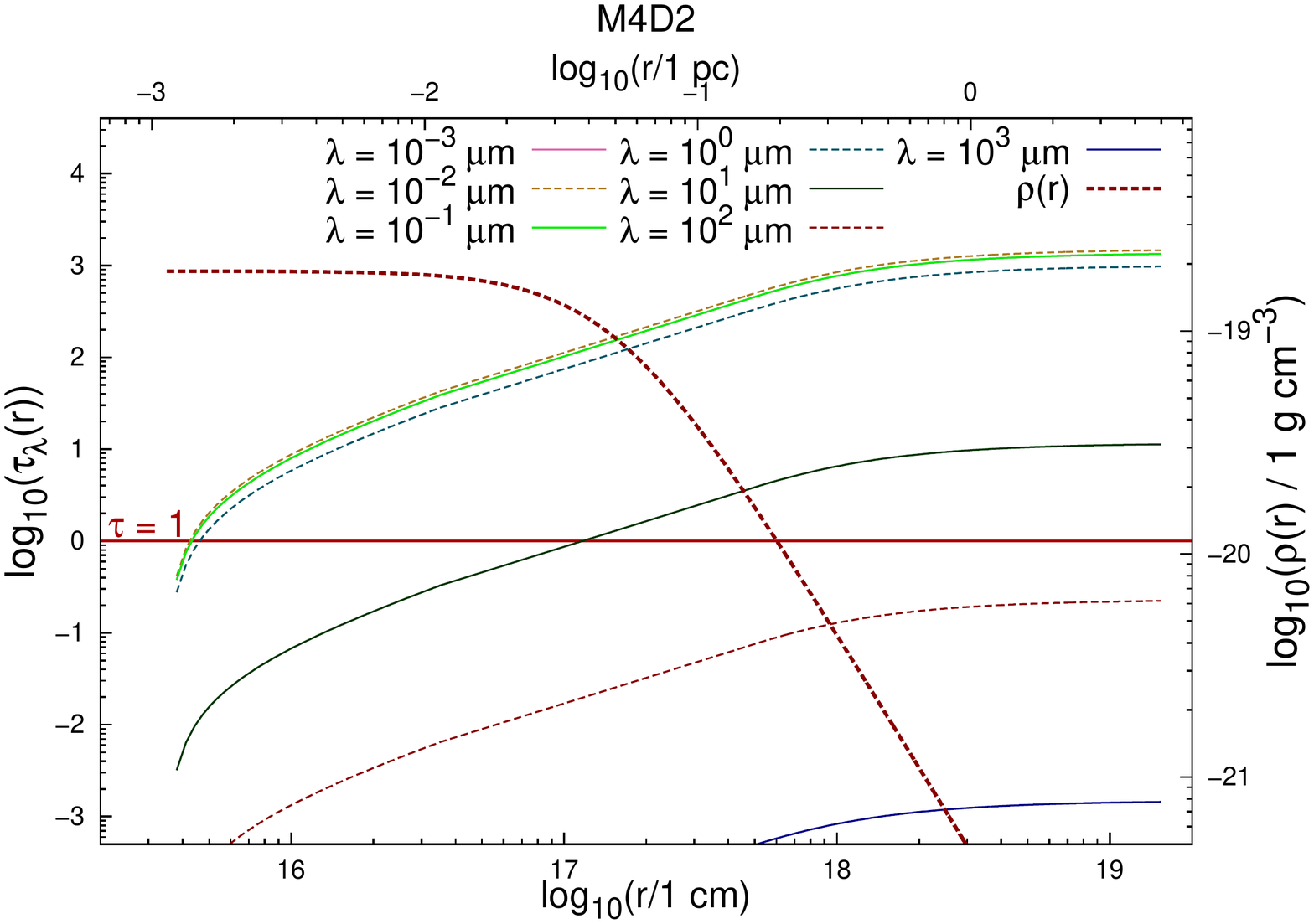}
    \end{center}
  \end{minipage}
  \begin{minipage}[c]{0.49\linewidth}
    \begin{center}
      \includegraphics[width=1.0\textwidth]{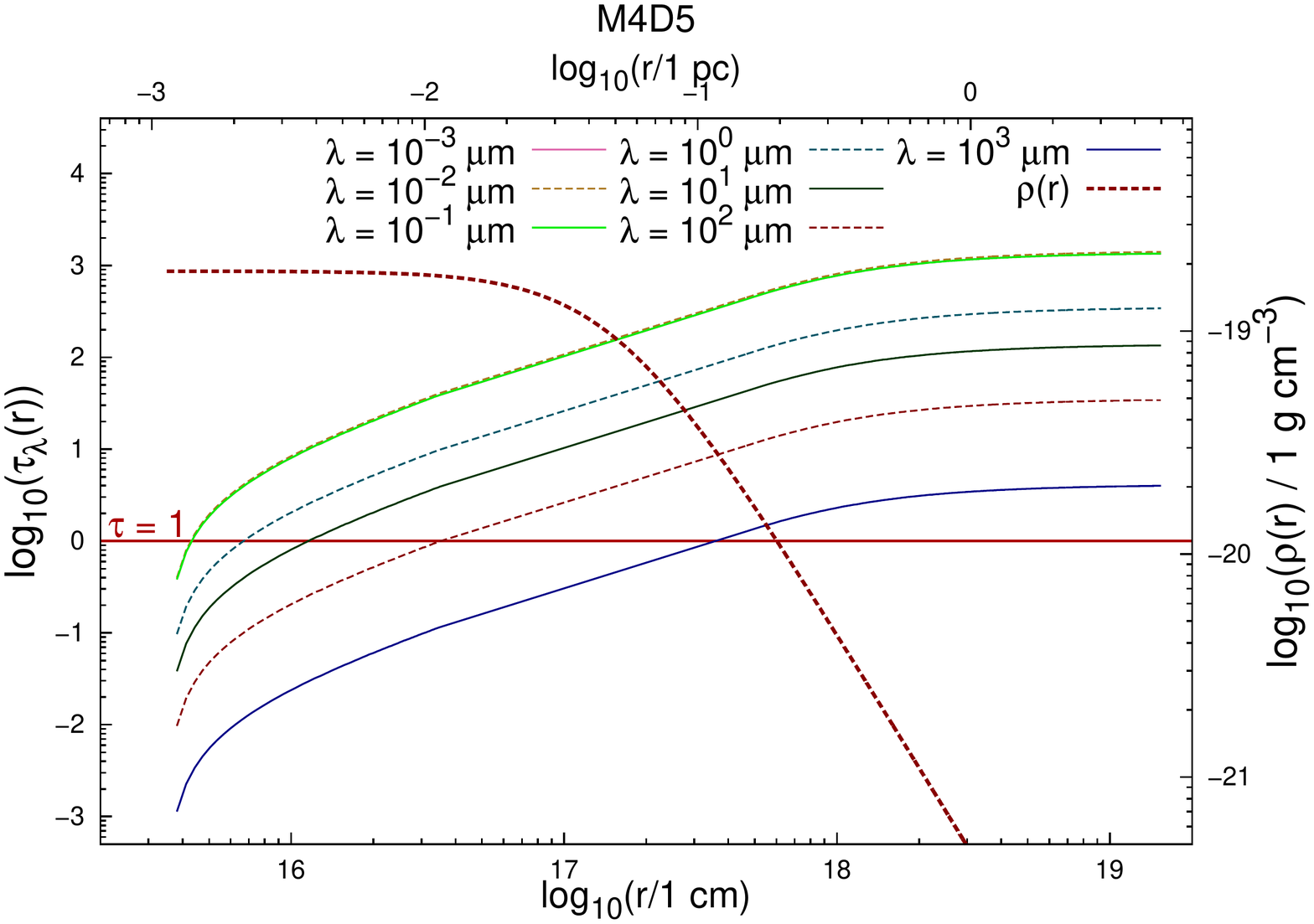}
    \end{center}
  \end{minipage}
\end{center}
\caption{Radial optical depth $\tau_\lambda$and density $\rho$ distribution for 
the ${\MM4}$ molecular 
cloud. The optical depth is calculated for different wavelengths $\lambda$ 
applying the dust models ${\DD2}$ (left panel) and ${\DD5}$ (right 
panel).}
\label{fig:OpticalDepth}
\end{figure*}
The right column of Figure~\ref{fig:DustProperties} shows the re-emission 
probabilities for a dust grain mixture corresponding to the panel on the left 
hand side.
The color scale gives the probability for a certain wavelength bin to 
be re-emitted from a size averaged dust grain with a constant
temperature from the given size distribution (see 
Eq.~\ref{eq:BW}).
As the comparison of the left column and the right column of
Figure~\ref{fig:DustProperties} reveals, the resulting re-emission
probabilities match  
the characteristic bumps of the cross sections of 
absorption. Furthermore, absorption is directly related to the optical 
properties of the assumed dust grain materials. Hence, the choice of dust grain 
materials is also of 
relevance for the re-emission spectrum of the dust and 
thus the resulting radiative force acting on the dust grains.
The optical depth $\tau_\lambda$ is related to the cross section of extinction 
by integration along a path. Figure~\ref{fig:OpticalDepth} shows the
optical depth as a function of radius and wavelength from the center to 
the edge of the ${\MM4}$ molecular cloud, applying two different dust grain 
models. Since all molecular cloud models have the same density distribution 
scaled by the central density $\rho_{\rm{MC}}$ (see 
Eq.~\ref{eq:CloudDensity}), 
the plotted optical depths and density distribution can easily be re-scaled for 
different molecular cloud 
masses. As shown in the left panel of Figure~\ref{fig:OpticalDepth}, we used the standard 
MRN 
dust grain model (${\DD2}$, see Table~\ref{tab:parameter}) to calculate the 
optical 
depth for seven different wavelengths covering the entire spectrum of the 
pre-calculated dust cross sections. Since the extinction declines toward 
longer 
wavelengths, wavelengths exceeding $100\ \rm{\mu m}$ do not become optically thick
at all for this size distribution. The right panel of 
Figure~\ref{fig:OpticalDepth} shows the optical depths resulting from the 
${\DD5}$ dust model. Since all dust grain models have similar cross sections in 
the ultraviolet and visible wavelength regime, the optical depth in those regimes 
is rather similar (see Figure \ref{fig:OpticalDepth}). However, in contrast to 
the ${\DD2}$ model, in the ${\DD5}$ model, all paths to distances larger than 
0.1$\,$pc become optically thick within this relatively modest-sized molecular 
cloud. This suggests the dust grain size distribution is a central parameter in 
determining the contribution of radiation to the net balance of forces.

\subsection{Radiative transfer}
\label{sect:RTModel}
\begin{figure}[ht]
\begin{center}
\includegraphics[width=0.49\textwidth]{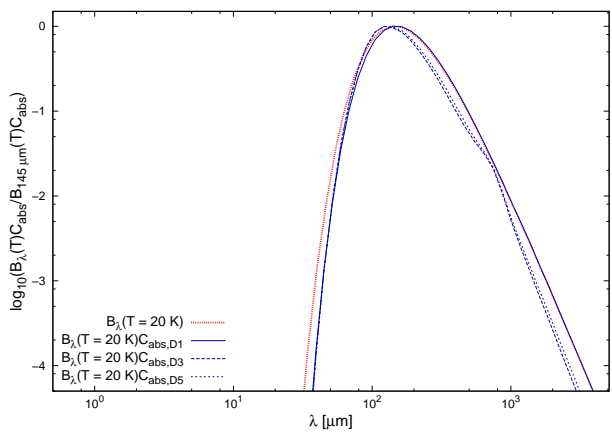}\\
\end{center}
\caption{Comparison of modified blackbody spectra for the three 
different dust models displayed in  Figure~\ref{fig:DustProperties} for a dust 
temperature of $T_{\rm d} =  20\,$K and upper values of $a_{\rm{max}}$ of  
$0.02\,\rm{\mu m}$ (${\DD1}$), $2\,\rm{\mu m}$ (${\DD3})$, and $200\,\rm{\mu m}$ 
(${\DD5}$), respectively. These spectra are more narrowly peaked and skewed 
toward smaller wavelengths than the corresponding Planck function 
$B_{\rm{\lambda}}(T_{\rm d})$, because the emission is weighted by the 
absorption cross section, $C_{\rm{abs,\lambda}}$, which drops significantly at 
sub-mm and mm wavelengths.}
\label{fig:modifiedBB}
\end{figure}

The simulations are performed using {\sc polaris}, a 3D, Monte Carlo 
RT code developed for dust temperature and polarization calculations 
\citep[see][for detail]{Reissl2016}. We 
   used spherical symmetry, with a grid divided into 
azimuthal and polar separations of eight distinct octants 
 in the azimuthal and polar directions
and 700 radial zones on a logarithmically 
increasing grid, to assure that the density and temperature differences 
between adjacent cells remain small. 
Dealing with single photons would not allow us to correctly treat the 
conservation of energy when the photon is reprocessed by dust. 
The code therefore uses distinct monochromatic photon packages to follow their 
transfer through the grid. For example, when a photon package representing a 
certain number of ultraviolet photons gets absorbed, it becomes reemitted in the 
infrared as photon package which carries a larger number of photons but contains 
the same energy as the absorbed one. 
As photon packages propagate through the grid they eventually interact with the 
dust grain mixture in a grid cell. The nature of the interaction is 
statistically governed by the physical laws. The probability of radiation-dust 
interaction is determined by the cumulative contributions from each cell the 
photon package has traveled through. 
The optical depth a photon package accumulates up to the 
$N$th cell of the grid is
\begin{equation}
        \tau_{\rm{ext}} = \sum_{i=1}^N C_{{\rm{ext,}}i}n_{{\rm{d,}}i}\ell_{{i}}\;,
\label{eq:TauPath}
\end{equation}
with the path length $\ell_i$,  the cross section of extinction 
$C_{{\rm{ext}},i}$ and  the number density $n_{{\rm{d}},i}$ of the dust in each 
cell $i$. Here, $\ell_{\rm{i}}$ is defined as the distance between the entrance 
and exit point of the ray through the boundaries of the cell. The statistical 
distribution of the optical depth can simply be sampled from $\tau_{\rm{st}}= 
-\ln(1-z)$ \citep{Wolf2003}, where $z \in [0,1[$ denotes any random number. A 
photon package interacts with the dust when the condition $\tau_{\rm{ext}}= 
\tau_{\rm{st}}$ is fulfilled in a particular cell $i$. In this special case the 
 path length $\ell_{{i}}$ is defined to be between the cell wall and the 
exact point of interaction along the trajectory of the photon package within 
that cell.\\
The nature of the interaction itself can be either scattering, or absorption 
and re-emission. The wavelength-specific dust grain albedo determines the 
probability of scattering,
\begin{equation}
        \alpha(\lambda) = 
\frac{C_{\rm{sca,\lambda}}}{C_{\rm{abs,\lambda}}+C_{\rm{sca,\lambda}}} \in 
[0,1].
\label{eq:albedo}
\end{equation}
If an interaction takes place, we drew a new random number $z$ 
uniformly in the interval $[0,1[$. In the case of $z>\alpha$ the photon 
package scatters on the dust grain. The scattering event changes the direction 
of the photon package by an angle of $\psi$, dependent on the characteristic 
phase function $F(\psi)$. For spherical dust grains the scattering angle can be 
sampled from \citep{Henyey-Greenstein1941,Hong1985}
\begin{equation}
F(\psi)=\frac{1-\left\langle \cos\psi 
\right\rangle^2}{4\pi\left[1+\left\langle \cos\psi \right\rangle^2-2 
\cos\psi\right]^{\frac{3}{2}}}\;,
\label{eq:HGPhaseFunctionDefinition}
\end{equation}
where $-1\leq \left\langle \cos\psi \right\rangle<0$ means that back
scattering  of photons is most likely, while $\left\langle \cos\psi
\right\rangle=0$ corresponds to isotropic scattering, and
$0<\left\langle \cos\psi \right\rangle \leq  1$ to forward
scattering. For $z<\alpha$ the photon package gets absorbed and instantaneously 
re-emitted by the dust grain in order to sustain local thermodynamic 
equilibrium. Here, the direction of the thermal radiation is isotropic. In 
contrast to scattering, the wavelength changes after re-emission. 
Assuming that the dust grain radiates thermally, the new wavelength 
has to be randomly sampled from a modified blackbody spectrum 
that is weighed by the dust absorption coefficient \cite[e.g.,][]{lucy1999},
\begin{equation}
        p(\lambda,T_{\rm{d}}) = \frac{\int_{\lambda}^{\lambda + d 
\lambda}{C_{\rm{abs,\lambda}} B_{\rm{\lambda}}(T_{\rm{d}})  d\lambda 
}}{\int_{0}^{\infty}{C_{\rm{abs,\lambda}} B_{\rm{\lambda}}(T_{\rm{d}})  
d\lambda}}\,,
        \label{eq:BW}
\end{equation}
where $B_{\rm{\lambda}}(T)$ is the Planck function. 
Figure~\ref{fig:modifiedBB} shows the resulting spectrum for our fiducial dust 
temperature of  $T_{\rm{d}} = 20\,$K for different grain size distributions. 
Because $C_{\rm{abs,\lambda}}$ drops significantly in the thermal regime 
longwards of $\lambda \approx 100\,\mu$m (see Figure~\ref{fig:DustProperties}) 
the modified blackbody is much more narrowly peaked and skewed toward smaller 
wavelengths than $B_{\rm{\lambda}}(T)$. As we discuss in 
Section~\ref{sect:SpectralShift} this has important consequences for the 
effective optical depth of the cloud and for the spectral energy distribution 
that is seen by an outside observer.
To  compute the radiative forces acting on dust, we extended the {\sc polaris} 
code with an additional RT mode implementing the equations introduced in 
Section\ \ref{sect:Force}. After each 3D simulation we 
average the results  among the eight octants 
in our grid in order to minimize the noise inherent in Monte Carlo RT simulations.

\section{Results}
\label{sect:Results}
To assess whether a cloud is contracting or expanding in the 
presence 
of radiative feedback from a central star cluster, we introduced the parameter 
\begin{equation}
\zeta(r) = \dfrac{F_{\rm{rad}}(r)}{F_{\rm{gra}}(r)} \;,
\end{equation}
 which is the ratio of radiative to 
gravitational forces acting on a fluid element (composed of gas and dust) at 
any 
given radius $r$ in the cloud. We calculated $\zeta$ for a wide range of model 
parameters, covering small star-forming clouds in the solar neighborhood as 
well 
as conditions expected in starburst systems. More precisely, we varied the mass 
and size of the molecular cloud, the mass and thus the luminosity of the 
cluster, as well as the cluster's spatial extension, and the dust grain size 
distribution, as summarized in Table~\ref{tab:parameter}. We considered the 
fiducial case of the dust temperature being fixed to a typical value of 
20$\,$K, 
but we also investigate models where it is allowed to vary with distance to the 
central cluster. We remind the reader that in each model we consistently 
account 
for the continuous shift of the spectrum toward longer wavelengths as the 
radiation propagates outwards to the boundary of the cloud. 

\subsection{Constant dust temperature}
\label{sect:ResultsConstantModels}
\begin{figure*}[ht]
\begin{center}

        \begin{minipage}[c]{0.49\linewidth}
                        \begin{center}
                                \includegraphics[width=1.0\textwidth]{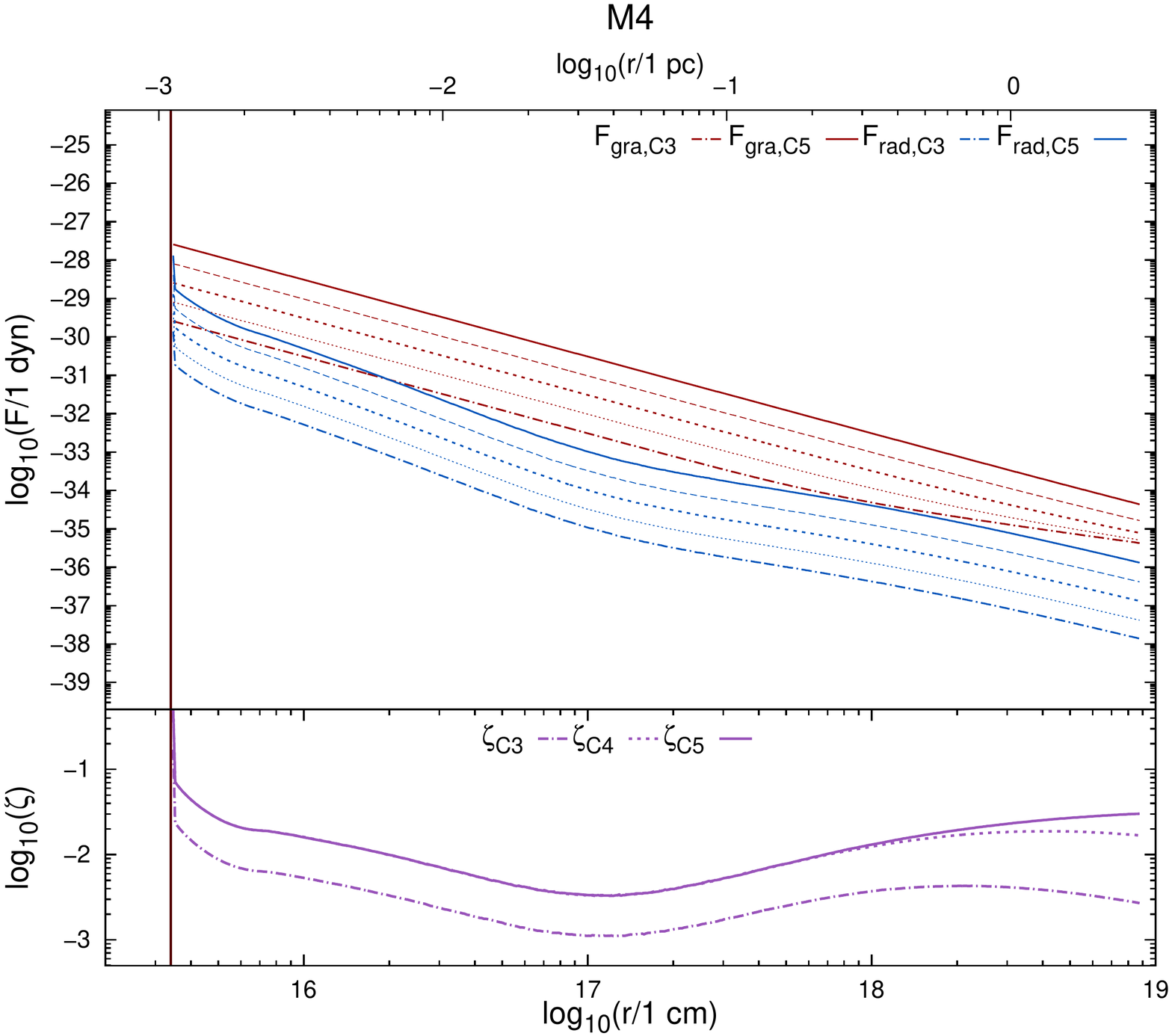}\\
                                \includegraphics[width=1.0\textwidth]{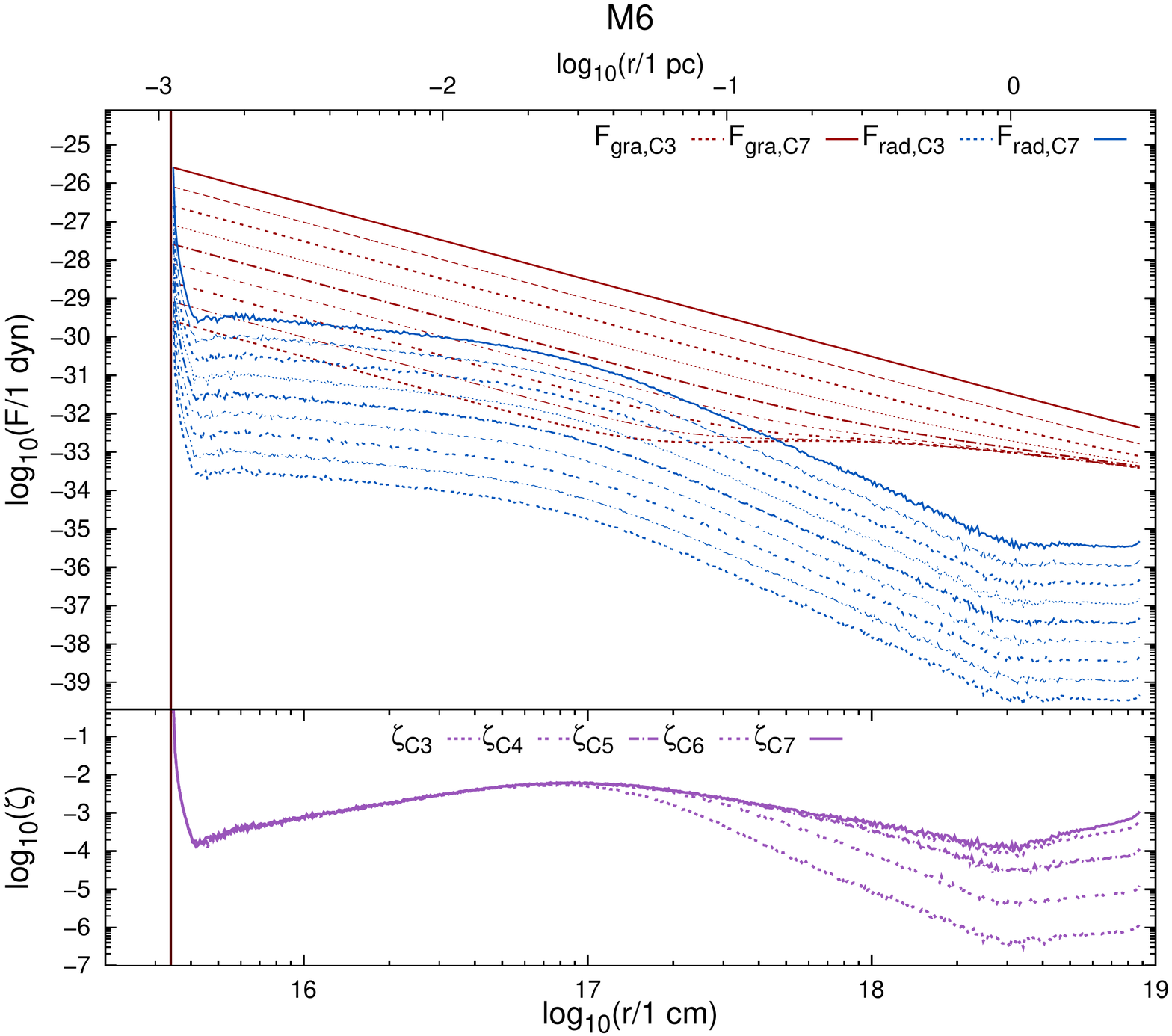}\\
                        \end{center}
                \end{minipage}
                \begin{minipage}[c]{0.49\linewidth}
                        \begin{center}
                                \includegraphics[width=1.0\textwidth]{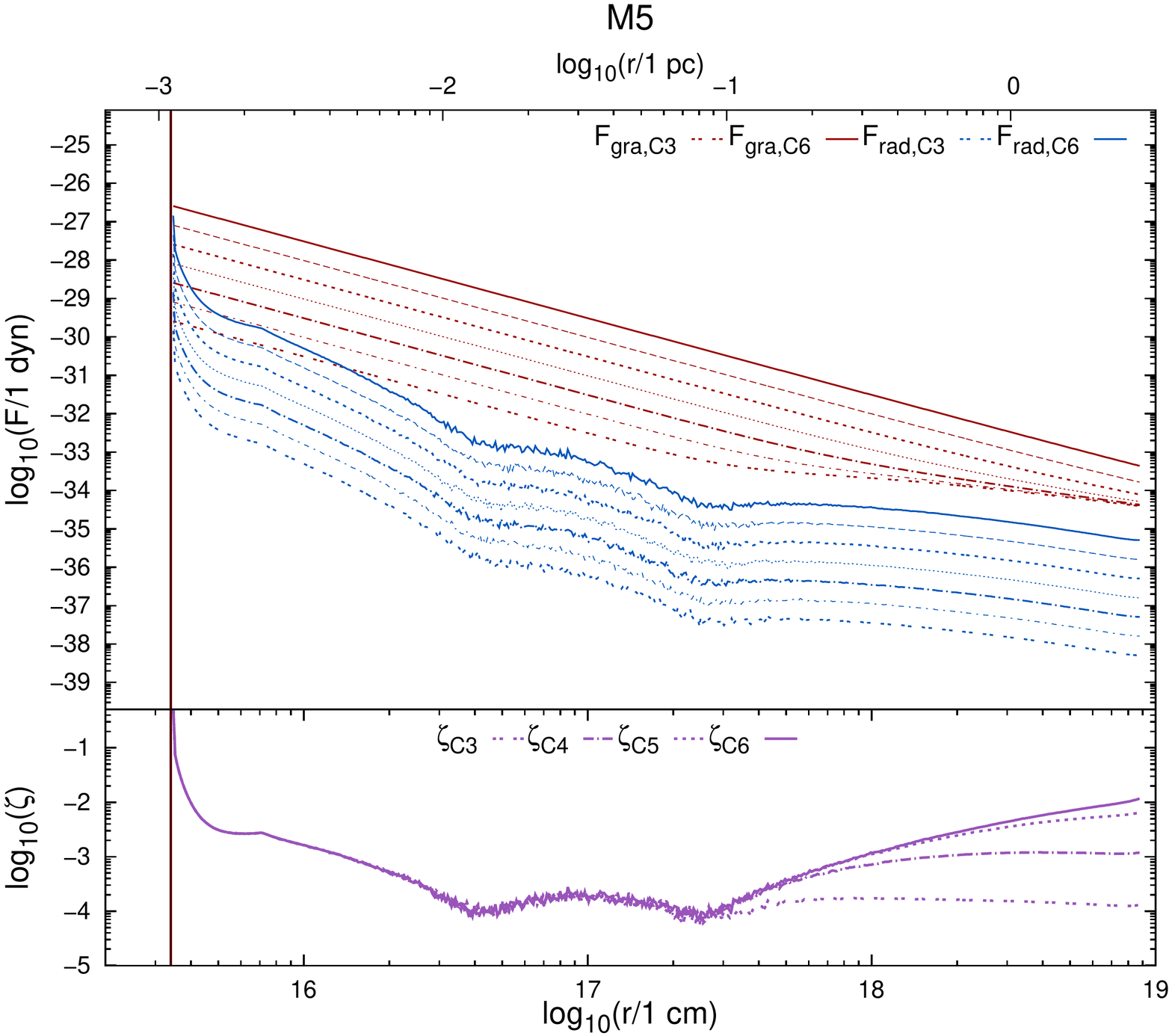}
                                \includegraphics[width=1.0\textwidth]{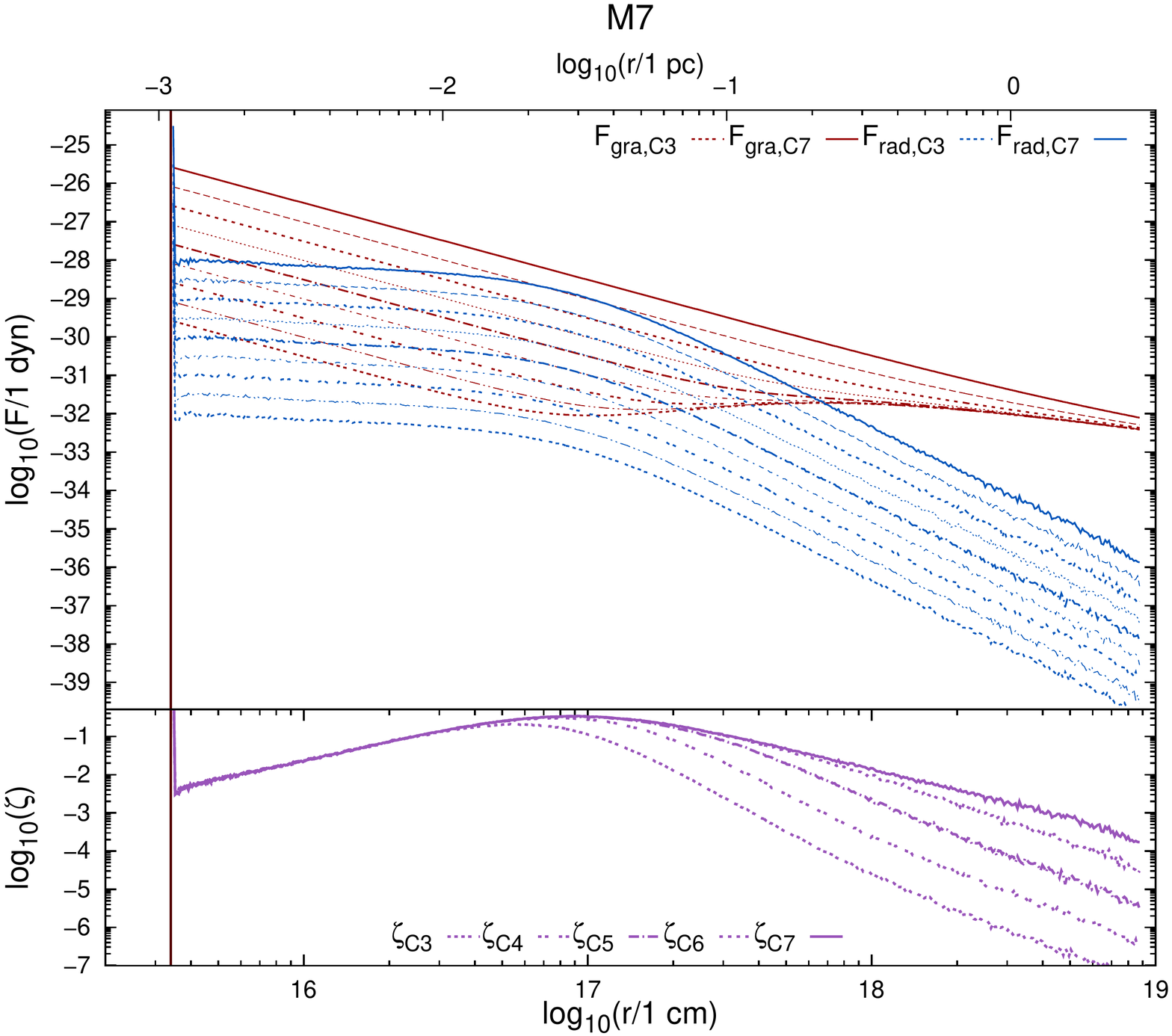}
                        \end{center}
                \end{minipage}  
\end{center}
        
\caption{Gravity ($F_{\rm{gra}}$, red 
lines) in comparison to radiative forces 
($F_{\rm{rad}}$, blue lines) for models ${\MM4}$ (top left), ${\MM5}$ (top right 
left), 
${\MM6}$ (bottom left), and ${\MM7}$ (bottom right). The ratio of forces is 
defined as $\zeta = F_{\rm{rad}}/F_{\rm{gra}}$ (purple lines). All cases have a
constant dust 
temperature of $T_{\rm{d}} = 20\ \rm{K}$, an outer radius of $R_{\rm{out}} = 5\ 
\rm{pc}$ and use dust model ${\DD2}$. We note that $\zeta < 1$ everywhere, 
implying that radiation 
pressure does not support the cloud against gravitational
contraction. The vertical brown line marks the sublimation radius. 
}
\label{fig:ForceCluster}
\end{figure*}

To touch base with the existing literature, we begin our discussion with 
calculations in which we artificially fix the dust temperature to a constant 
value of $T_{\rm{d}} = 20\ \rm{K}$. We  focus on the fiducial model
with cloud radius $R_{\rm{out}} = 5\,\rm{pc}$, point cluster geometry,
and MRN dust model ${\DD2}$ (see Table~\ref{tab:parameter}). 
Furthermore, we assume a sharp dust sublimation radius $R_{\rm{sub}} = 
1.1\times 10^{-3}\ \rm{pc}$  (see Section \ref{sect:DustModel}),
within which 
each photon package propagates without interaction. 
Figure~\ref{fig:ForceCluster} shows the resulting radiative force in
comparison to gravity and their ratio $\zeta$ for models
${\MM4}$--${\MM7}$.  For each cloud we study clusters ranging from 
our minimum cluster mass ${\CC3}$ with $10^3\,$M$_{\odot}$, up to ten times the 
cloud mass, in steps of 0.5 dex. The exception is the ${\MM7}$ model were 
the 
considered cluster mass is only between ${\CC3}$ and ${\CC7}$. This is 
equivalent to  
total star formation efficiencies of $\epsilon_{\rm tot} = M_{\rm{CL}} / 
(M_{\rm{CL}} + M_{\rm{MC}})= 9.1 \ \% \-- 91\ \%$ in the case of the ${\MM4}$ 
cloud model and $\epsilon_{\rm tot} = 0.009\ \% \-- 50\ \%$ in the case of the 
${\MM7}$ cloud model.\\
Figure~\ref{fig:ForceCluster} shows that radiation pressure cannot 
disperse any of these clouds. In every case the outward radiative
force is overwhelmed by the inward gravitational force, mostly by 
factors of $\zeta \sim$100 or more, only reaching even as high as $\zeta \sim 0.1$ for
the $10^7$~M$_\odot$ cloud {\MM7}.
Figure~\ref{fig:ForceCluster} also shows that gravitational force scales similarly 
with cloud and cluster masses. A similar scaling also holds for the radiative 
force, although there are also marked differences in the overall
radial profiles. This results from the strong radial variations in the 
number of 
scattering and absorption events for the different cloud and cluster models, as 
we discuss in more detail below. We find that higher mass clouds have a larger 
radius within which scattering is still relevant, while the maximum of 
absorption and re-emission events remains at the same distance. As a 
consequence 
the radiative force declines less steeply with radius. This can lead to a 
distinct bulge, as in model ${\MM5}$ between 0.02$\,$pc and 0.1$\,$pc, or even 
to a complete flattening of the radiative force out to radii of $\sim0.05\,$pc, 
as noticeable in models ${\MM6}$ and ${\MM7}$.  Here, the dilution of photons 
with larger distance is fully compensated by an increase in radiation-dust 
interactions.

\subsection{Spectral shift}
\label{sect:SpectralShift}

\begin{figure*}[tp]
\begin{center}
        \begin{minipage}[c]{0.49\linewidth}
          \begin{center}
            \includegraphics[width=1.05\textwidth]{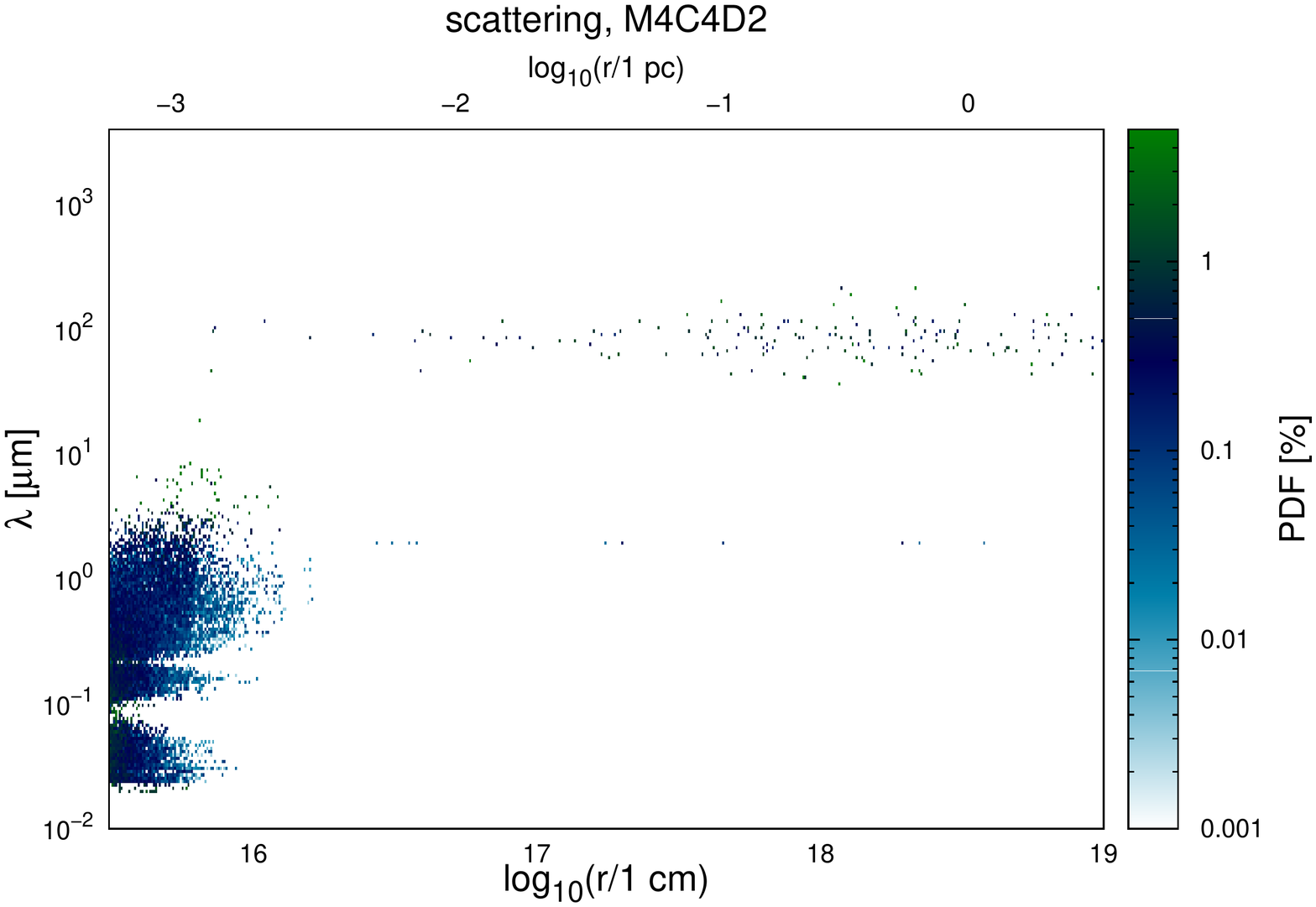}\\
            \includegraphics[width=1.05\textwidth]{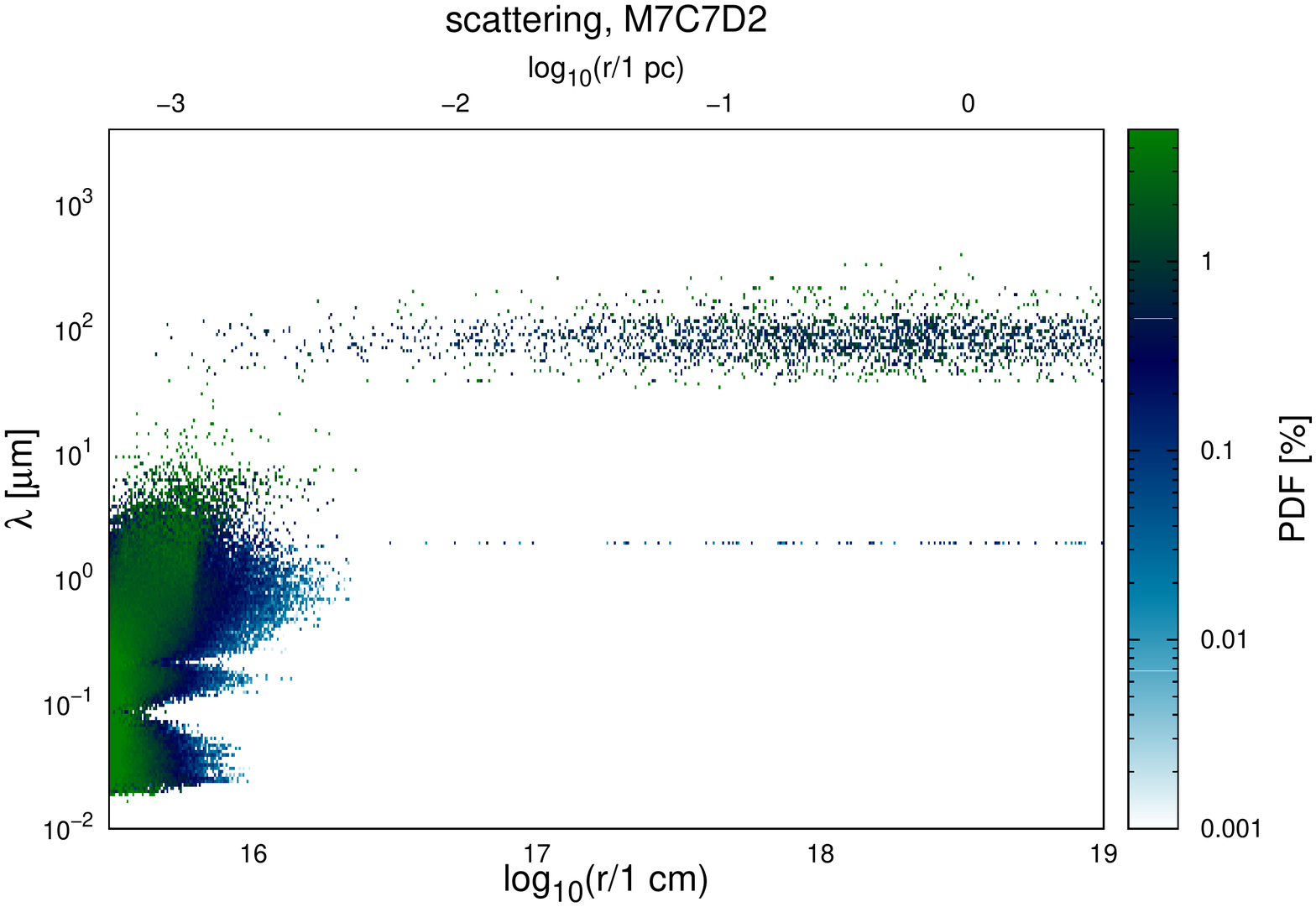}
          \end{center}
        \end{minipage}
        \begin{minipage}[c]{0.49\linewidth}
          \begin{center}
            \includegraphics[width=1.05\textwidth]{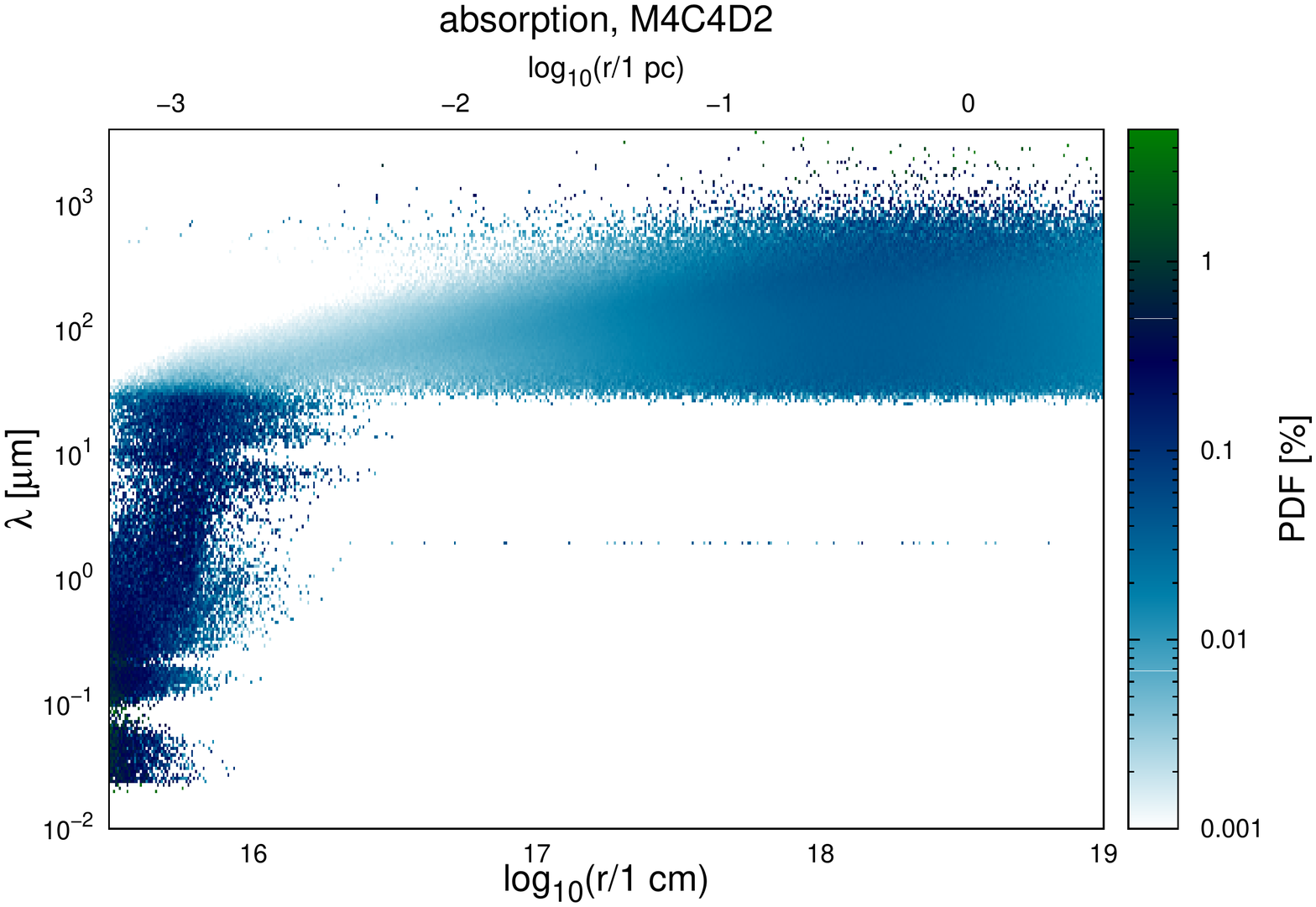}\\
            \includegraphics[width=1.05\textwidth]{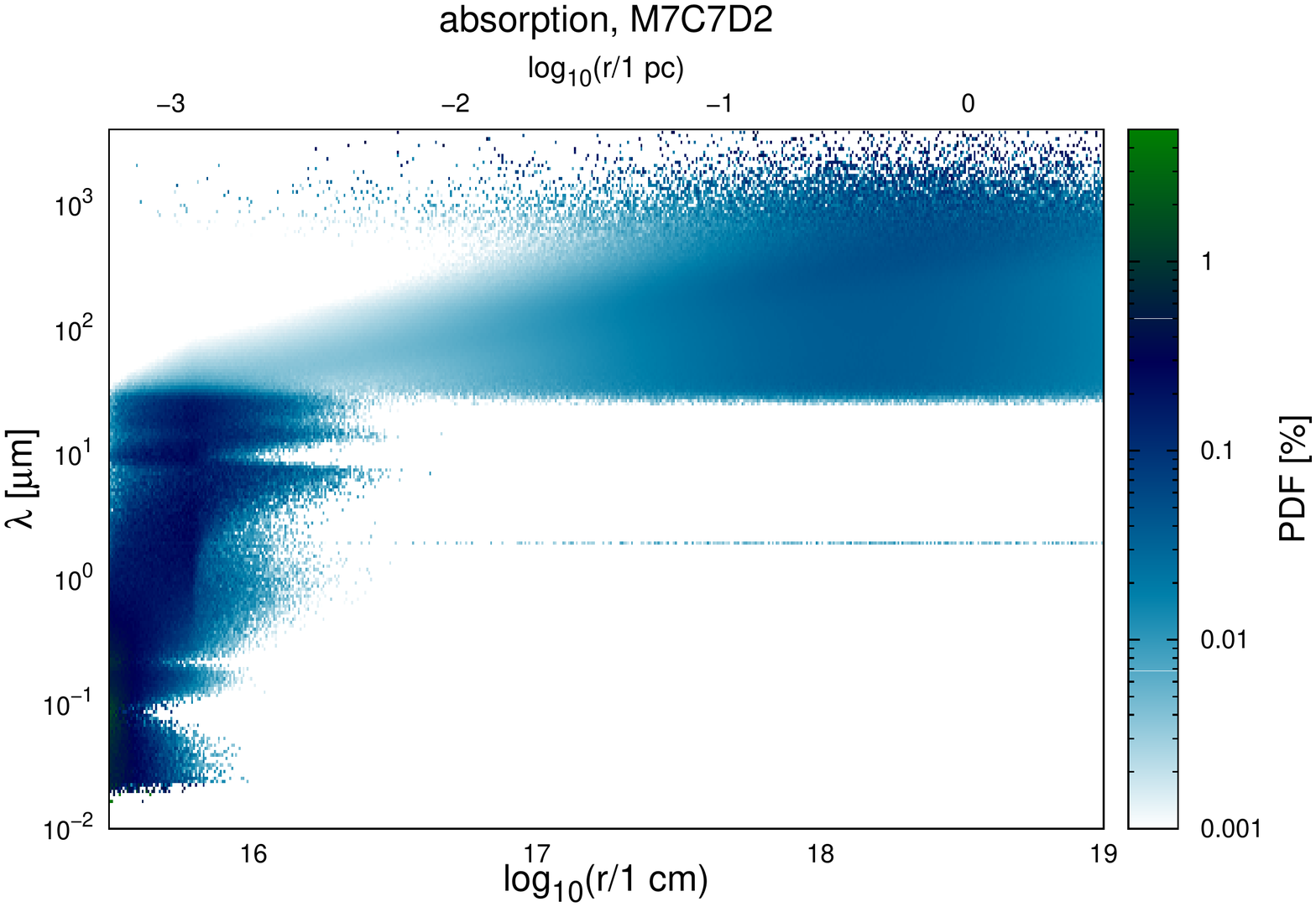}
          \end{center}
        \end{minipage}  
\end{center}
\caption{Radial and wavelength distribution of the scattering 
(left) and 
absorption (right) probability distribution for models ${\MM4}{\CC4}{\DD2}$ 
(top) and 
${\MM7}{\CC7}{\DD2}$ (bottom). The $x$-axis gives the radius of interaction 
while the 
$y$-axis is the local wavelength of the photon package at the point of 
interaction.}
\label{fig:RadialInteractions}
\end{figure*}
\begin{figure*}[t]
\begin{center}
        \begin{minipage}[c]{0.49\linewidth}
          \begin{center}
            \includegraphics[width=1.05\textwidth]{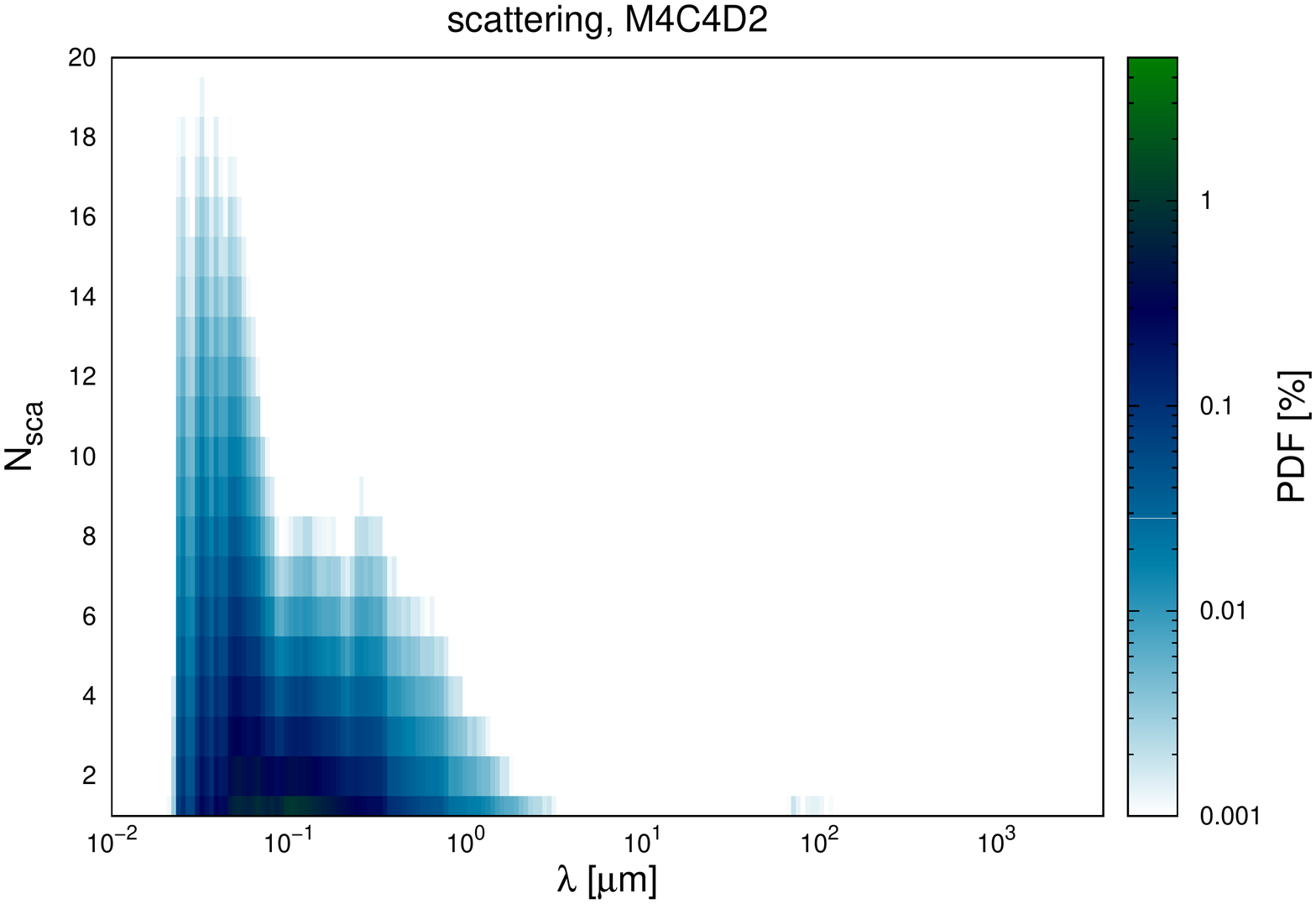}\\
            \includegraphics[width=1.05\textwidth]{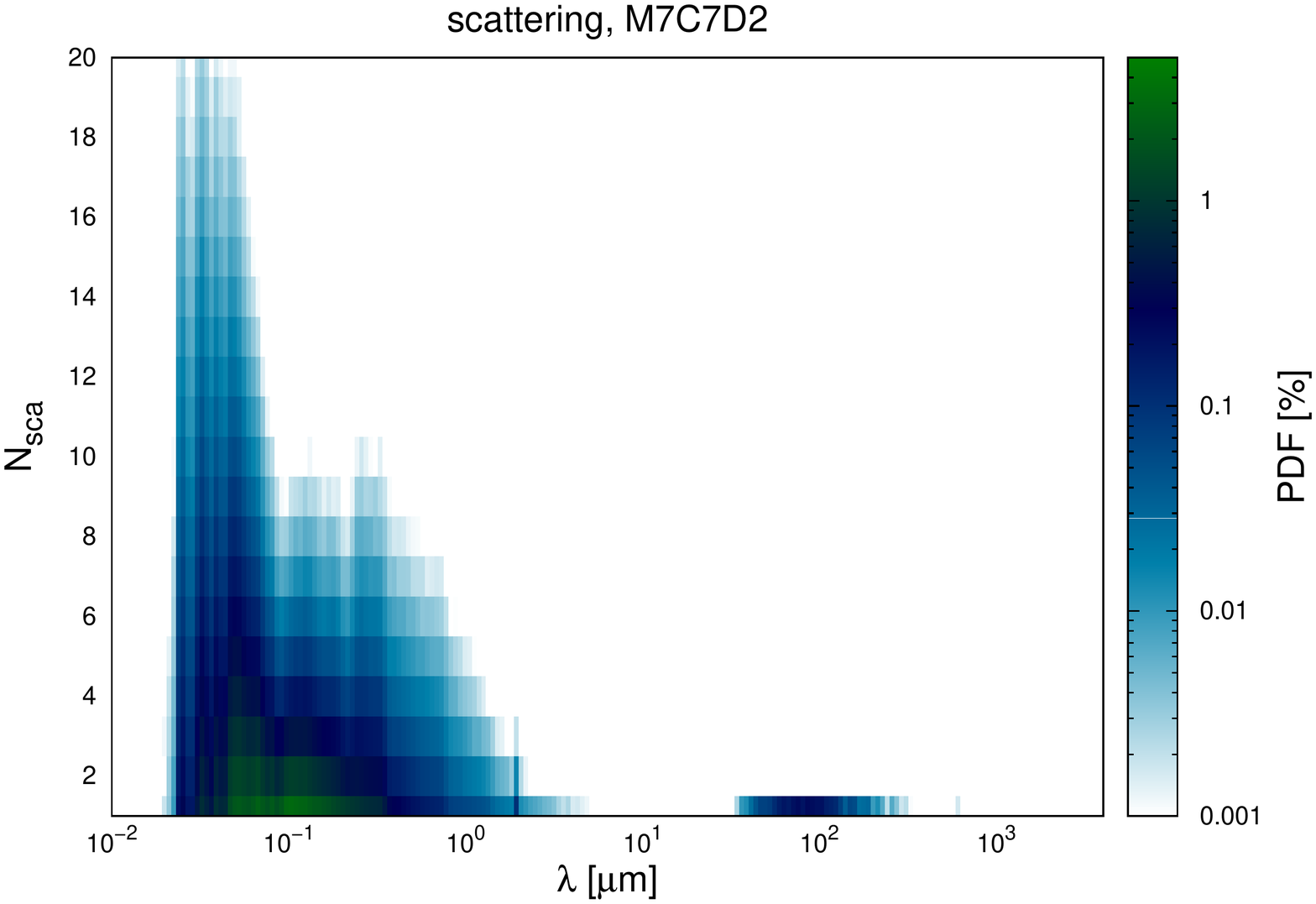}
          \end{center}
        \end{minipage}
        \begin{minipage}[c]{0.49\linewidth}
          \begin{center}
            \includegraphics[width=1.05\textwidth]{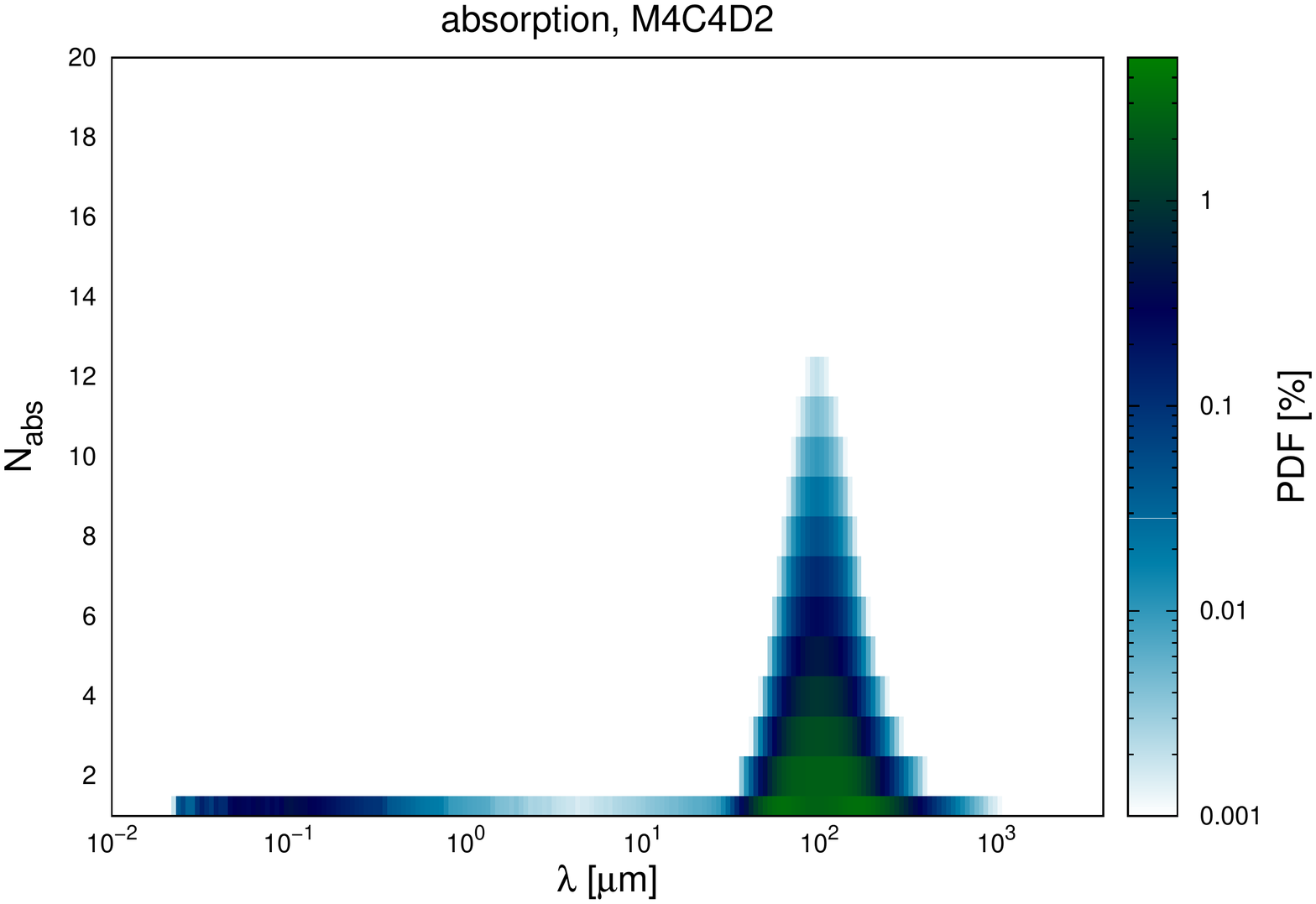}\\
            \includegraphics[width=1.05\textwidth]{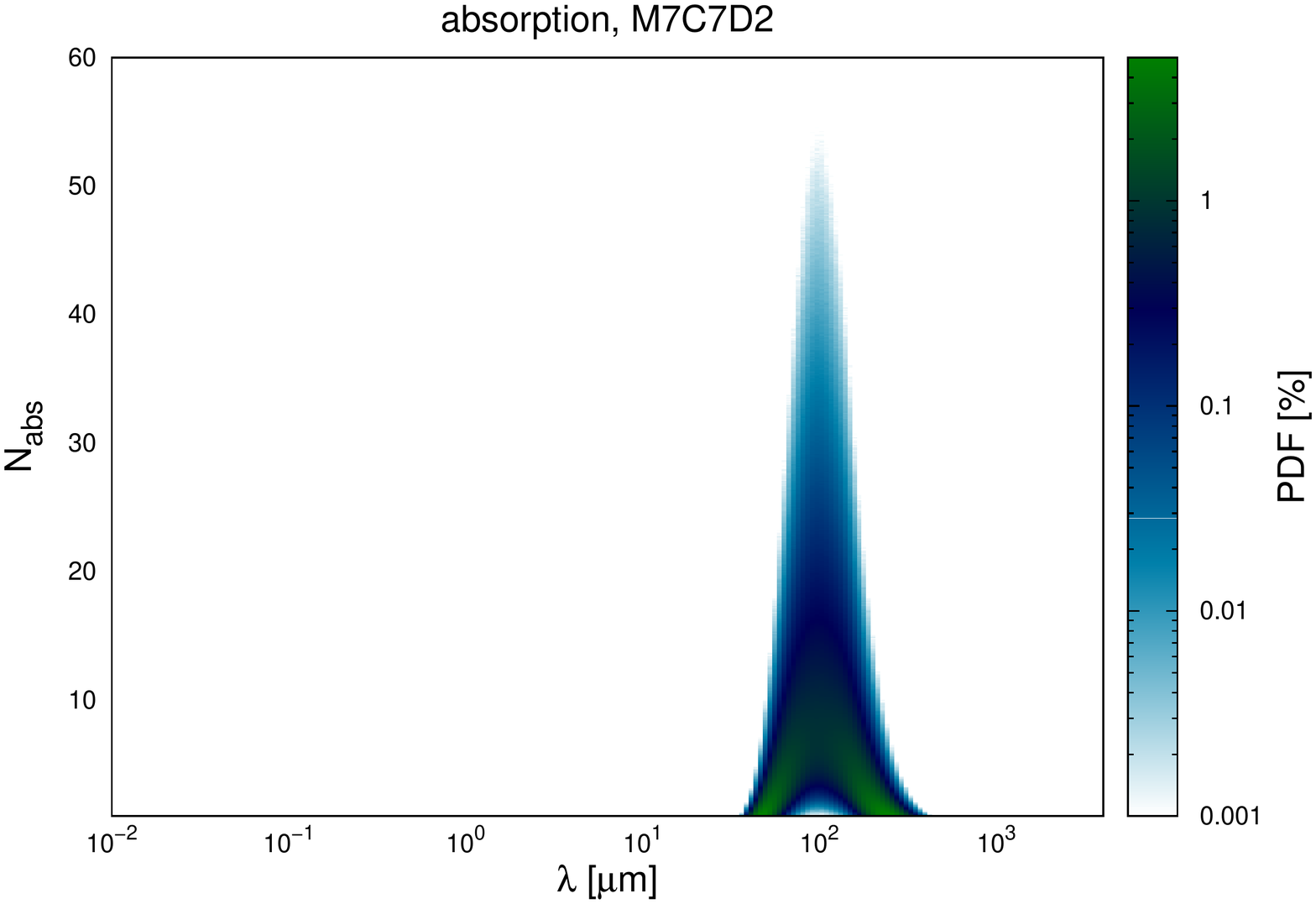}
           \end{center}
        \end{minipage}    
\end{center}
\caption{Number of scattering events (left column) and 
absorption and 
re-emission events (right 
column) experienced by a photon package before escaping the molecular 
cloud.  
The same models are shown as in Figure~\ref{fig:RadialInteractions}, 
${\MM4}{\CC4}{\DD2}$ 
(top) and  ${\MM7}{\CC7}{\DD2}$ (bottom). The $x$-axis gives the local wavelength of 
the photon package at the point of 
interaction, and the $y$-axis gives the number of interactions of 
each type.}
\label{fig:NumberOfInteractions}
\end{figure*}

In order to analyze the radial profile of the radiative force $F_{\rm rad}$ in 
more detail, we track the position of each scattering and absorption 
event in the Monte Carlo RT simulations. Figure~\ref{fig:RadialInteractions} 
shows the resulting maps of interactions in our lowest and highest mass clouds 
for a cluster with 50\% star formation efficiency (${\MM4}{\CC4}$ model in the 
top row and ${\MM7}{\CC7}$ model in the bottom row). We find that scattering is 
only efficient within $R < 3\times 10^{-3}\ \rm{pc}$, in fairly distinct 
wavelength ranges, while absorption and re-emission occur mostly at distances 
larger than about $3\times 10^{-2}\,$pc for essentially all wavelengths. 
However, even further in, absorption and re-emission can not be entirely 
neglected. Closely related, Figure~\ref{fig:NumberOfInteractions} illustrates the number of 
scattering events as well as absorption and re-emission events occurring for the 
same two models, ${\MM4}{\CC4}$ (top) and  ${\MM7}{\CC7}$
(bottom). Cluster radiation longwards of $\lambda \gtrsim 100 \mu$m
barely scatters once, since the low-mass cloud is  
optically thin at these wavelengths (Figure~\ref{fig:OpticalDepth}, left
panel) Even for the shortest 
wavelength emitted by the cluster, the most likely number of
scatterings is under five. Since longer wavelengths are unlikely to
scatter and shorter  
wavelengths quickly shift after a few interactions, all photons finally end up at 
infrared wavelengths. With increasing molecular cloud mass, the optical depth at 
any given radius increases. As a result the radius within which scattering is 
relevant gets larger while the maximum of absorption and re-emission events 
remains at the same distance. 

\begin{figure*}[ht]
  \begin{center}
    \begin{minipage}[c]{0.49\linewidth}
      \begin{center}
        \includegraphics[width=1.0\textwidth]{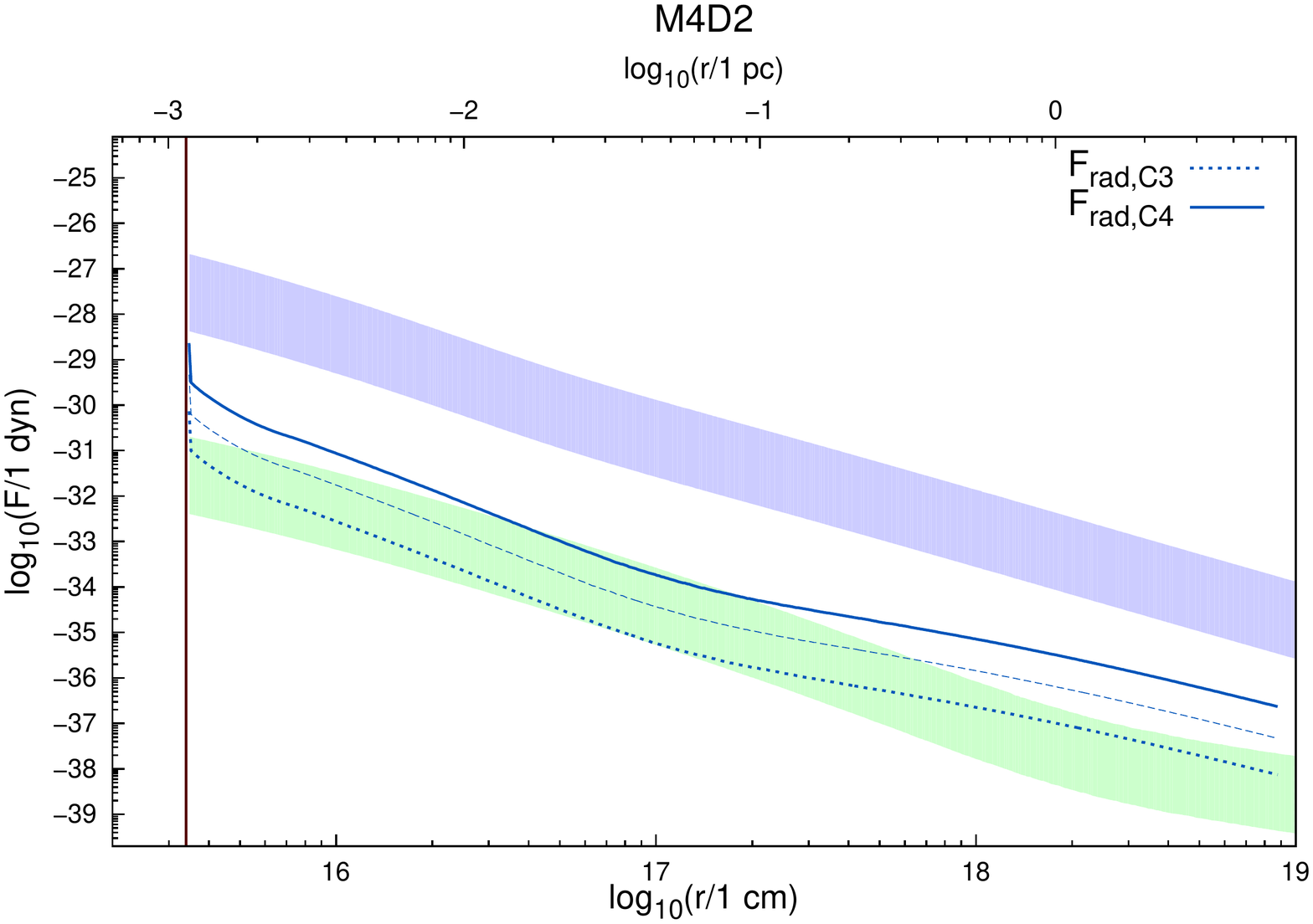}
      \end{center}
    \end{minipage}
    \begin{minipage}[c]{0.49\linewidth}
      \begin{center}
        \includegraphics[width=1.0\textwidth]{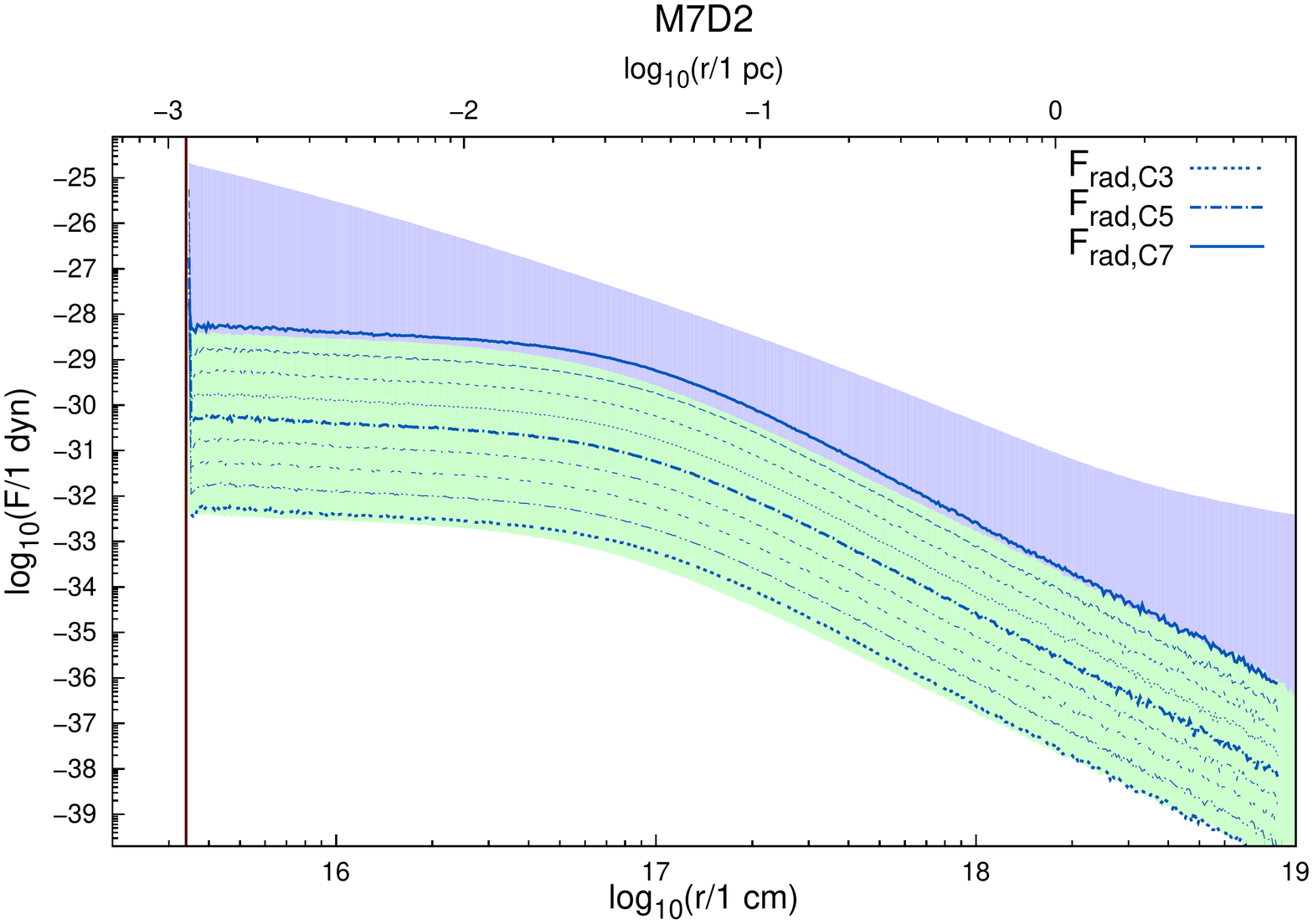}
      \end{center}
    \end{minipage}  
  \end{center}        
  \caption{Radial profile of radiative force (blue lines) for different 
molecular cloud models ${\MM4}$ (left panel) and ${\MM7}$ (right panel) with an 
outer radius of $R_{\rm{out}} = 5\ \rm{pc}$ and a constant dust temperature of 
$T_{\rm{d}} = 20\,\rm{K}$ simulated with cluster models ${\CC3}$ through 
${\CC7}$ in half steps. The blue band represents the same simulations with the 
absorption and reprocessing of radiation switched off ($C_{\rm{abs}} = 0 \mbox{ 
cm}^2$), while the green bands result from the same simulations with an 
instantaneous redistribution of radiation ($C_{\rm{sca}} = 0\mbox{ 
cm}^2$).}
\label{fig:ForceLimits}
\end{figure*}
The rapid reprocessing of the light from the central star cluster and
the deviation of the resulting radiation field from a 
single-temperature blackbody has important consequences for the  overall 
optical depth and for the radiative force. As seen in Figure~\ref{fig:ForceLimits}, 
we examined how this force would change if we adopted the two most extreme 
assumptions possible. First, we neglected absorption and re-emission at thermal 
wavelengths ($C_{\rm{abs}} = 0\ \rm{cm^2}$) and only rely on scattering of the 
incident stellar radiation field. The cloud is extremely optically think for 
radiation in ultraviolet and optical bands, and consequently, the radiative 
force is boosted by orders of magnitude, easily exceeding  gravity. Second, we 
took the other extreme and neglect scattering ($C_{\rm{sca}} = 0\ 
\rm{cm^2}$). That is each stellar photon is 
immediately absorbed and the corresponding energy is emitted as thermal photons 
with frequencies sampled from a $20\;$K modified blackbody 
(Eq.~\ref{eq:BW}). We find that the resulting radiative force is very 
similar to the full model with both scattering as well as absorption and 
re-emission. This demonstrates again that scattering is rather unimportant for 
the momentum transfer from the radiation field to dust for 
the type of clouds considered here and that the spectral shift associated with 
absorption and re-emission is the key to understanding the impact of radiation 
pressure. 
\subsection{Varying dust size}
\label{sect:ResultsDustModel}
To better understand which parameters can increase the ratio 
$\zeta$ of radiative to gravitational force, we now turn our attention to the 
dust grain model,  continuing to hold dust temperature constant at $T_{\rm d} = 
20\;$K. We assumed a constant dust mass fraction of 1\%, but consider different 
values for the maximum grain size. This is a key parameter as the size 
distribution controls the spectral dependence of the scattering and absorption 
cross sections (see Section~\ref{sect:DustModel}). 
\begin{figure*}[ht]
\begin{center}

        \begin{minipage}[c]{0.49\linewidth}
                        \begin{center}
                                
\includegraphics[width=1.0\textwidth]{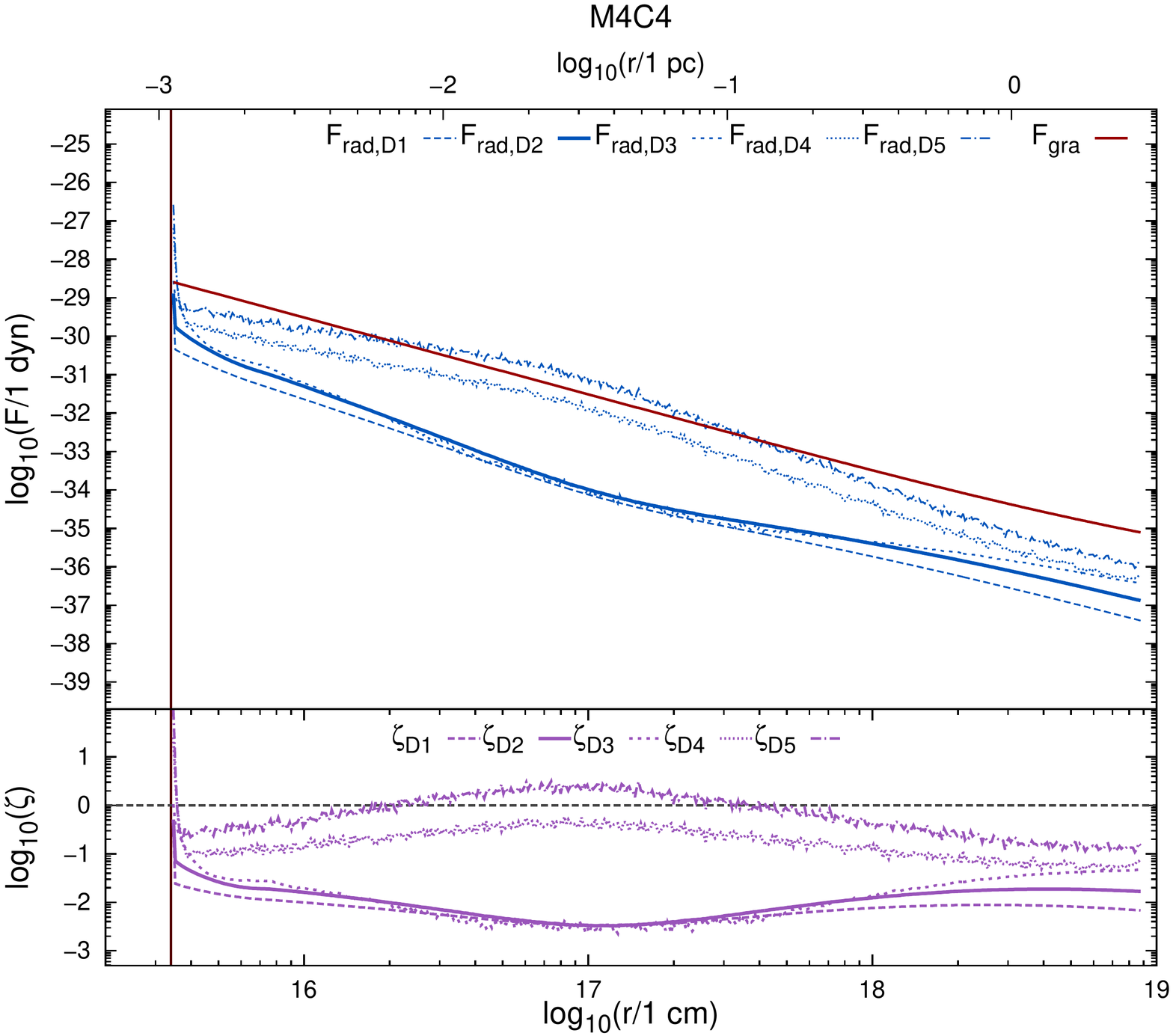}
                        \end{center}
                \end{minipage}
                \begin{minipage}[c]{0.49\linewidth}
                        \begin{center}
                                
\includegraphics[width=1.0\textwidth]{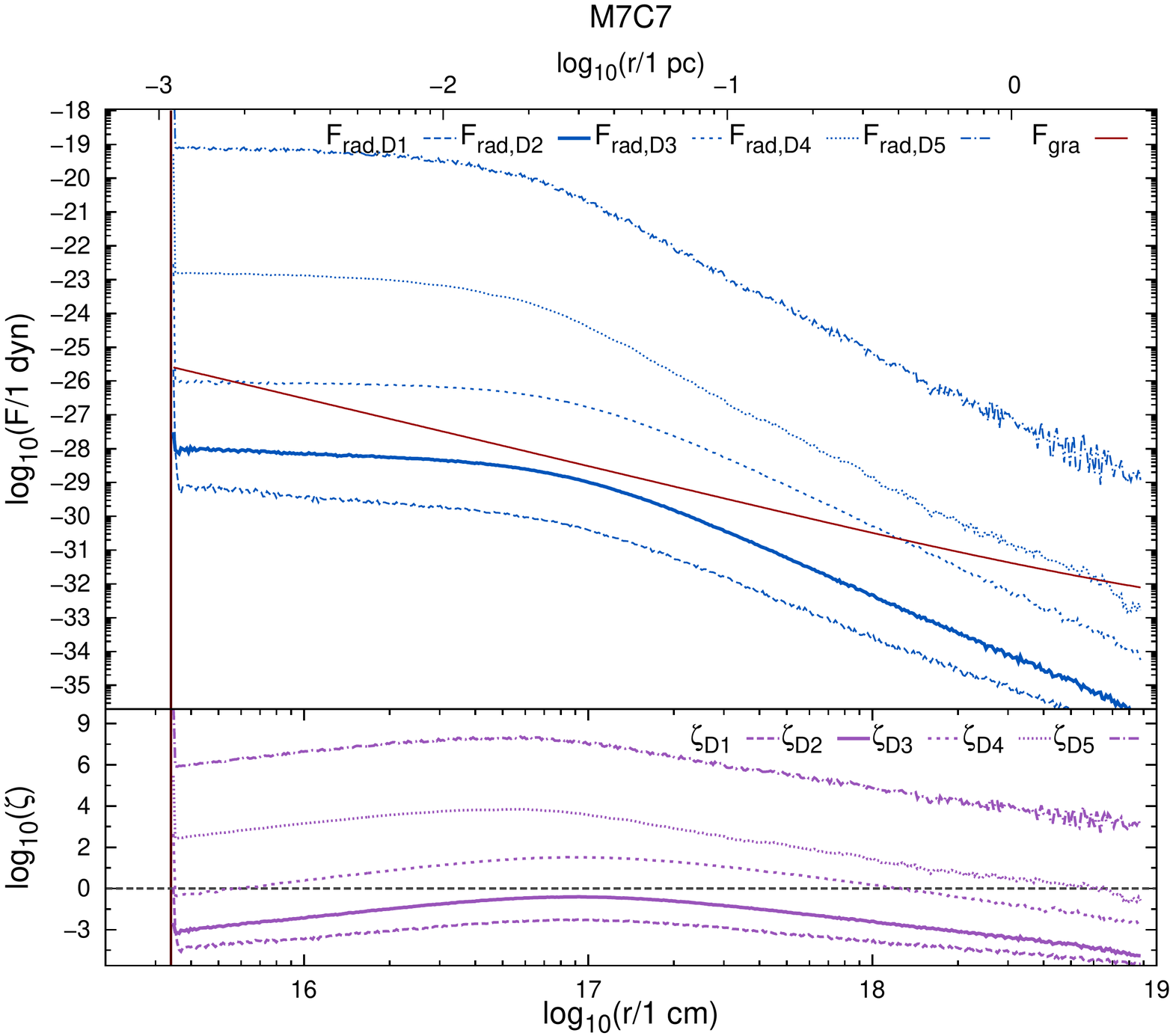}
                        \end{center}
                \end{minipage}  
\end{center}
\caption{Comparison of radiative and gravitational forces for varying dust size 
distribution
upper cutoffs using dust grain models ${\DD1}$, ${\DD2}$, ${\DD3}$, ${\DD4}$, 
and ${\DD5}$ with 
models ${\MM4}{\CC4}$ (left panel) and ${\MM7}{\CC7}$ (right panel). Notation 
as in 
Figure~\ref{fig:ForceCluster}.}
\label{fig:ForceDust}
\end{figure*}
We used the different dust grain size distributions presented in Table 
\ref{tab:parameter} to repeat a subset of the simulations discussed in 
Section~\ref{sect:ResultsConstantModels}. 
Since the radiative force scales with the cluster mass,
we focus on the extreme models ${\MM4}{\CC4}$ and 
${\MM7}{\CC7}$. In Figure~\ref{fig:ForceDust} 
we present the resulting radial distribution of radiative force and gravity. 
When calculating the radiative force, we consistently took the frequency shift 
into account as radiation propagates outwards through the cloud, as was done in 
Figure \ref{fig:ForceCluster}. 
For the low-mass cloud ${\MM4}{\CC4}$ (Figure~\ref{fig:ForceDust} left panel) 
the radiative forces resulting from the ${\DD1}$, ${\DD2}$, and ${\DD3}$ dust 
models are quite similar. The radial radiative force profile deviates slightly 
from a $r^{-2}$ law and the ratio $\zeta$ remains in the gravitationally 
dominated regime, varying between $10^{-3}$ and $10^{-1}$. However, with larger 
dust 
grains in the system the scattering cross section increases and the cloud 
becomes optically thick even for longer wavelengths (see also Figure 
\ref{fig:OpticalDepth}). Here, an effect occurs similar to that in the ${\MM7}$ 
simulations presented in Section~\ref{sect:ResultsConstantModels}. The radius 
where scattering events can take place shifts outwards toward the cloud 
boundary. Therefore, a bulge forms in the radiative force at a distance of 
about 
$0.03\ \rm{pc}$ for the ${\DD4}$ and ${\DD5}$ models. Here, the radiative force 
and gravity are of the same order of magnitude so the ratio $\zeta$ reaches 
unity for model ${\DD5}$ at a distance of $\approx 0.01$--0.1~pc. 
The same trend occurs in the ${\MM7}{\CC7}$ model (Figure~\ref{fig:ForceDust} 
right panel). Since this cloud has a higher mass and thus a higher density, it 
is already optically thick even for the standard MRN (${\DD2}$) dust model. 
Hence, the bulge and the slow decrease in radiative force near the center are 
more pronounced and already become clearly visible for small grain sizes. The 
radiative force exceeds gravity for dust grain models with an upper 
cutoff radius of $a_{\rm{max}} \gtrapprox 2 \mu$ m. In the most extreme 
case of the ${\DD5}$ dust model, the number of scattering events increases 
dramatically even for the longest wavelengths, and the ratio $\zeta$ reaches 
values up to $10^6$. Hence, in the {\MM7}{\CC7}{\DD3} \-- {\MM7}{\CC7}{\DD5} 
models radiation may disrupt the molecular cloud. However, considering the 
processes of grain growth in the ISM \cite[see, e.g.,][]{zhukovska2008, 
zhukovska2016} dust with a radius exceeding $a_{\rm{max}} = 200\ \rm{\mu m}$ 
seems improbable.

\subsection{Variable dust temperature}
\label{sect:ResultsTempModel}
Next we explored how the assumption of a constant dust temperature affects our 
results. In the previous sections the dust temperature was fixed to $T_{\rm{d}} 
= 20\,\rm{K}$ as in previous work on this subject. 
In reality, however,  $T_{\rm{d}}$ varies by over an order of 
magnitude from the inner to the outer edges of the cloud. 
Since the dust temperature determines the redistribution of the cluster 
spectrum 
toward longer wavelengths, we need to investigate the impact of this 
assumption 
and explore how variations in the radial dust temperature change the radiative 
force. To connect with the discussion in 
Section~\ref{sect:ResultsConstantModels}, we focused again on the standard MRN 
dust model ({$\DD2$}), and calculate the parameter $\zeta$ for the full range 
of cloud and cluster models listed in Table~\ref{tab:parameter}. This time we 
used the full {\sc polaris} Monte Carlo RT scheme to self-consistently calculate 
the dust temperature \citep[see][for details]{Reissl2016}.

\begin{figure}[th]
\begin{center}
  \begin{minipage}[c]{0.99\linewidth}
    \begin{center}
      \includegraphics[width=1.0\textwidth]{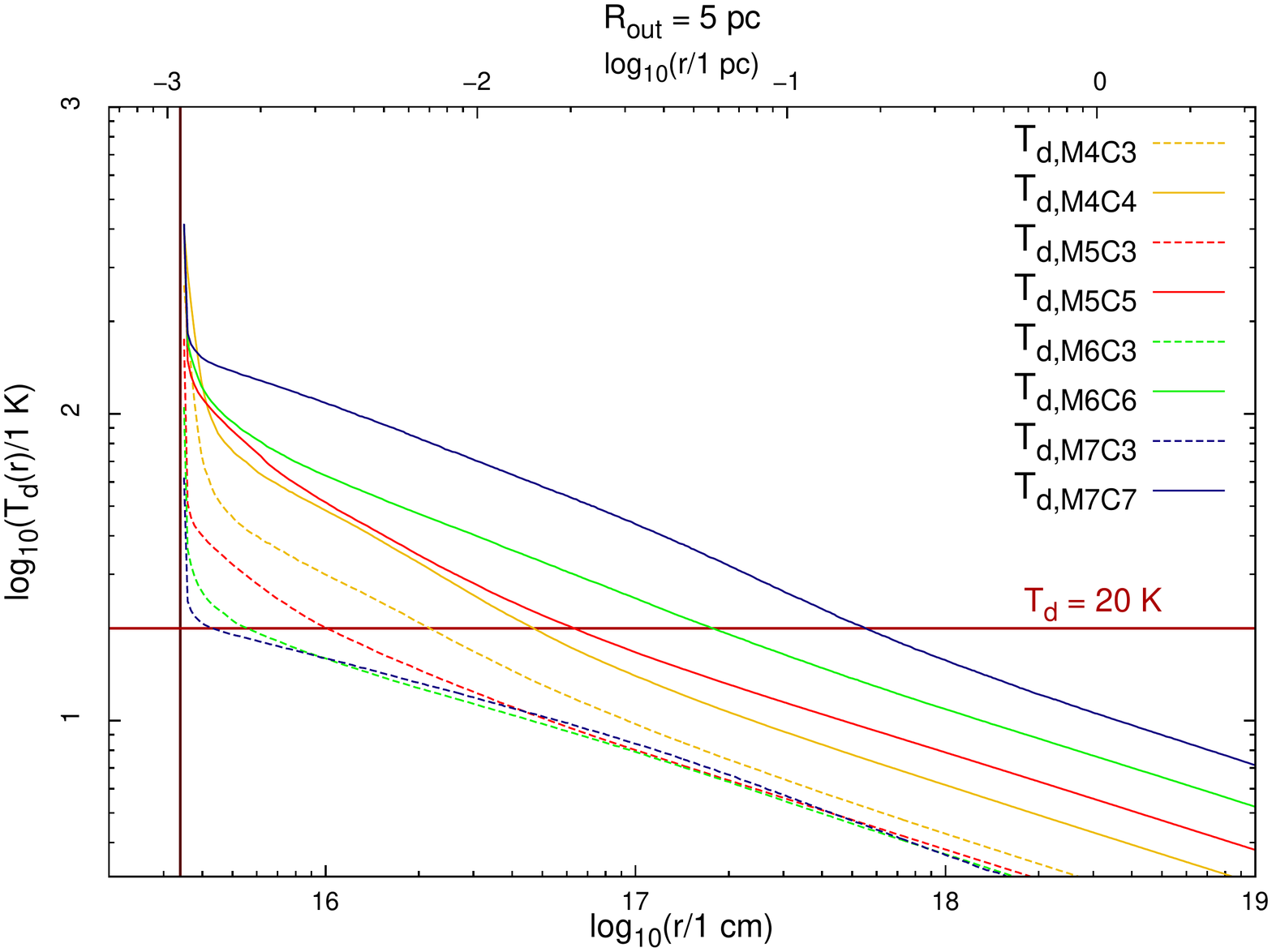}
    \end{center}
  \end{minipage}
\end{center}
\caption{Grain-size averaged dust temperature distribution for different cluster 
and molecular cloud models with an outer radius of $R_{\rm{out}} = 5\ \rm{pc}$. 
The vertical dark brown line marks the sublimation 
radius.} 
\label{fig:TempDistribution}
\end{figure}
\begin{figure*}[th]
\begin{center}
  \begin{minipage}[c]{0.49\linewidth}
    \begin{center}
      \includegraphics[width=1.0\textwidth]{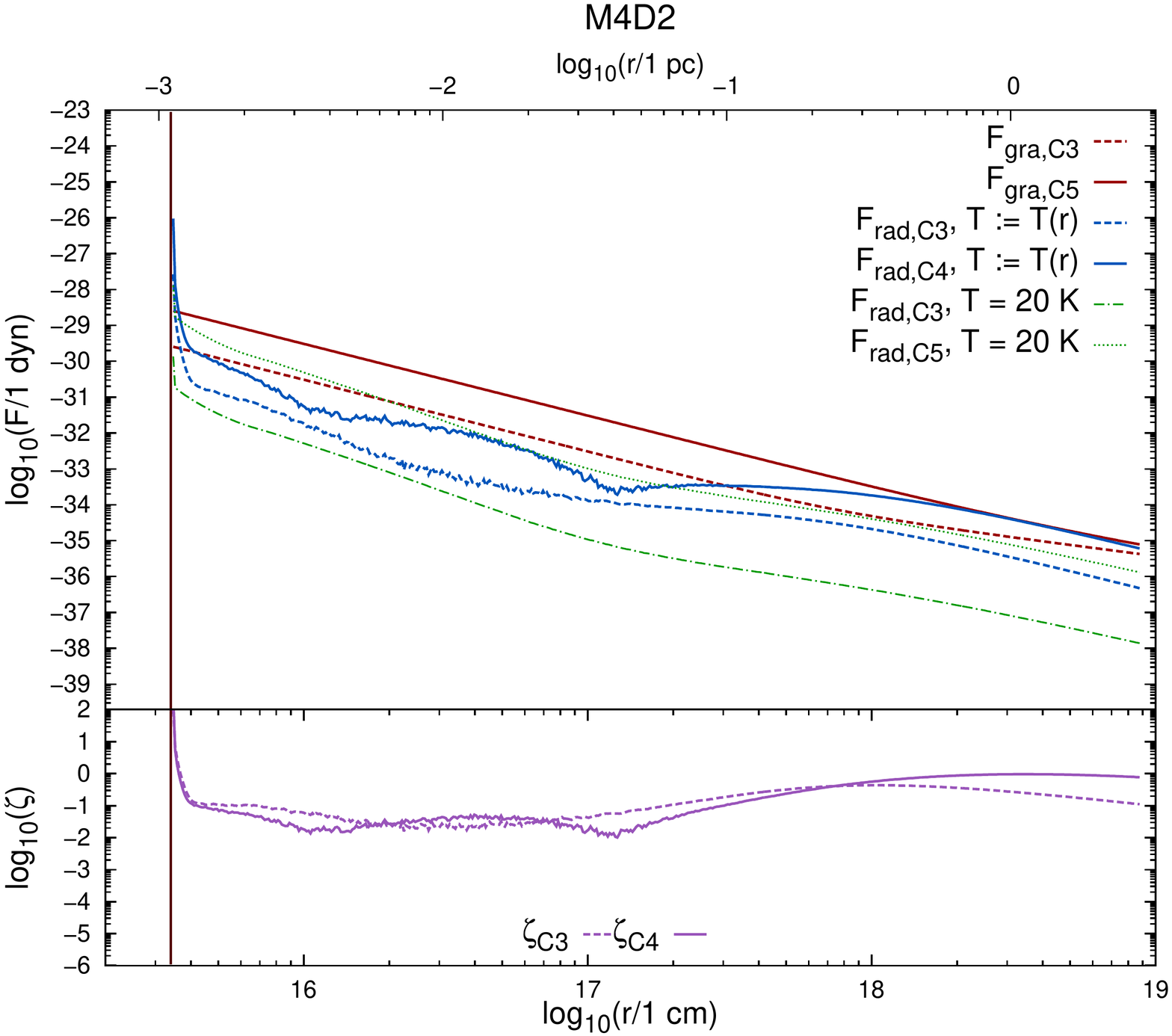}                  
    \end{center}
  \end{minipage}
  \begin{minipage}[c]{0.49\linewidth}
    \begin{center}
      \includegraphics[width=1.0\textwidth]{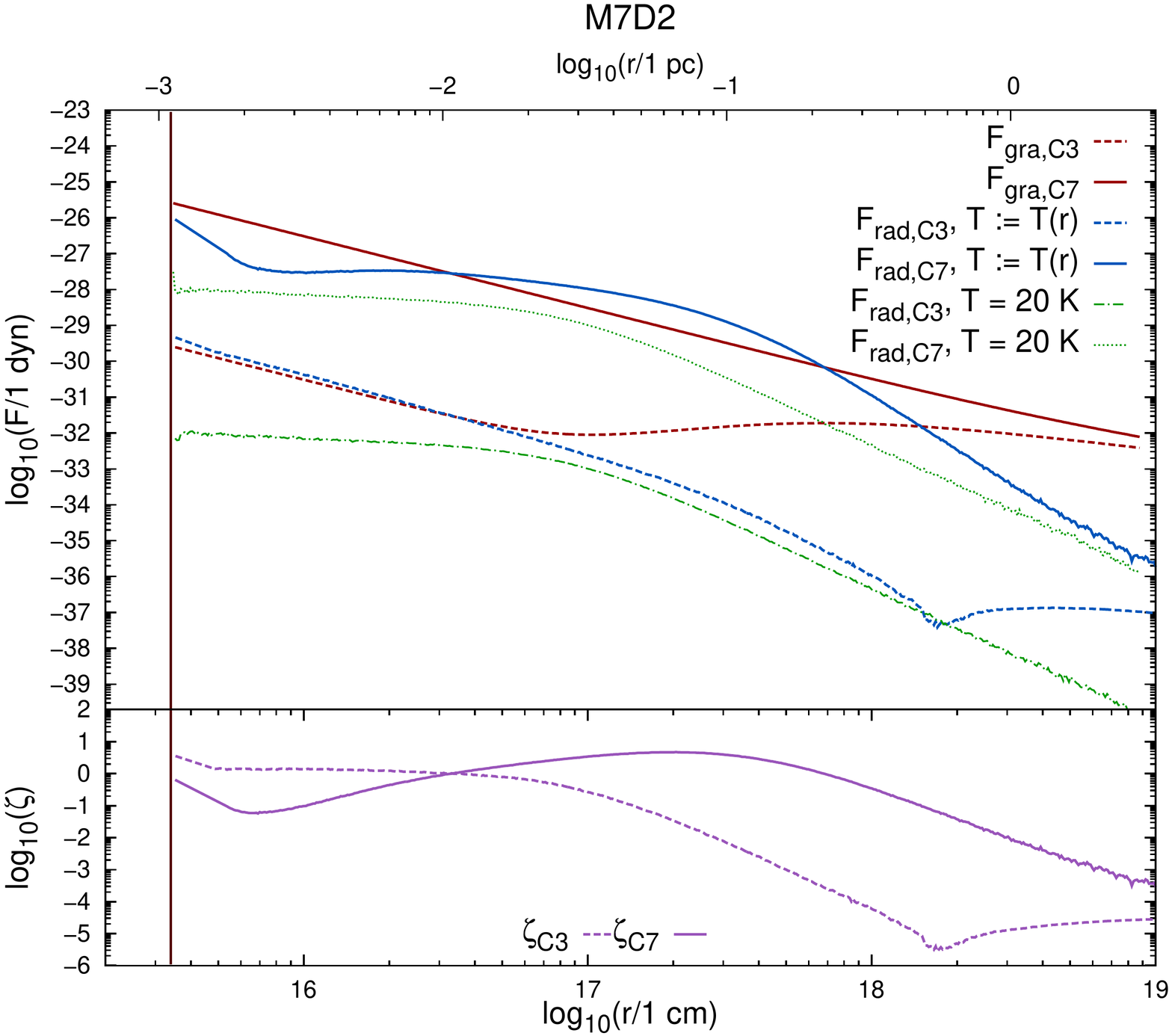}
    \end{center}
  \end{minipage}  
\end{center}
\caption{Comparison of radiative and gravitational forces in models with 
self-consistent (blue lines) and constant  (green lines) dust temperature.
Cloud 
models are ${\MM4}$ {\em (left)} and ${\MM7}$ {\em (right)}, and radiative 
forces were calculated for cluster models ${\CC3}$, ${\CC5}$, and ${\CC7}$ and 
the ${\DD2}$ 
dust model. The radial dust temperature distributions follow the temperatures 
shown in Figure~\ref{fig:TempDistribution}. The ratio $\zeta$ 
is shown below {\em (purple)} for the variable dust temperature models.}
\label{fig:ForceTemp}
\end{figure*}

Figure~\ref{fig:TempDistribution} presents the angle-averaged dust temperatures 
for different molecular cloud models and clusters. The plot demonstrates that 
the dust temperature can indeed reach several hundred Kelvin near the cloud 
center. However, it also indicates that $T_{\rm{d}}$ drops quickly with 
increasing radius and falls significantly below $20\ \rm{K}$ at 
0.01--0.7~pc, in the outskirts of the cloud. 
Figure~\ref{fig:ForceTemp} shows the resulting radiative forces and gravity for 
the molecular cloud models ${\MM4}$ and ${\MM7}$. Similar to the models with 
large grain sizes presented in 
Figure~\ref{fig:ForceDust}, calculations that take radial temperature variation 
into account also show an  increase in radiative force compared to simulations 
with a constant temperature. \\
However,  in the case of variable temperature this 
is not due to an increase in optical depth (see Figure \ref{fig:OpticalDepth}). 
A constant temperature of $20\ \rm{K}$ leads to an efficient distribution of 
the initial cluster spectrum toward a wavelength regime where lower extinction 
allows the photons to escape more easily. With higher temperatures close to the 
center this shift in wavelength is smaller. Consequently, the photons remain 
trapped for longer close to the cloud center. For model ${\MM4}$ 
(Figure~\ref{fig:ForceTemp}, left panel) this results in an additional bulge in 
the radial force at a distance of about $0.02\,$pc similar to that observed in 
Section~\ref{sect:ResultsDustModel}. This increased photon trapping is even 
more effective in the ${\MM7}$ (Figure \ref{fig:ForceTemp} right panel). 
Compared to the calculations assuming a constant dust temperature shown in 
Figure~\ref{fig:ForceCluster} (lower right panel) a higher inner temperature 
leads to a steeper slope. This allows the radiative force to be of the same 
order as gravity and even to exceed it out to a radius of $\sim0.1\,\rm{pc}$. 
We note, however, that the typical size scale of massive star clusters is about 
1$\,$pc
\cite[see, e.g.,][and references therein]{portegieszwart2010}, considerably larger than the region of radiative acceleration
seen here, so values of 
$\zeta \gtrsim 1$ are an artifact of our assumption of point-like clusters 
(model ${\FF0}$ in Table \ref{tab:parameter}).  
 
\begin{figure}[ht]
        \begin{minipage}[c]{1.0\linewidth}
                        \begin{center}
                               
\includegraphics[width=1.0\textwidth]{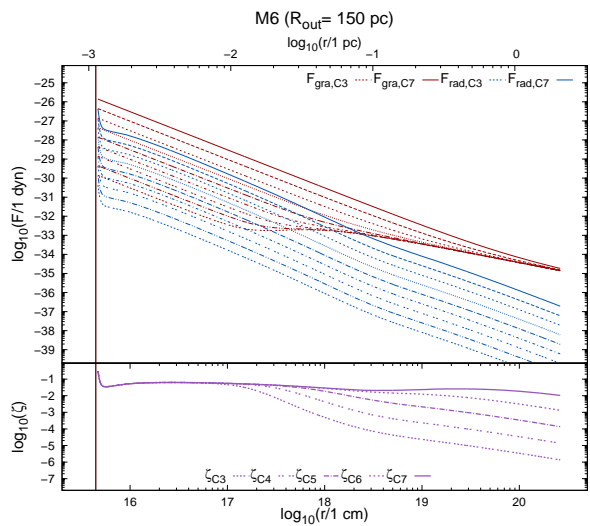}
                        \end{center}
         \end{minipage}                         
\caption{Comparison of radiative and gravitational forces for model ${\MM6}$, 
except with an outer radius of $R_{\rm{out}} = 150\,$pc instead of the $5\,$pc 
used 
in Figure~\ref{fig:ForceCluster}. Again, we considered point-like cluster models 
${\CC4}$ to ${\CC7}$ with masses of $10^4\,$M$_{\odot}$ to $10^7\,$M$_{\odot}$. 
}
\label{fig:CloudLarge}
\end{figure}
  That is, concentrating the radiation source to a point in the center
  of the cloud rather than a distributed region increases the
  radiation pressure in the innermost regions.  
When we relaxed this assumption in Section~\ref{sct:Result3DStellarDistribution}, 
we indeed saw lower values of $\zeta$ in the central regions, as expected. 

\begin{figure}[th]
\begin{center}
  \begin{minipage}[c]{0.9\linewidth}
    \begin{center}
      \includegraphics[width=1.0\textwidth]{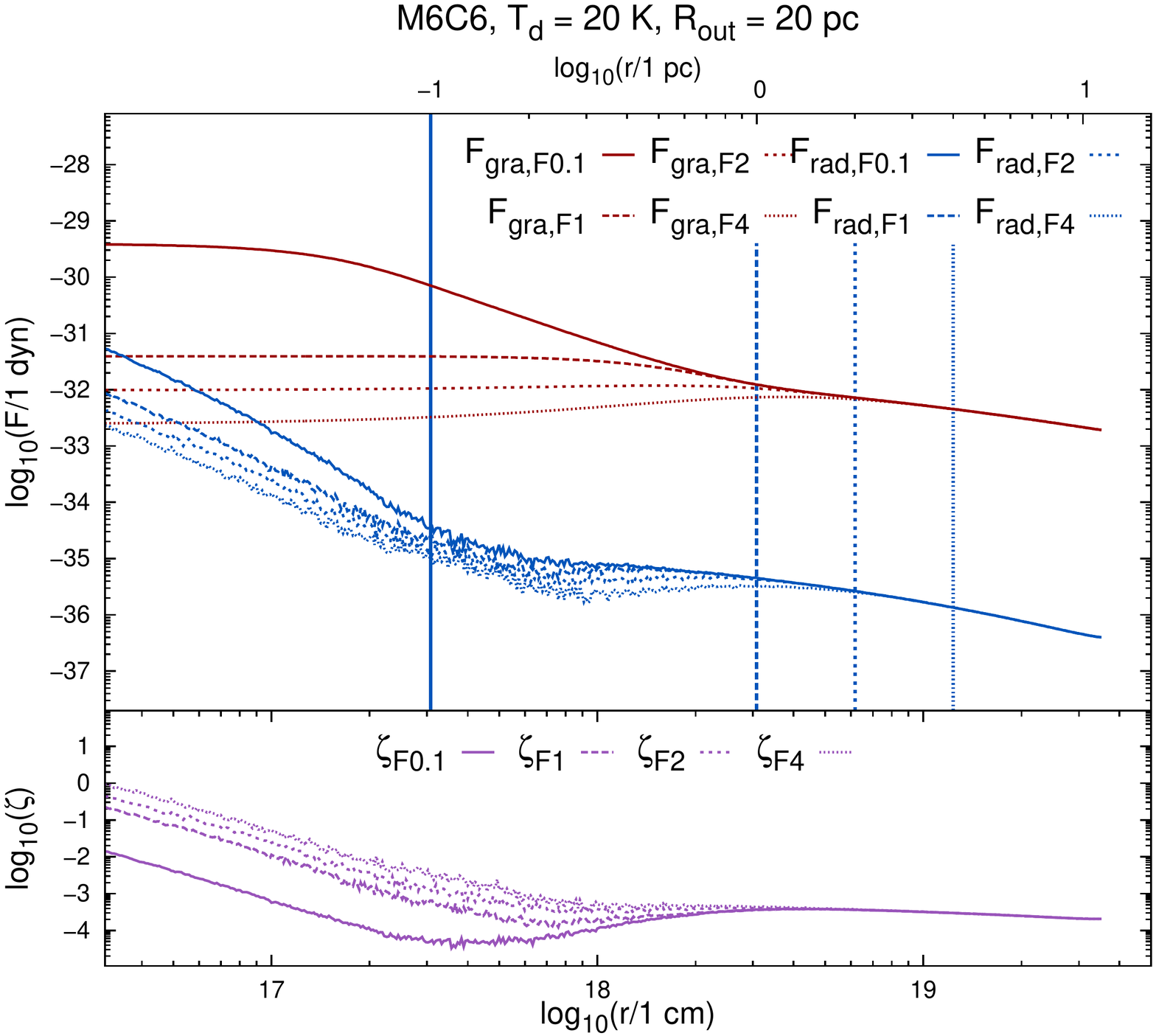}\\
      \includegraphics[width=1.0\textwidth]{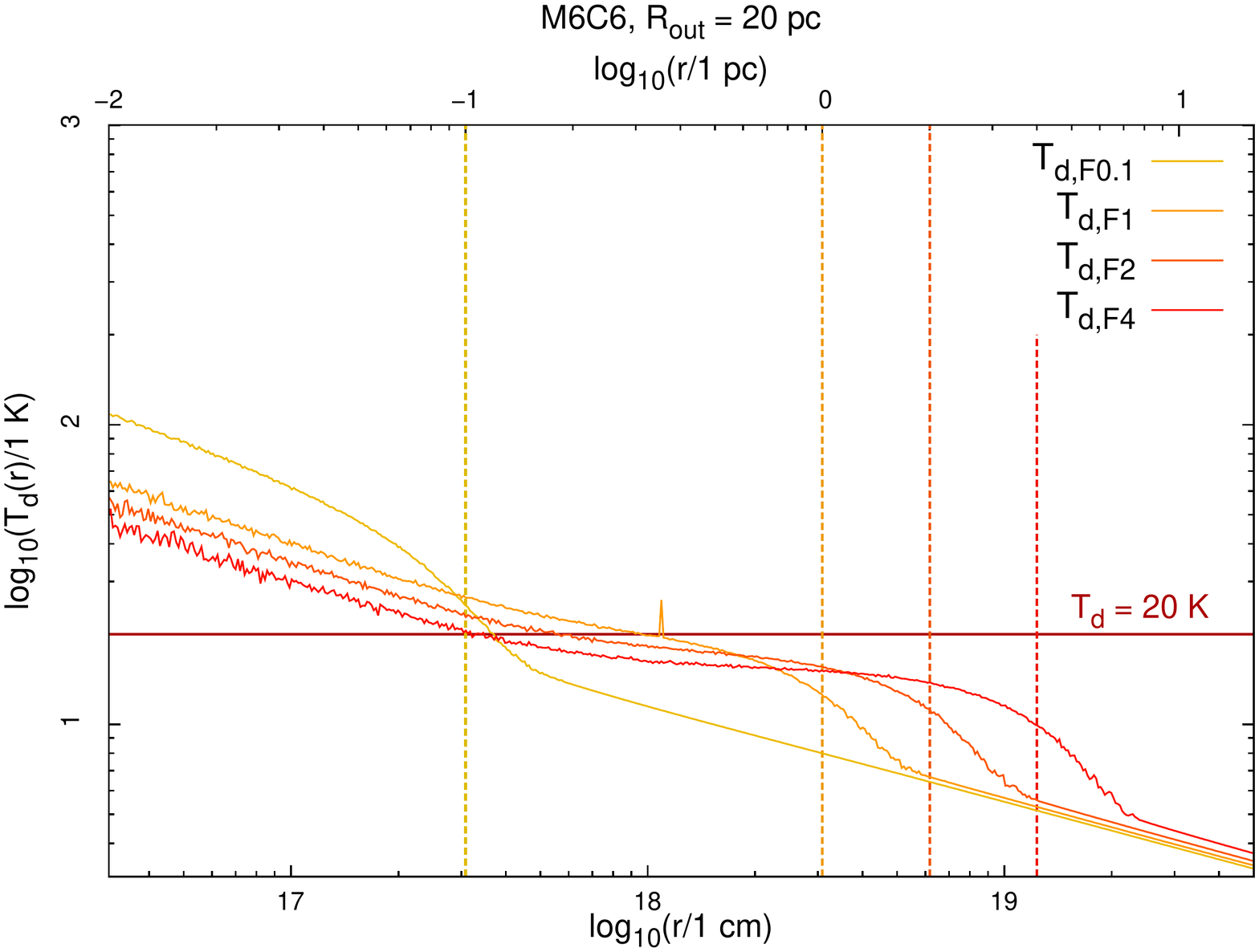}\\
      \includegraphics[width=1.0\textwidth]{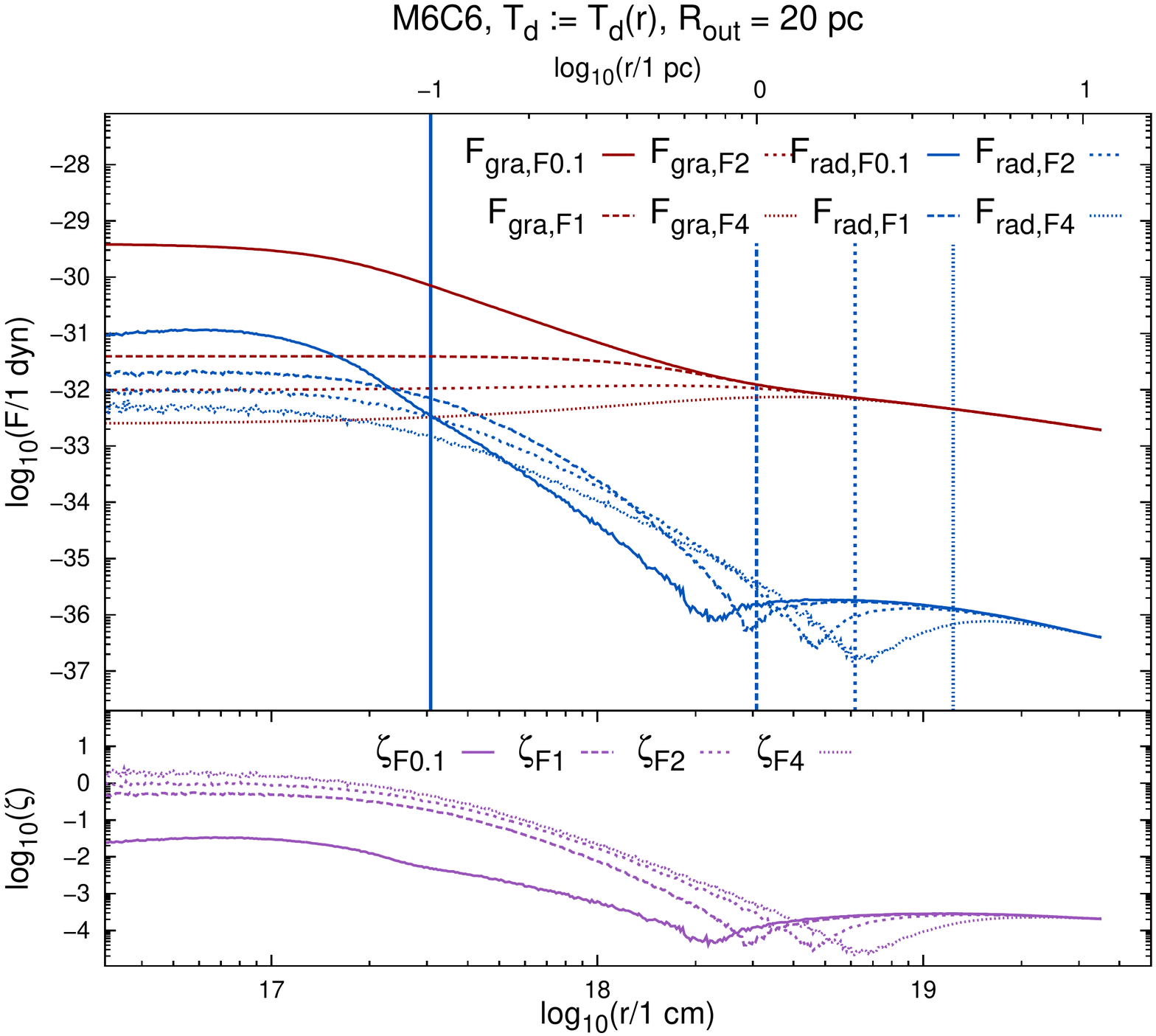}
    \end{center}
  \end{minipage}
\end{center}
\vspace{-5mm}
\caption{{\em Top:} Gravity in comparison to radiative force for varying 
stellar cluster 
size in the ${\MM6}{\CC6}$ model with a constant temperature of $T = 20\ 
\rm{K}$ and an outer radius of $R_{\rm{out}} = 5\ 
\rm{pc}$. Vertical dotted lines give the FWHM of the Gaussian 
spatial distribution of the cluster, with values $0.1$, $1$, $2$, and $4\ 
\rm{pc}$.
{\em Middle:} Grain-size averaged dust temperature distribution for the 
${\MM6}{\CC6}$ model dependent on the different FWHM.
{\em Bottom:} Same as the top panel with the radial dependent dust temperature 
shown in the middle.} 
\label{fig:ForceExtendedCluster}
\end{figure}

\subsection{Impact of cloud and cluster size}
\label{sct:Result3DStellarDistribution}
Finally, we investigated the impact of different cloud and cluster sizes, relaxing 
the simplifying assumption of point-like clusters. We began with cloud size, and 
repeated our simulations for an ${\MM6}$ model with an outer radius of 
$R_{\rm{out}}= 150\,$pc, rather than $5\,$pc, with the usual MRN (${\DD2}$) dust 
model, and with a fixed dust temperature of $T_{\rm d} = 20\,$K. For the dust 
sublimation radius we still assumed a value of $R_{\rm{sub}} = 1.1\times
10^{-3}\,$pc. As before, we varied the cluster mass from $10^4$ to 
$10^7$ solar masses, corresponding to star formation efficiencies from 1\% to an 
unrealistically high 91\%. This choice of parameters includes a cloud very 
similar to the model of the Milky Way molecular cloud G298.4-0.3
considered by \citet{murray2010},  who adopted a star formation efficiency of 3\%.
Figure~\ref{fig:CloudLarge} shows the radiative and gravitational forces and 
their ratio for this larger cloud. In contrast to the ${\MM6}$ model plotted in 
Figure~\ref{fig:ForceCluster} in the top right panel, the larger molecular 
cloud  has a far lower surface density, and subsequently, a slower increase in 
optical depth. Hence, it is optically thin at much shorter wavelengths, and the 
probability of interaction between radiation and dust is reduced. However, the 
radiative force is down by only two orders of magnitude in this case because 
the 
redistribution in wavelength is also less efficient, so the ratio $\zeta$ remains at 
an almost constant value of $\sim10^{-2}$ in the inner regions of the cloud, 
rather than dropping below that as in the smaller cloud model.  The larger 
ratio 
toward the border of the cloud is solely a result of the increasing influence 
of the cloud mass. Indeed, after its sharp decline near the sublimation radius 
the radiative force follows an almost exact $r^{-2}$ power law. The resulting 
forces of radiation and gravity scale exactly with the considered cluster mass 
in all of our models. All simulations have also in common that the ratio 
$\zeta$ 
remains around or below $10^{-2}$, making gravity the dominant considered force 
in these simulations, independent of the combination of cloud model and cluster 
mass. We now turn our attention to cluster size.  As discussed above, some 
${\FF0}$ cluster models show values of $\zeta \gtrsim 1$ in the innermost 
$0.1\,$pc of the cloud, where the intense radiation field can lead to multiple 
scatterings per photon. This overestimates the strength of radiation pressure 
for three reasons. First, a point-like cluster maximizes the radiation flux per 
unit mass in the interior of the cloud, and hence the radiative force. Second, 
all photons from the cluster are emitted from a common origin. The resulting 
forces exerted 
by every photon from absorption and back scattering are therefore purely 
additive, meaning no  component of the radiative force from one star ever 
cancels another. Third, the dust temperatures near the cluster are maximal 
for a point-like cluster. A spatially extended cluster produces less radiation 
energy per dust grain near the center, 
and so will result in comparatively lower dust temperatures. 
To investigate these effects we again considered the ${\MM6}$ model, but now 
with an outer radius of $R_{\rm{out}}= 20\,$pc and with a fixed cluster mass of 
$10^6\,$M$_{\odot}$. The model can be compared to the models of the
W49 giant molecular cloud in the Milky Way  and the M82 starburst galaxy
in  \citet{murray2010}, which have cloud masses of 
$7.5\times10^5\,$M$_{\odot}$ and $3\times10^6\,$M$_{\odot}$ as well as
cluster masses  of $4.3\times10^4\,$M$_{\odot}$ and
$7\times10^5\,$M$_{\odot}$, respectively, with cloud radii of $22\,$pc
and $30\,$pc. This corresponds to star formation  
efficiencies of 6\% and 24\%, and so our choice of 50\% leads to comparatively 
stronger radiative feedback than \citet{murray2010}.
Figure~\ref{fig:ForceExtendedCluster} presents our results for clusters with 
varying sizes. As outlined in Table \ref{tab:parameter}, the stellar 
distribution is now sampled from a 3D Gaussian distribution with a characteristic 
full-width half maximum (FWHM) of $0.1$, $1$, $2$, or $4\,$pc. This changes both 
the calculation of the gravitational potential and of the radiative output of 
the cluster (see Section~\ref{sect:ClusterModel}). The inward gravitational 
force no longer follows a $r^{-2}$ law. It is almost constant within the 
cluster FWHM with values that are several hundred times smaller than for the 
point-like cluster. Further out, the forces converge when the potential is 
dominated by cloud material. The cluster photons are no longer emitted  from a 
single point, but instead the distance is randomly sampled from a 3D Gaussian 
distribution with the corresponding FWHM.  Consequently, the radiative 
force also changes compared to the previous point-like
models. Figure~\ref{fig:ForceExtendedCluster}  
demonstrates that the  combined impact of these three effects
decreases the effective radiation pressure by a factor of 10--30  
as the cluster size increases. As a result, gravity dominates at all radii, 
by factors of a few in the center and by three to four orders of magnitude in the 
cloud outskirts. 

\subsection{Results summary}
\begin{figure*}[tp]
                \begin{minipage}[c]{0.47\linewidth}
                        \begin{center}
                                
\includegraphics[width=0.99\textwidth]{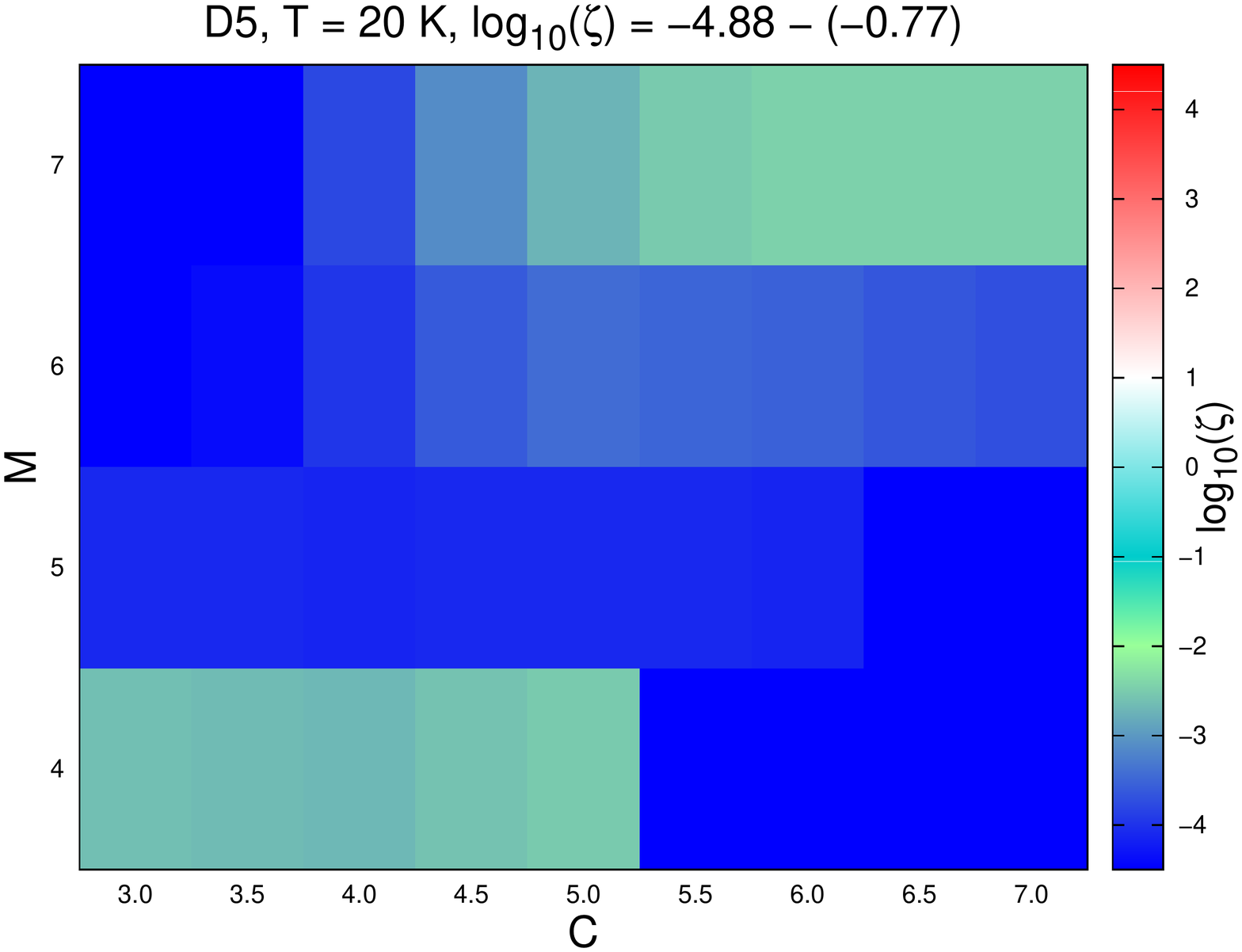}\\
\includegraphics[width=0.99\textwidth]{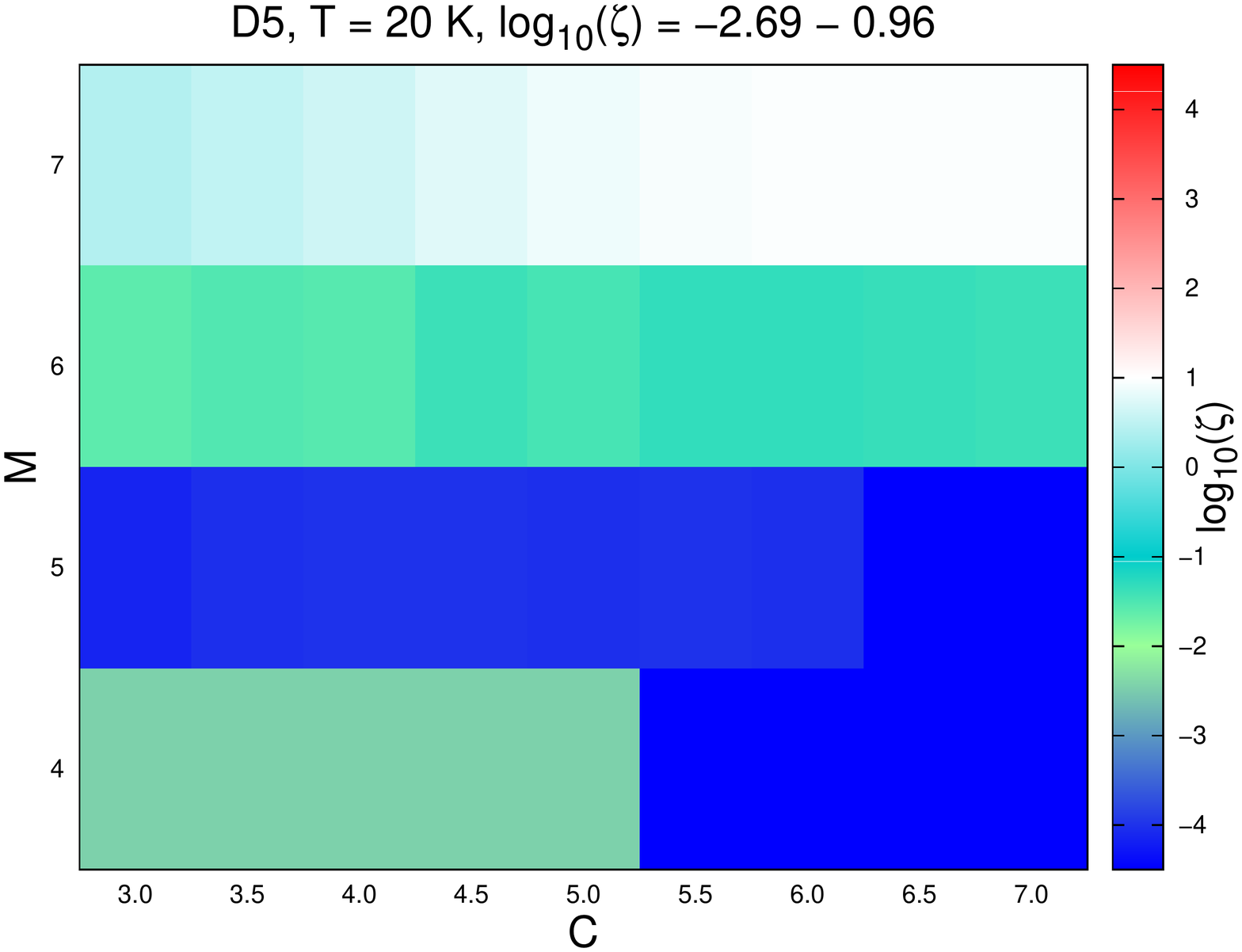}\\
\includegraphics[width=0.99\textwidth]{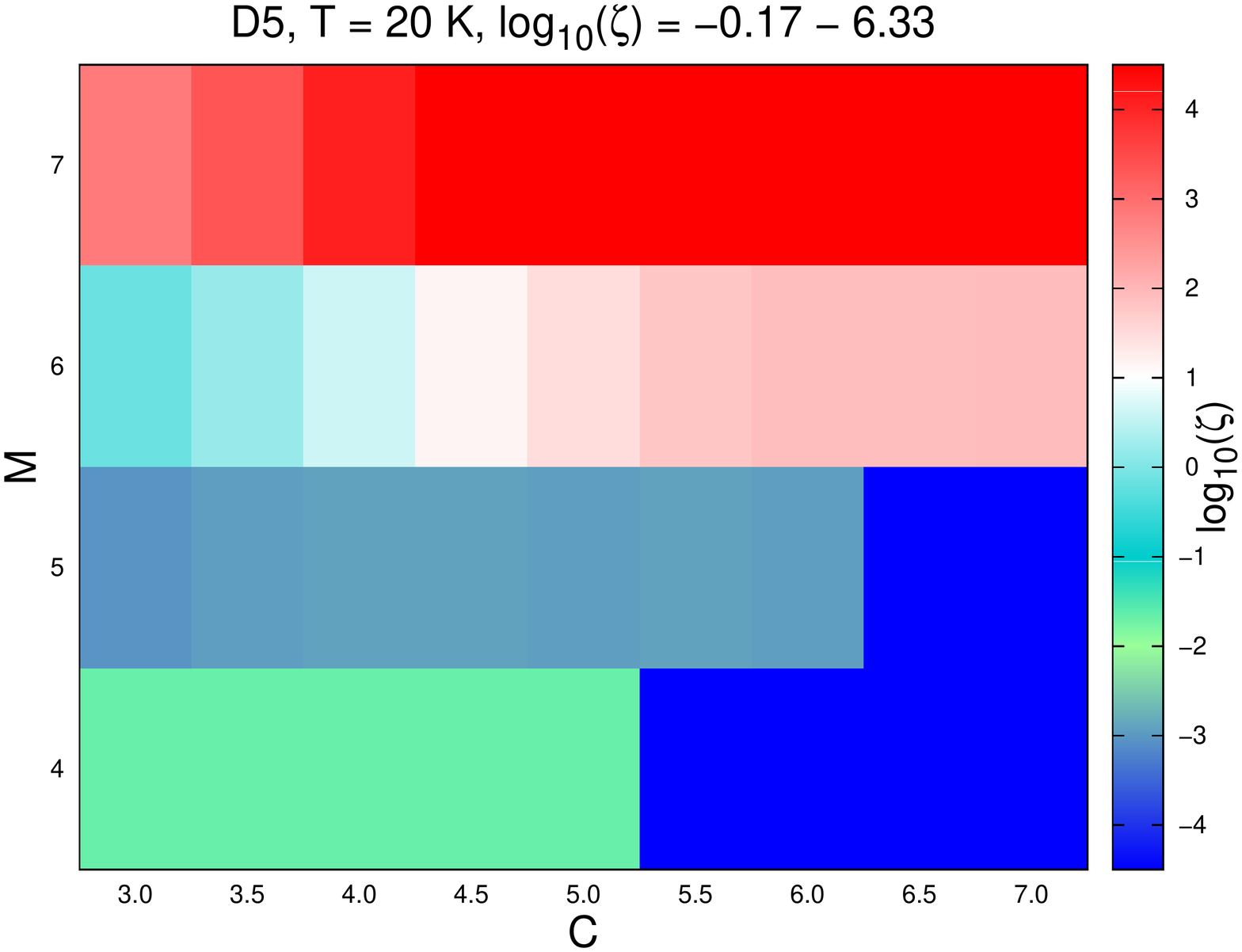}
                        \end{center}
                \end{minipage}
                \begin{minipage}[c]{0.47\linewidth}
                        \begin{center}
                                
\includegraphics[width=0.99\textwidth]{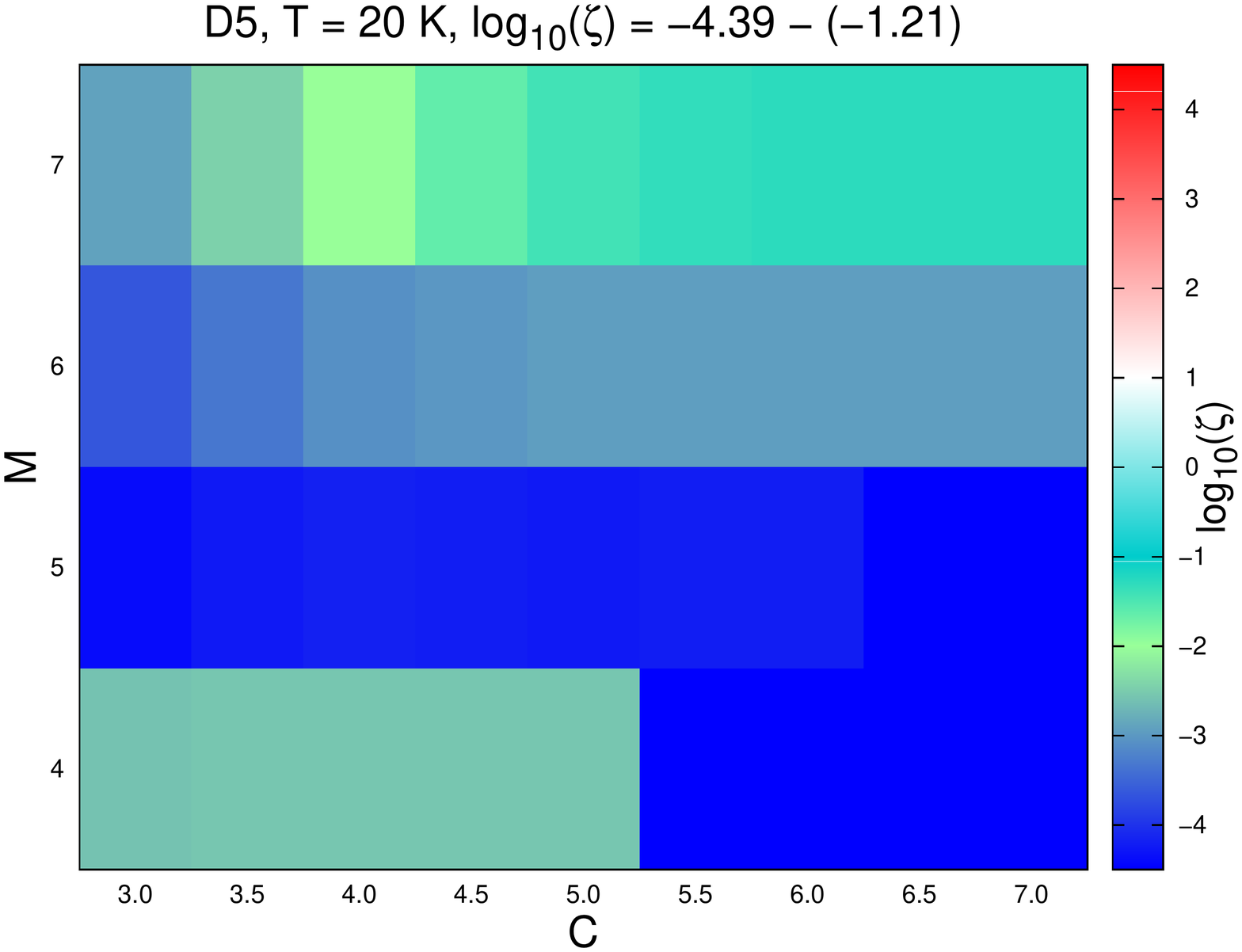}\\
\includegraphics[width=0.99\textwidth]{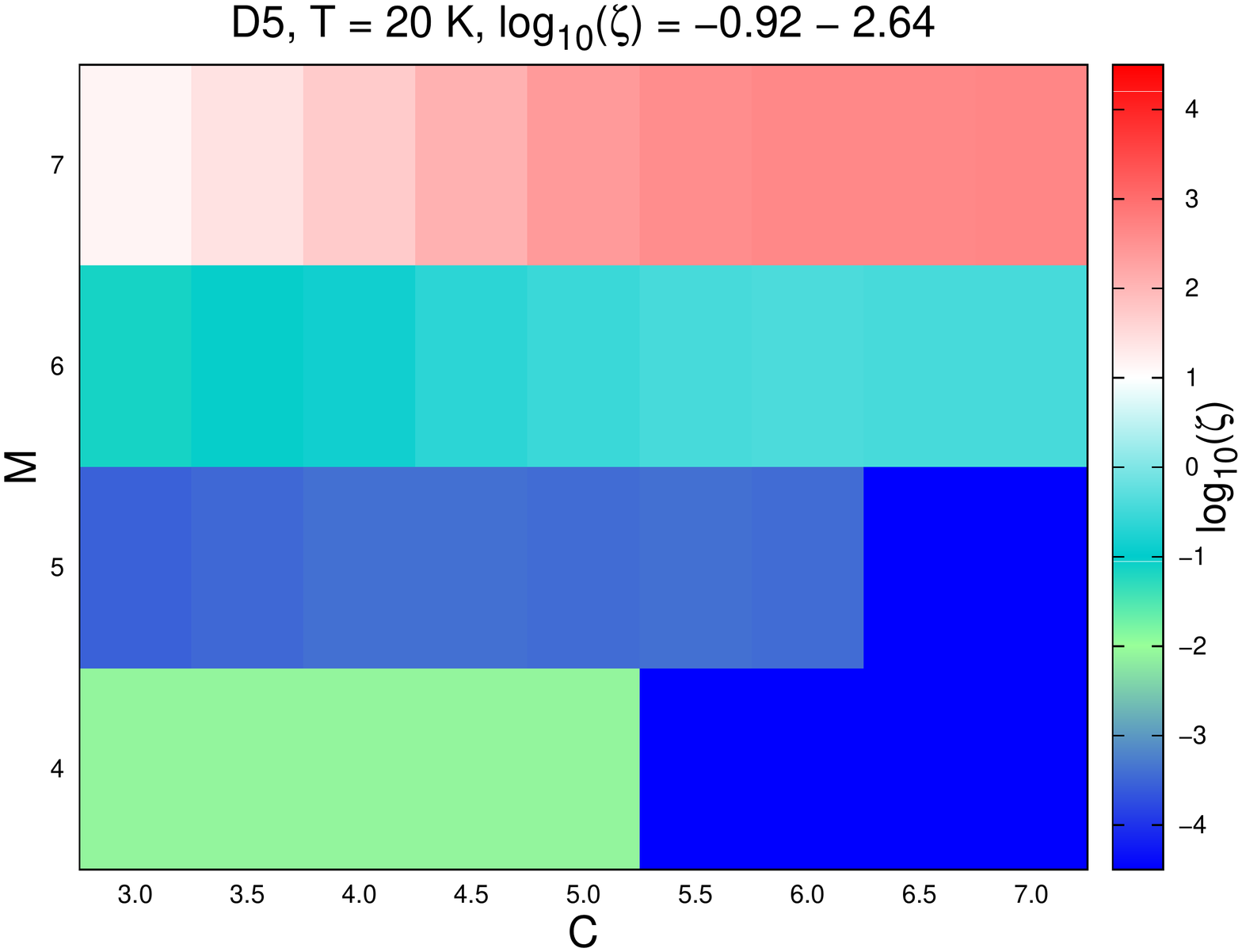}\\
\includegraphics[width=0.99\textwidth]{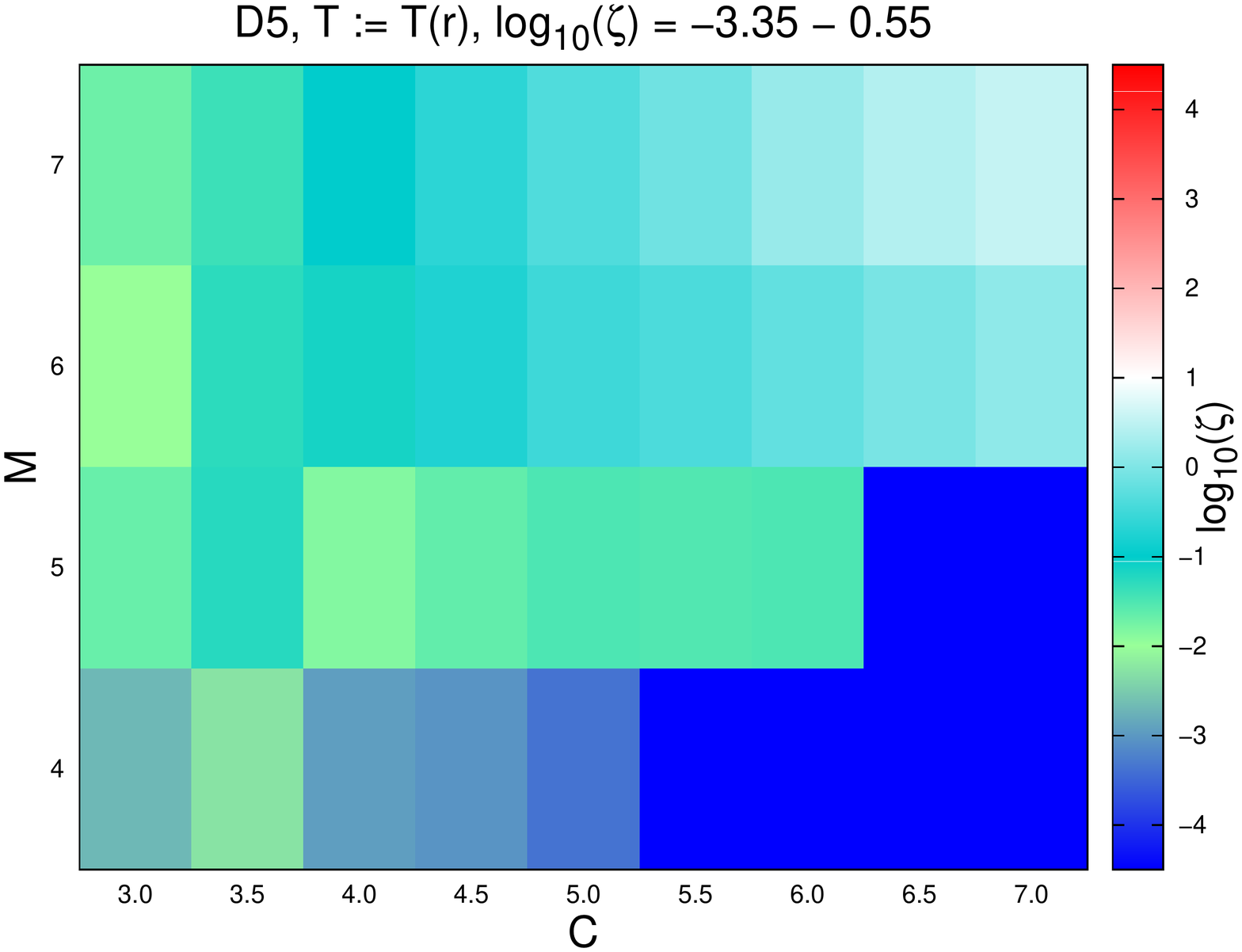}
                        \end{center}
                \end{minipage}
\caption{Parameter space of the ratio of radiative force to gravity $\zeta$ 
dependent on cluster model ${\textsf{C}}$, molecular cloud model 
${\textsf{M}}$, and the different dust models ${\DD1}$, ${\DD2}$, ${\DD3}$, 
${\DD4}$, and ${\DD5}$. The ratios are calculated
    for a distance of $r = 0.1\,$pc from the cluster center. The panel on 
the bottom
    right panel is for the radially dependent temperature distribution
    shown in the left panel of Figure \ref{fig:TempDistribution} while
    the other panels are for a constant dust temperature of $T_{\rm{d}} = 
    20\,$K. We note that this image is smoothed when reading it with MacOS {\sc preview}.}
\label{fig:ZetaRatios}
\end{figure*}

A summary of the full parameter range covered for point-like clusters is 
presented in Figure \ref{fig:ZetaRatios}.  It shows the ratio $\zeta$ for all 
calculations discussed so far, and it demonstrates that for any reasonable 
permutation of molecular cloud and cluster masses, and for typical Galactic 
dust grain size distributions the ratio of outward radiative force and inward 
gravitational force barely reaches unity. 
Including the effects of extended stellar clusters or larger clouds 
reduces the ratio further, leaving radiation pressure dominant only in 
unrealistic cases requiring star formation efficiencies approaching 
$\epsilon_{{tot}} \sim 100\ \%$ and maximum dust sizes $a_{\rm max} \gtrapprox 
2\ \rm{\mu m}$.
For typical conditions in the Milky Way or in other Local Group galaxies, 
radiation pressure on dust grains from young clusters cannot disrupt the 
parental molecular cloud, and thus this process is not likely to contribute 
strongly to the (self) regulation of the star formation process.

\section{Discussion}
\label{sect:Discussion}
To better connect our results to the observational domain, we  calculate d
the resulting spectral energy distribution that an external observer would see 
when looking at the cloud. We then compared our models with the existing 
literature and discuss the limitations of our approach. 
\subsection{Observational consequences}
\label{sect:Observations}
\begin{figure}[ht]
  \begin{center}
    \includegraphics[width=0.49\textwidth]{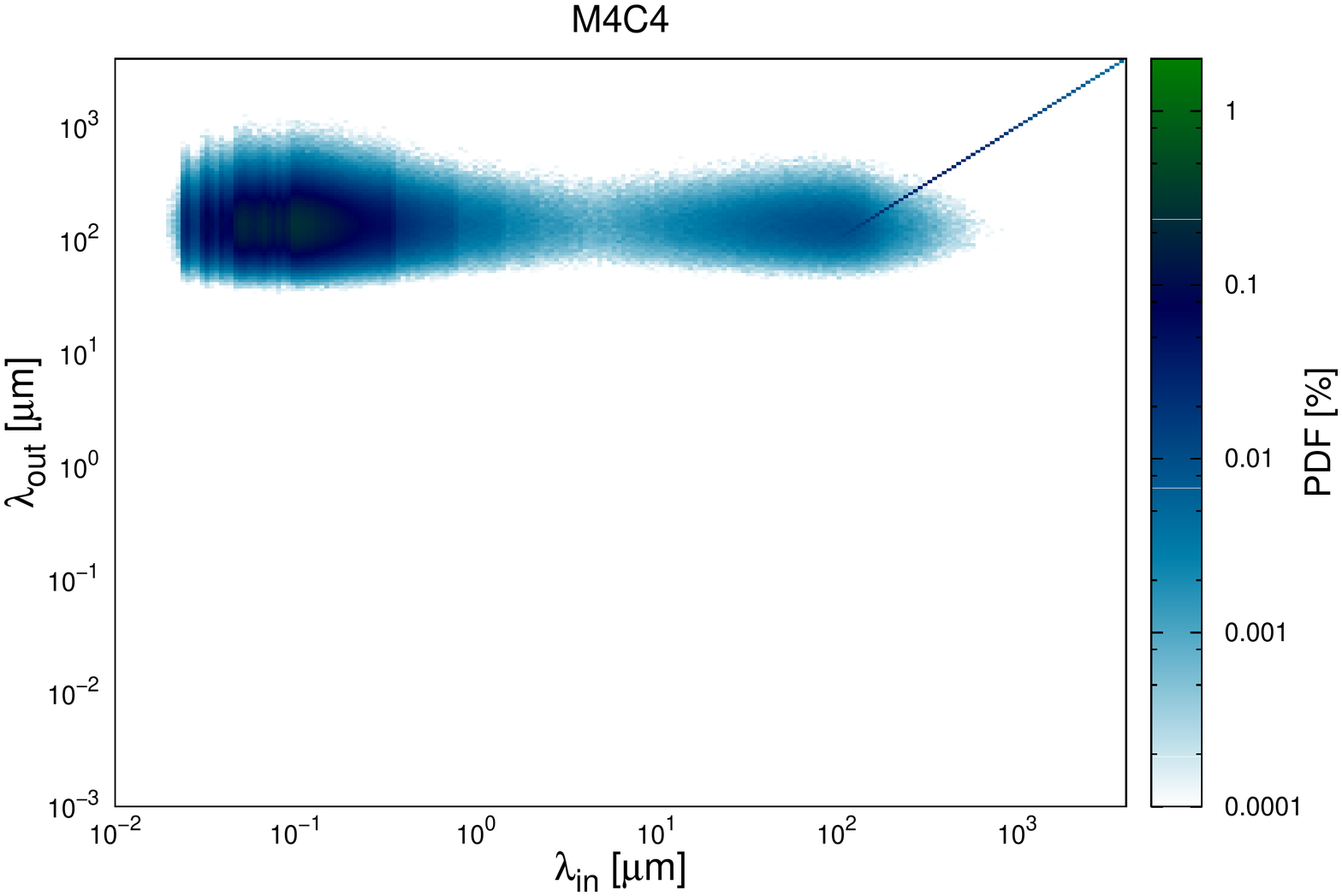}
  \end{center}
\caption{Probability for each permutation of shift between initial wavelength 
and final wavelength for the ${\MM4}{\CC4}$ model. The abscissa shows the  
wavelength of the photon packages initially  emitted by the cluster while the 
ordinate shows the final observable wavelength leaving  
the molecular cloud.  
}
\label{fig:RatioSpectrumShift}
\end{figure}

\begin{figure*}[th]
  \begin{center}
    \includegraphics[width=0.49\textwidth]{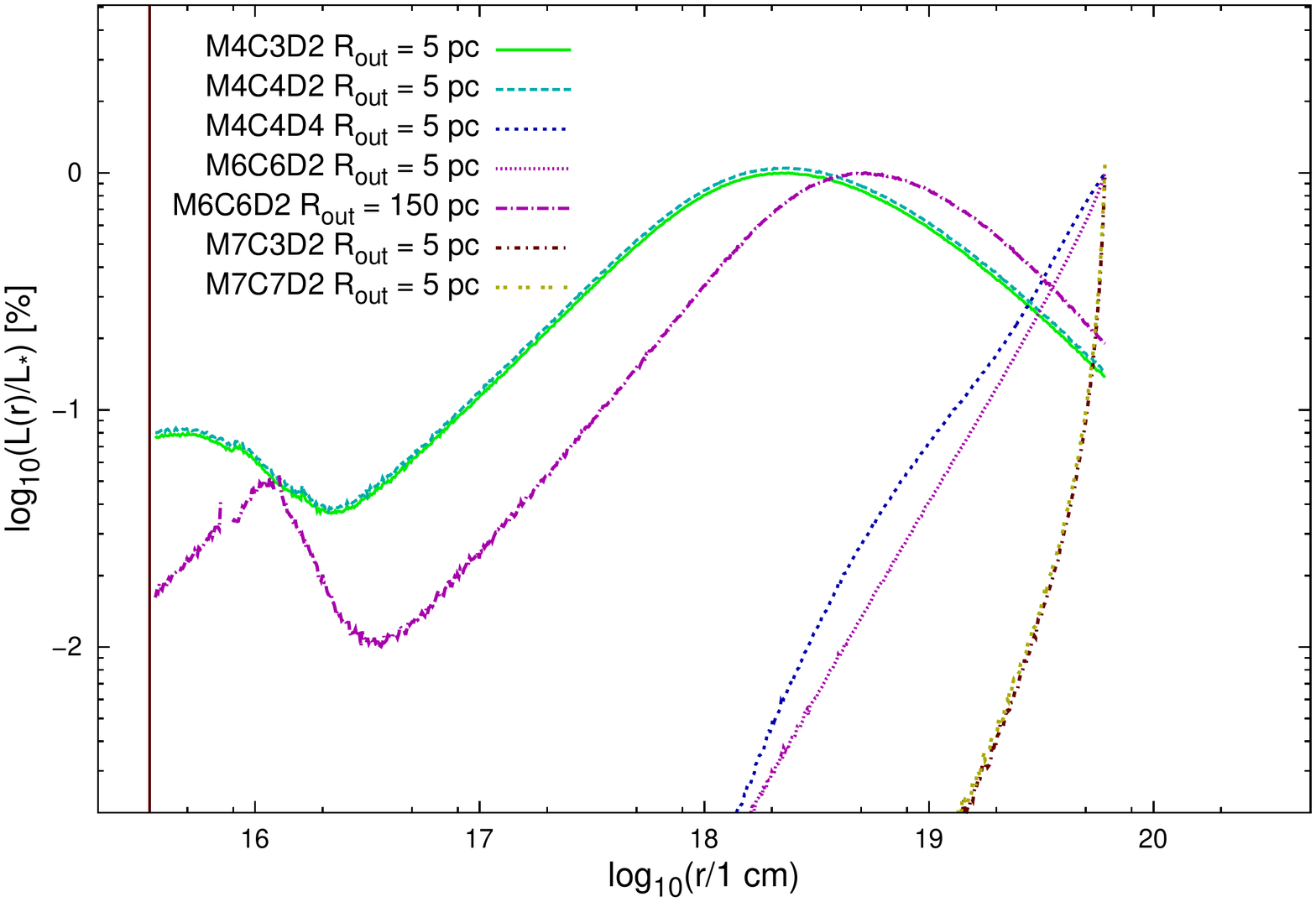} 
   \includegraphics[width=0.49\textwidth]{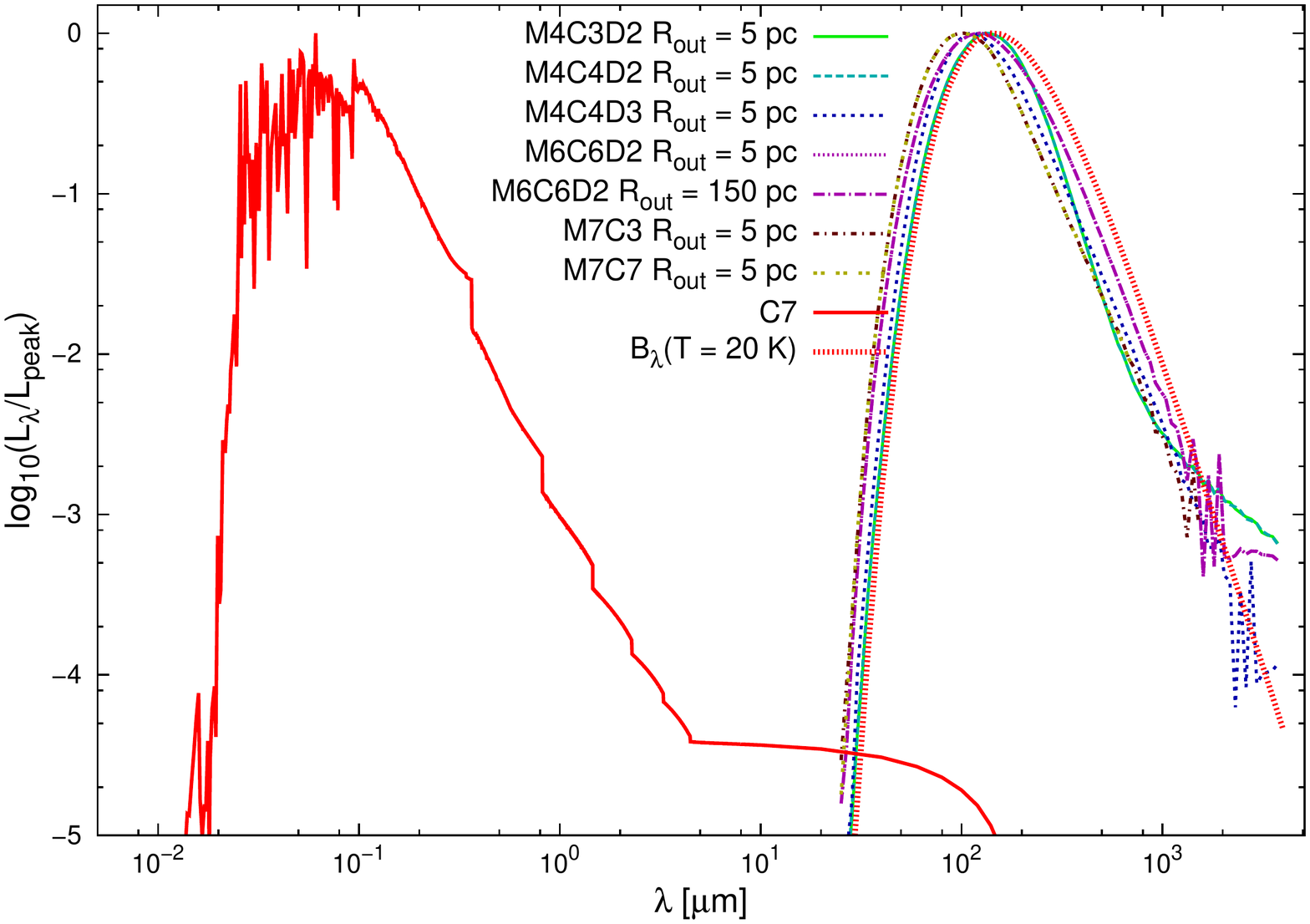}
  \end{center}
 \caption{{\em Left:} Escape fraction of photons as a function 
of cloud radius 
for various models. 
All the models are modeled with constant dust temperature $T_{\rm{d}}=20\ K$. 
{\em Right:} 
Corresponding spectra visible to an outside observer compared to the normalized 
stellar input spectrum. All output spectra are very similar and are reasonably 
well described as a blackbody with a temperature of $20\,$K.}
\label{fig:ObservationsConst}
\end{figure*}

\begin{figure*}[ht]
  \begin{center}
    \includegraphics[width=0.49\textwidth]{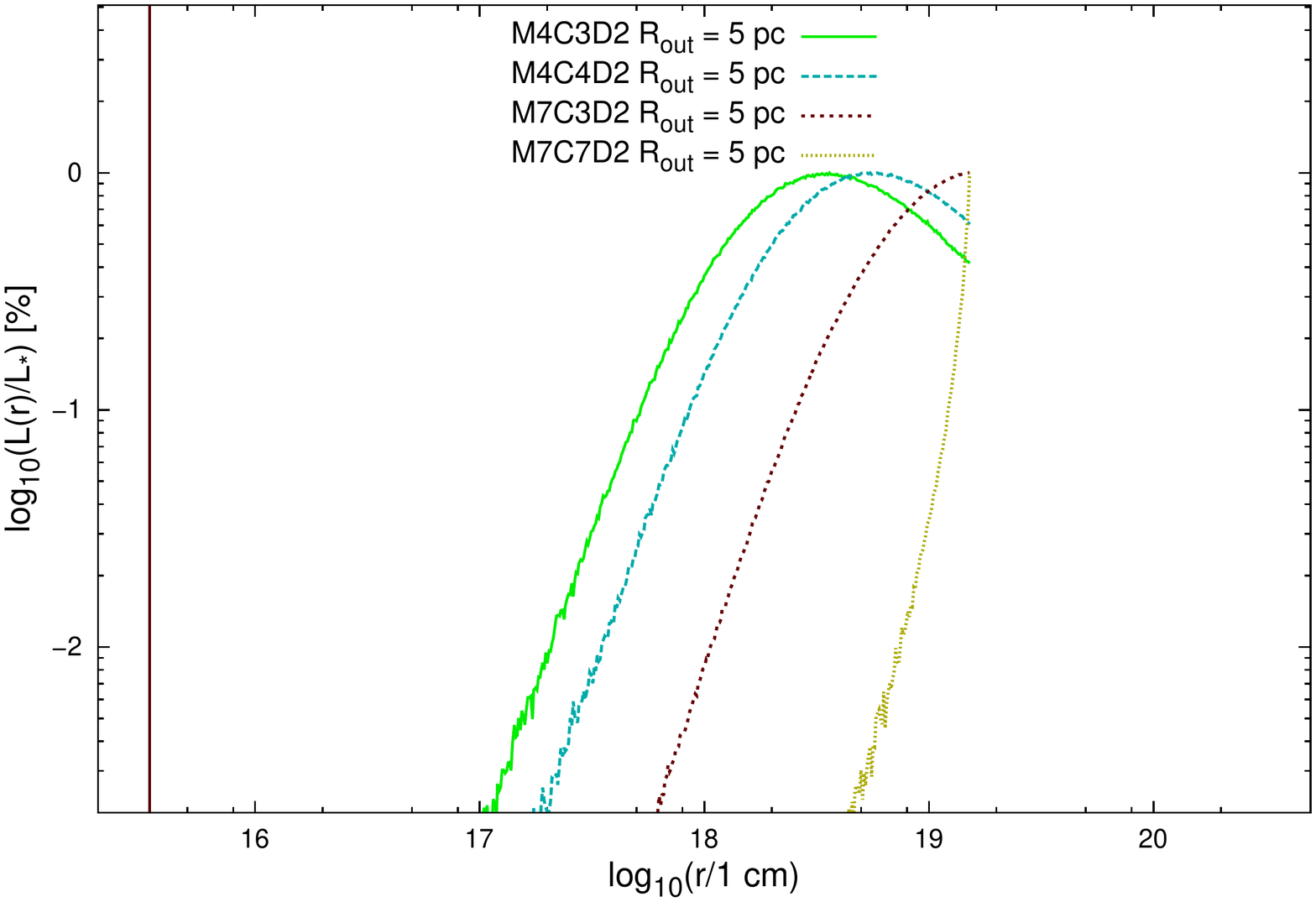} 
   \includegraphics[width=0.49\textwidth]{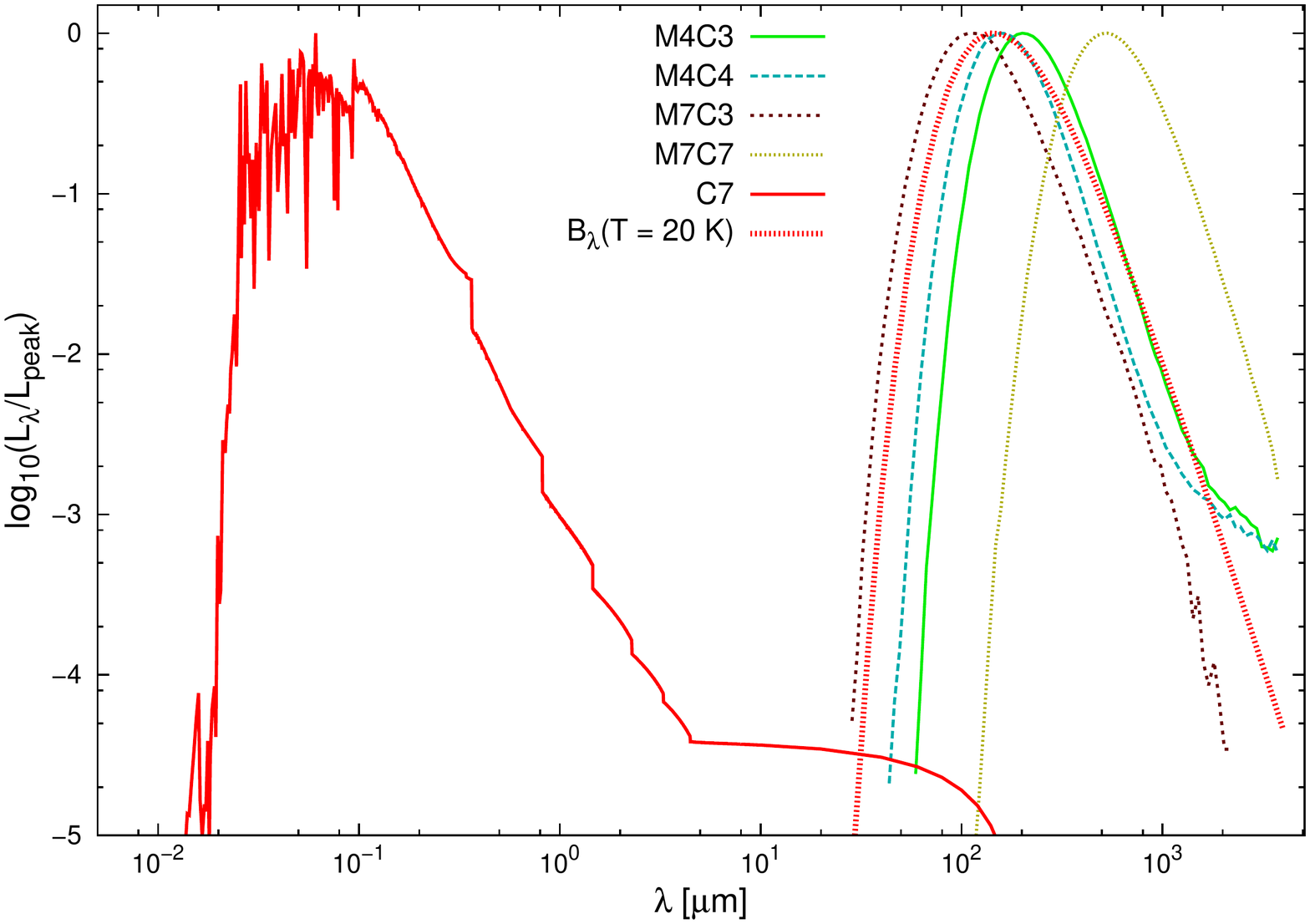}
  \end{center}
 \caption{Same as Fig. \ref{fig:ObservationsConst} for selected 
models with a radially dependent dust temperature $T_{d}(r)$ as shown in 
the middle panel of Figure \ref{fig:ForceExtendedCluster}. We note the shifted peak wavelengths in the different models.}
\label{fig:ObservationsTemp}
\end{figure*}

The rapid absorption of ultraviolet and optical photons by dust 
grains followed by their re-emission at longer wavelengths (Section
\ref{sect:SpectralShift}),  
has important observational consequences. The high efficiency 
of wavelength 
redistribution by dust is illustrated in Figure \ref{fig:RatioSpectrumShift}, 
which plots the probability distribution of the final wavelength of photon 
packages at the edge of the cloud as a function of their initial wavelength of 
emission from the central cluster. (The figure shows the photons 
emitted by the cluster with a wavelength between the far-infrared to
mm that escape the cloud without interaction.)
For simplicity we only show results of model ${\MM4}{\CC4}$ 
with standard grain size distribution ${\DD2}$ and a cloud radius of $5\,$pc, 
but we note that a similar result holds for all clouds and clusters considered 
here. As explained in Section\ref{sect:SpectralShift} an outside observer will 
see emission at sub-mm and mm wavelengths. The figure 
demonstrates that it is highly unlikely for ultraviolet, visible or 
near-infrared photons to escape the molecular cloud unaltered even for the lowest 
mass cloud with the least massive cluster inside. Thus, this finding holds even 
more strongly for higher mass molecular clouds. Further detail is provided in Figures \ref{fig:ObservationsConst} and 
\ref{fig:ObservationsTemp}, where we plot the frequency-averaged 
escape probability of photons as a function of cloud radius (left) and the 
spectrum emerging from the cloud (right). In Figure \ref{fig:ObservationsConst} 
we show a range of models. For ${\MM4}$ and ${\MM7}$ cluster models we take the  
fiducial case with standard MRN dust (${\DD2}$) and fixed $T_{\rm d} = 20\,$K, 
and we consider a larger maximum grain size (${\DD4}$). For ${\MM6}{\CC6}$ we 
adopt the  standard radius of $5\,$pc as well as an inflated one of $150\,$pc. 
We see that for models with increasing dust grain sizes and cluster masses, 
photons escape mostly from the cloud outskirts with only small contributions 
coming from close to the central star cluster. The exception is the model 
${\MM6}{\CC6}$ with $R_{\rm{max}} = 150\,$pc where the cloud mass is distributed 
over a large volume leading to an decrease in optical depth. We emphasize that all these rather 
different models result in similar spectra in the case of a constant 
temperature. They peak at  $\lambda \approx 
145\,\rm{\mu m}$, corresponding to a modified blackbody with a temperature of 
$20\,$K, and in most cases, they show an extended tail of emission at mm 
wavelength. The distribution is more narrowly peaked than a single temperature 
Planck function, while it has an extended tail at larger wavelength, in 
particular for those models with dust where IR radiation can easily escape from 
the inner cloud layers. Consequently, the observable emission typically falls 
below 
the blackbody for wavelengths $\lambda \lesssim 1000\,\rm{\mu m}$ and lies 
above it for longer wavelengths. This is well explained by the relative 
contributions of scattering and absorption events as discussed in 
Section~\ref{sect:SpectralShift} and by the corresponding different optical 
depths at different wavelength. Assuming a constant dust temperature, this 
behavior is quite insensitive to the adopted cloud mass and size 
and independent of the star formation efficiency. \\
In Figure \ref{fig:ObservationsTemp} we show models complementary to the models 
presented in Figure \ref{fig:ObservationsConst} with $R_{\rm{max}} = 
5\,$pc and ${\DD2}$ dust. By applying a radially dependent dust temperature the 
observable spectrum is mostly determined by the properties of the (cold) dust in 
the outer layers of the cloud sufficiently far away from the cluster. Any 
spectral characteristics 
of the stars in the center is entirely veiled from an outside observer in our 
spherical, static approximation. We may only notice that the dust emission 
appears colder for more massive clouds since almost all of the radiation 
escapes from the outer layers. Furthermore, we caution that 
this insensitivity to the parameters of cluster and cloud is in part caused by 
our assumption of spherical symmetry. In reality, the cluster may be extended 
and have substructure while the cloud is likely to be clumpy and to exhibit a 
degree of porosity. This means that certain lines of sight allow for a deeper 
look 
into the cloud closer to the emitting stars, so that the resulting spectrum will 
correspond to a larger range of dust temperatures than the one dimensional model 
predicts and that it may even contain contributions from the central cluster 
itself at the late phases of the evolution (see also 
Section~\ref{sect:Limitations} below). 

The extended power-law tail at $\lambda \gtrsim 1000\,\rm{\mu m}$ occurs because, at these 
wavelengths, all our clouds with standard MRN dust size distribution
are essentially transparent (see left side of
Figure~\ref{fig:OpticalDepth} for the example of  
${\MM4}$ clouds). Therefore, every photon package produced at mm 
wavelength by thermal dust emission (with probabilities according to
Eq.\ \ref{eq:BW}) contributes to the local radiation field inside the
cloud and  
eventually to the spectrum seen by an outside observer. These photons neither 
get scattered nor absorbed. They are not in local thermodynamic equilibrium 
with the cloud material and simply add up in number to build up a non-thermal 
tail in the overall spectrum. This non-thermal 
tail results in the solid diagonal line seen in the upper right corner of 
Figure~\ref{fig:RatioSpectrumShift}. We note that the the right-hand side of 
the Figures~\ref{fig:ObservationsConst} and~\ref{fig:ObservationsTemp}, 
respectively, show the cluster spectrum in luminosity while 
Figure~\ref{fig:RatioSpectrumShift} represents the number of photon packages. 
Hence, the difference of contrast between the figures.

\subsection{Comparison to existing studies}
\label{sect:OtherWork}
The calculations presented here strongly suggest that radiation pressure acting 
on dust cannot be the dominant feedback agent terminating star formation and 
disrupting the parental cloud around young star clusters. It seems that
this process cannot significantly contribute to the self-regulation of stellar 
birth on either local cloud scales or on global galactic scales except under 
implausible assumptions about the maximum size of dust grains in molecular clouds.
This immediately raises the question of why other  studies arrived at
the opposite conclusion that radiation pressure on dust is  
important for regulating the star formation process, and how the models 
presented here differ from previous calculations in the literature. 
As outlined in Section~\ref{sec:Intro}, the effects of radiation pressure 
have been studied for a wide range of different environments ranging from 
galaxies as a whole down to scales of planetary systems. The models most similar 
to the calculations presented here are those of \citet{murray2010}, as well as 
\citet{thompson2008} and \citet{thompson2015}. These authors also studied 
the impact of star clusters of different masses in the centers of spherical 
clouds with varying properties. However, unlike us they arrived at the 
conclusion that radiation pressure on dust is the key to understand cloud 
disruption. The differences in our studies can be summarized as follows.
While we investigate clouds with Plummer-like radial density profiles for the 
calculation of the ratio between radiative pressure 
forces and gravitational attraction at any radius, the cited authors
assume that jets and winds from the  
cluster dominate the early evolution of the cloud and create an evacuated bubble with the 
swept up cloud material accumulated in a dense shell. They then study the impact 
of radiation illuminating the inner edge of this shell. However, because 
radiation pressure on dust is mostly a function of the total column density, it 
is unlikely that the difference in our results comes from  deviations in the 
radial density profiles of the clouds. 
 Sweeping material up into a shell moves the same column density of
    material to larger radii, reducing its 
    temperature, and thus its opacity. This reduces rather than
    increases the effectiveness of radiation pressure. Therefore, dynamical effects
    seem unlikely to assist in driving an outflow.  Furthermore, even in models
    such as shown in Figure~\ref{fig:ForceTemp} where there is a small range of
    radii in the inner regions with $\zeta > 1$, the resulting shell
    will quickly cease to be accelerated as it is moved outwards \cite[see, 
e.g.,][]{rahner2017} and
    its temperature corresponding drops. A similar argument holds for the cloud sizes and
masses, as well as for the luminosity of the embedded clusters. As 
mentioned in Section~\ref{sct:Result3DStellarDistribution}, our suite of models 
includes cloud and cluster parameters quite similar to those considered by 
\citet{murray2010}. As far as we can tell, their assumed dust model resembles 
our fiducial case with an MRN size distribution (model ${\DD2}$).
The divergence in results could potentially arise from different assumptions about 
the dust temperature. For example,  the temperature in the shell in the model of 
\citet{murray2010}  is about 100$\,$K throughout the entire evolution (see their 
Appendix A.3.2). This is larger than our fiducial choice of 20$\,$K, but 
comparable to the values we find close to the star cluster in the models 
discussed in Section \ref{sect:ResultsTempModel}. Recall, though, that we 
find that the dust temperature quickly drops even below 20$\,$K further out in 
the cloud. Nevertheless, we conclude that the dust temperature is not the 
primary cause of the difference between our models. Instead, we suspect that the main culprit is the different 
treatment of radiative transport. We employ the multi-frequency Monte Carlo RT 
code {\sc Polaris} (see Section~\ref{sect:RTModel}) with a detailed 
treatment of the microphysics of dust scattering and absorption 
(Section~\ref{sect:DustModel}). This allows us to correctly follow the 
frequency shift as radiation moves outwards through the cloud. 
    That this shift reduces the effective radiation pressure was first noted by
    \citet{Habing1994} in the context of outflows from cool giant stars, a
    physically analogous situation.
\citet{murray2010} follow a simpler approach and assume blackbody radiation at 
each radius of their shell with a typical temperature of order of $100\,$K  for 
their calculations of the optical depth (see their Appendix A.3.2). We tested 
the impact of an artificially enhanced dust temperature of $90\,$K and indeed 
find a stronger radiative force. However, gravity still dominates. We note 
in Sections~\ref{sect:SpectralShift} and~\ref{sect:Observations} that the true 
spectrum shows noticeable deviations from a single temperature Planck function. 
It is typically more narrow around the peak and exhibits an extended tail 
toward longer wavelengths above $\sim 1\,$mm. If one assumes that the local 
radiation field at each radius is described by a single-temperature blackbody, 
then one overestimates the optical depth and consequently the resulting 
radiative pressure force. A further effect that leans in the same direction, but is not
accounted for in our models, is that the radiation field effectively
acts as a rarefied fluid accelerating a dense fluid when it
accelerates the gas. The result is that drives strong Rayleigh-Taylor
instabilities that fragment the gas, opening paths for photons to
quickly escape the cloud, for example as it was modeled by
\citet{2012ApJ...760..155K}. They found that this effect alone might
preclude the disruption of clouds by radiation pressure. Although
their quantitative result has been called into question based on
calculations with an improved RT algorithm \citet{davis2014} which
suggested that cloud disruption might still be able to occur, the
qualitative effect remains important, and will combine with the
spectral shift we have modeled to further reduce the effectiveness of
radiation pressure. We conclude that a realistic assessment of the balance between gravity and 
radiative forces requires multi-frequency radiative transfer 
calculations that take the spectral shift toward longer wavelength into account 
consistently. By doing so, we find boost factors of the momentum transfer from 
multiple scatterings of order unity rather than several tens to a 
hundred as has been suggested in the literature. 

\subsection{Limitations of the model}
\label{sect:Limitations}
There are several ways that the 3D radiative transfer model with azimuthally symmetric density
   distribution 
presented here could be improved. First, we neglect the momentum input
from stellar outflows and winds. 
In the models of \citet{murray2010} or
\citet{rahner2017}, these dominate over all other feedback mechanisms including radiation pressure, and sweep up the inner parts of cloud into a shell.
However,
stellar jets and collimated bipolar outflows are likely to pierce through the cloud and deposit their energy and momentum at distances far from the star, so their ability to stop accretion and remove gas from the cluster is subject to debate in the literature \cite[e.g.][]{banerjee2007, matzner1999, offner2011}.
Spherical winds from young stars are more efficient in evacuating a bubble and 
sweeping up cloud material into a dense shell surrounding the central cluster 
\cite[see, e.g.,][]{rahner2017}. In any case, the total momentum transfer from 
radiation pressure on dust only depends on the column density through the cloud, 
and so it is independent of the radial density structure as long as dust and gas 
are well coupled and as long as the temperature profile remains the
same. As noted above, sweeping material farther from the star can even
   reduce its temperature. 
Second, we focus on the effects of radiation pressure acting only on
dust grains, which has been proposed to be the dominant feedback term  
\cite[e.g.,][]{murray2010, thompson2008}. 
We do not include the momentum transfer from the interaction of photons with gas \citep[see e.g][]{KimOstriker2016}, though we argue in Section~\ref{subsec:radiation} that this is indeed a good approximation, since it leads to errors of at most $\sim 10$\% for our most massive cloud and cluster models. Therefore we neglect ionization and dissociation as well as other chemical processes, except the destruction of dust grains close to the central star cluster (see Section \ref{sect:DustModel}).
The coupling between photons 
and dust grains only depends on the total column density of dust grains of a 
given size and the temperature along the line of sight, independent of the 
details of the radial density profile. We note in this context that the momentum 
transfer from ionizing radiation would also depend on the local density $\rho$, 
because the ionization degree of the material is a function of the recombination 
rate, which goes as $\rho^2$ \cite[e.g.,][]{tielens2010, draine2011}. 
Third, we neglect the effects of an external radiation field acting on the dust 
grains. Such a radiation field could alter our results in two ways. On the one 
hand, the temperature distributions shown in Figure~\ref{fig:TempDistribution} 
would decrease less steeply in the outer regions of the cloud since the external 
radiation field would also heat up the dust component. Depending on the mass of 
the molecular cloud model and the spectral distribution of the external 
radiation field, one might observe an outer shell with an increased dust 
temperature. In this case, the peak values of the output spectrum shown in 
Figures~\ref{fig:ObservationsConst} and~\ref{fig:ObservationsTemp}
would be shifted toward shorter wavelength. Since higher dust 
temperatures are associated with increased trapping of radiation, this could 
also result in a small increase in radiative force at the very outer edges of 
the molecular clouds. On the other hand, the external radiation field would also 
result in an additional force pushing on the outer shells of the molecular 
clouds from the outside. This effect would counterbalance the outwards acting radiative 
force and possibly even reverse the direction of the radiative force.  In the 
most extreme case, it could even lead to a radiation driven implosion 
\citep[e.g.,][]{klein1980,bertoldi1989}. 
Finally, we also neglect cloud dynamics, including Rayleigh-Taylor instabilities driven by the radiation
   pressure itself, 
and any deviations from spherical symmetry, such as filaments or
density fluctuations in the cloud material, or  
substructure in the stellar distribution.  However, since the radiative force on 
dust is a function of column density and temperature only, and since we are 
interested in a global average, we are confident that our calculations of 
$\zeta$, the ratio between gravitational attraction and radiation pressure, can 
answer the question of whether this form of feedback is able to disrupt star-forming clouds with high fidelity. 

\section{Summary and conclusion}
\label{sec:summary}
We have employed detailed, multi-frequency, Monte Carlo, RT 
calculations to study the impact of radiation pressure acting on dust on the 
evolution of star-forming clouds for a wide range of physical parameters, 
covering conditions ranging from the solar neighborhood up to massive starburst systems.   
In order to investigate the competition between outward radiative
forces and inward gravitational attraction, 
   we calculated the 3D propagation of radiation emitted from point or
   extended central clusters embedded in azimuthally
   symmetric density distributions.
Our RT method correctly calculates all the 
contributions of backscattering radiation to the total net radiative
force.  This approach includes a detailed treatment of scattering and absorption 
processes when photons interact with dust grains, so we can follow the radial 
evolution of the local spectrum as it is absorbed and re-emitted at longer 
wavelengths. We have further self-consistently computed the temperature profile through
     our clouds in order to determine its effect on the effective
     radiation pressure, as well as computing constant temperature
     models for comparison to previous work.
Our model clouds are Plummer spheres with masses of 
$10^4$--$10^7\,$M$_{\odot}$ harboring central star
clusters also having mass $10^4$--$10^7\,$M$_{\odot}$,
corresponding to star formation efficiencies from 1\% to 91\%. We have studied the 
impact of different dust properties with maximum grain sizes up to $200\,\mu$m. 
We find that radiation pressure acting on dust is almost never able to disrupt 
star-forming  clouds. The overall momentum deposition in the 
clouds is small. This is because  ultraviolet and optical 
photons from young stars, to which the cloud is optically thick, do not scatter 
sufficiently often. Instead, they get absorbed and reemitted by the dust at 
thermal wavelengths.  These photons promptly escape, as the cloud is typically 
optically thin to  sub-mm and mm radiation. We conclude that 
radiation pressure acting on dust cannot disrupt star-forming molecular clouds 
in galaxies such as we see in the Local Group. This process is therefore unable 
to regulate the star-formation process on local as well as global scales, 
contrary to what has been proposed in the literature. We have calculated the resulting spectral energy 
distribution that an external observer would see when looking at the cloud to 
better link our models to observations. It
is more narrowly peaked than a single-temperature Planck function 
and exhibits an extended tail of emission at mm wavelengths. More specifically we find
\begin{itemize}
\item In clouds a MRN grain size
 distribution extending to maximum grain radius of 2~$\mu$m, 
the strength of gravitational attraction exceeds the force from radiation 
pressure on dust by 1--2 orders of magnitude. Similar results are
found for clouds with properties ranging from the Milky Way to
moderate starbursts. The momentum transfer from the stellar radiation 
field coupling to dust cannot disrupt star-forming clouds in most
galaxies.

\item We have varied the cloud size and the cluster size and find that the 
ratio of radiation pressure to gravity drops as cloud size increases or the cluster becomes more extended.
Only for maximum grain sizes  exceeding 2--20~$\mu$m, 
in combination with unrealistically high star 
formation efficiencies, well above 50\%, and point-like clusters
do we find that radiative forces can dominate over gravity
in clouds with masses exceeding $10^6 \mbox{ M}_{\odot}$.
\item We find that the number of scattering events for cluster photons is typically of order unity, so 
that the absorption-reemission process dominates photon transfer. Only under 
circumstances where the cloud is optically thin to infrared radiation, and so 
scattering becomes important at thermal wavelengths, is the momentum transfer 
from the radiation field onto the cloud material important in the force 
equation. However, this is only the case for 
unrealistically large grain sizes in very massive clouds.  

    \item
As the photon packages propagate outwards through the cloud, the 
original spectrum of the central cluster is quickly absorbed and replaced by 
thermal emission from dust. In our models with a fixed dust 
temperature of $20\,$K, the resulting spectrum peaks at $\lambda \approx 145\ 
\rm{\mu m}$. When we allow the dust temperature to 
consistently adjust to the local radiation field, we find that the peak 
wavelength of the local spectrum shifts to larger wavelengths as the dust 
temperature drops toward larger radii. The resulting
spectrum is somewhat narrower than a single-temperature blackbody and exhibits a power-law tail at wavelengths $\lambda \gtrsim 
1000\,\mu$m. 
\item In all cases 
considered here, the dust in the outer layers of the cloud drops below $20\,$K, 
so the approximation of a fixed value of $T_{\rm d} = 20\,$K provides
a good estimate of the resulting radiation pressure force.
\end{itemize}
Although our RT calculations are fully 3D, our 
model is limited in the sense that it treats the cloud and its embedded cluster 
in spherical symmetry only. Realistic clouds have filamentary morphology and a 
high degree of substructure that allows radiation to more easily escape along 
channels of low density.  \\
In that sense, our calculations constitute an upper limit on the importance of 
radiation pressure. On the other hand, our models neglect other potentially 
important feedback processes, such as stellar winds and supernovae. We also 
ignore ionizing radiation, which can destroy dust grains close to the cluster 
and form a hot bubble of ionized gas in the center. Taking these processes into 
account consistently would increase the outward forces acting on the cloud. In 
the current study, we focused only on one aspect of the overall problem, the 
importance of radiation pressure acting on dust. A full assessment of the effect 
of all stellar feedback processes in realistic three-dimensional cloud models 
remains to be done. 

\begin{acknowledgements}
We thank Xander Tielens for motivating this paper at the 2015 IMPRS Heidelberg 
Summer School. Furthermore, we thank Paul Clark, Sam Geen, and Simon Glover for 
stimulating discussions. 
S.R., E.P., and R.S.K. acknowledge support from the Deutsche 
Forschungsgemeinschaft in the Collaborative Research Center (SFB 881) ``The 
Milky Way System'' (subprojects B1, B2, and B8) and in the Priority Program SPP 
1573 ``Physics of the Interstellar Medium'' (grant numbers KL 1358/18.1, KL 
1358/19.2). RSK further thanks the European Research Council for funding in  
the 
ERC Advanced Grant ``STARLIGHT'' (project number 339177).  M.-M.M.L. was partly 
supported by NSF grant AST11-09395 and by the Alexander-von-Humboldt Stiftung.
\end{acknowledgements}

\end{document}